\documentclass{aa}  

\newcommand{\msun}{\ensuremath{M_{\odot}}}

\newcommand{\beq}{\begin{equation}}
\newcommand{\eeq}{\end{equation}}
\newcommand{\beqa}{\begin{eqnarray}}
\newcommand{\eeqa}{\end{eqnarray}}

\newcommand{\nbeq}{\begin{equation*}}
\newcommand{\neeq}{\end{equation*}}

\newcommand{\vel}{Vela~X-1\xspace}

\usepackage{graphicx}
\usepackage[varg]{txfonts}
\usepackage{xspace} 
\usepackage{hyperref}
\usepackage{siunitx}

\newcommand{\mcii}[1]{\multicolumn{2}{c}{#1}}
\newcommand{\los}{LOS\xspace}%
\newcommand{\muas}{\mbox{$\mu$as}}%

\usepackage{amstext}

\begin{document} 
 
\title{Revisiting the archetypical wind accretor Vela X-1 in depth}

\subtitle{Case study of a well-known X-ray binary and the limits of our knowledge}

   \author{P. Kretschmar
          \inst{1}
          \and
          I. El Mellah\inst{2,3,4}
          \and
          S. Mart\'inez-N\'u\~nez\inst{5}
          \and
          F. F{\"u}rst\inst{6}
          \and
          V. Grinberg\inst{7}
          \and
          A. A. C. Sander\inst{8}
          \and
          J. van den Eijnden\inst{9,10}
          \and
          N. Degenaar\inst{9}
          \and
          J. Ma\'iz Apell\'aniz\inst{11}
          \and
          F. Jim\'enez Esteban\inst{11}
         \and
          M. Ramos-Lerate\inst{12}
          \and
          E. Utrilla\inst{13}
}

   \institute{European Space Agency (ESA), European Space Astronomy Centre (ESAC), Camino Bajo del Castillo s/n, 28692 Villanueva de la Cañada, Madrid, Spain
         \email{Peter.Kretschmar@esa.int}
      \and
          Centre for Mathematical Plasma Astrophysics, KU Leuven, Celestijnenlaan 200B, 3001 Leuven, Belgium
      \and 
         Institute of Astronomy, KU Leuven, 3001 Leuven, Belgium
      \and 
          Institut de Plan\'{e}tologie et d'Astrophysique de Grenoble, 414 Rue de la Piscine, 38400 Saint-Martin-d'Hères, France
          \and
          Instituto de F\'isica de Cantabria (CSIC-Universidad de Cantabria), E-39005, Santander, Spain
      \and 
        Quasar Science Resources S.L for European Space Agency (ESA), European Space Astronomy Centre (ESAC), Camino Bajo del Castillo s/n, 28692 Villanueva de la Cañada, Madrid, Spain
      \and
      Institut f\"ur Astronomie und Astrophysik (IAAT), Universit\"at T\"ubingen, Sand 1, 72076 T\"ubingen, Germany.
      \and 
      Armagh Observatory and Planetarium, College Hill, Armagh BT61 9DG, Northern Ireland, UK
      \and 
      Anton Pannekoek Institute for Astronomy, University of Amsterdam, Science Park 904, NL-1098 XH Amsterdam, the Netherlands
      \and
      Department of Physics, Astrophysics, University of Oxford, Oxford, UK
      \and
      Centro de Astrobiolog{\'\i}a, CSIC-INTA, Campus ESAC, Camino bajo del castillo s/n, E-\num{28692}, Villanueva de la Cañada, Spain
      \and
      Vitrociset Belgium for European Space Agency (ESA), European Space Astronomy Centre (ESAC), Camino Bajo del Castillo s/n, 28692 Villanueva de la Cañada, Madrid, Spain
      \and
      Aurora Technology BV for European Space Agency (ESA), European Space Astronomy Centre (ESAC), Camino Bajo del Castillo s/n, 28692 Villanueva de la Cañada, Madrid, Spain
    }

   \date{Received 31 December 2020; accepted 22 April 2021}
 
\abstract
{The Vela~X-1 system is one of the best-studied X-ray binaries because it was detected early, has persistent X-ray emission, and a rich phenomenology at many wavelengths. The system is frequently quoted as the archetype of wind-accreting high-mass X-ray binaries, and its parameters are referred to as typical examples. Specific values for these parameters have frequently been used in subsequent studies, however, without full consideration of alternatives in the literature,  even more so when results from one field of astronomy (e.g., stellar wind parameters) are used in another (e.g., X-ray astronomy). The issues and considerations discussed here for this specific, very well-known example will apply to various other X-ray binaries and to the study of their physics.}
{We provide a robust compilation and synthesis of the accumulated knowledge about \vel as a solid baseline for future studies, adding new information where available. Because this overview is targeted at a broader readership, we include more background information on the physics of the system and on methods than is usually done. We also attempt to identify specific avenues of future research that could help to clarify open questions or determine certain parameters better than is currently possible.}
{We explore the vast literature for \vel and on modeling efforts based on this system or close analogs. We describe the evolution of our knowledge of the system over the decades and provide overview information on the essential parameters. 
We also add information derived from public data or catalogs to the data taken from the literature, especially data from the \textit{Gaia} EDR3 release.}
{We derive an updated distance to \vel and update the spectral classification for HD~77518. At least around periastron, the supergiant star may be very close to filling its Roche lobe. Constraints on the clumpiness of the stellar wind from the supergiant star have improved, but discrepancies persist. The orbit is in general very well determined, but a slight difference exists between the latest ephemerides. The orbital inclination remains the least certain factor and contributes significantly to the uncertainty in the neutron star mass. Estimates for the stellar wind terminal velocity and acceleration law have evolved strongly toward lower velocities over the years. Recent results with wind velocities at the orbital distance in the range of or lower than the orbital velocity of the neutron star support the idea of transient wind-captured disks around the neutron star magnetosphere, for which observational and theoretical indications have emerged. Hydrodynamic models and observations are consistent with an accretion wake trailing the neutron star.}
{With its extremely rich multiwavelength observational data and wealth of related theoretical studies, \vel is an excellent laboratory for exploring the physics of accreting X-ray binaries, especially in high-mass systems. Nevertheless, much room remains to improve the accumulated knowledge. On the observational side, well-coordinated multiwavelength observations and observing campaigns addressing the intrinsic variability are required. New opportunities will arise through new instrumentation, from optical and near-infrared interferometry to the upcoming X-ray calorimeters and X-ray polarimeters. Improved models of the stellar wind and flow of matter should account for the non-negligible effect of the orbital eccentricity and the nonspherical shape of HD~77581. There is a need for realistic multidimensional models of radiative transfer in the UV and X-rays in order to better understand the wind acceleration and effect of ionization, but these models remain very challenging. Improved magnetohydrodynamic models covering a wide range of scales are required to improve our understanding of the plasma-magnetosphere coupling, and they are thus a key factor for understanding the variability of the X-ray flux and the torques applied to the neutron star. A full characterization of the X-ray emission from the accretion column remains another so far unsolved challenge.}


\maketitle
%


\section{Introduction}\label{sec:intro}

\begin{figure}[hbt]
    \centering
\includegraphics[width=0.9\linewidth]{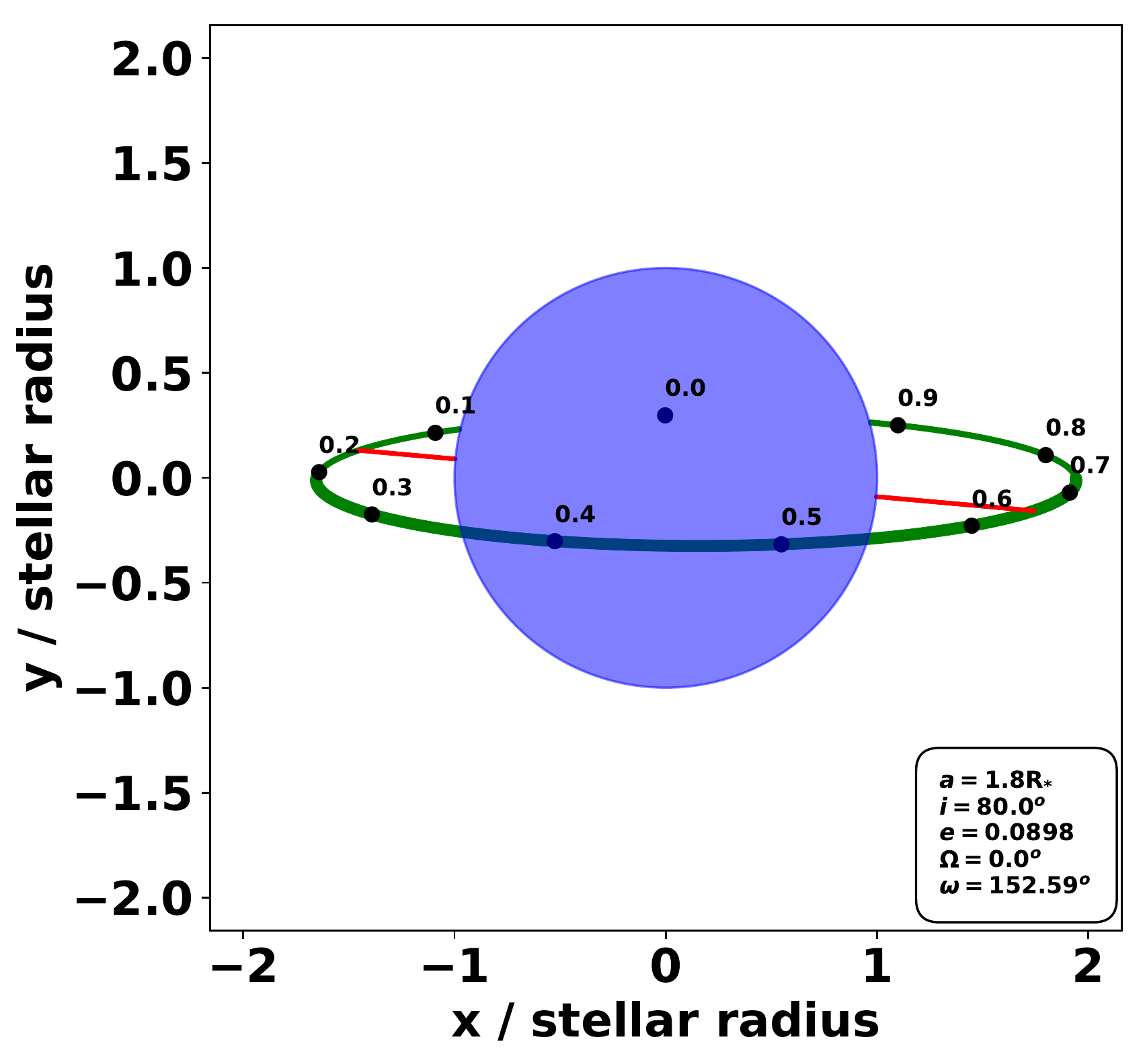}

\includegraphics[width=0.9\linewidth]{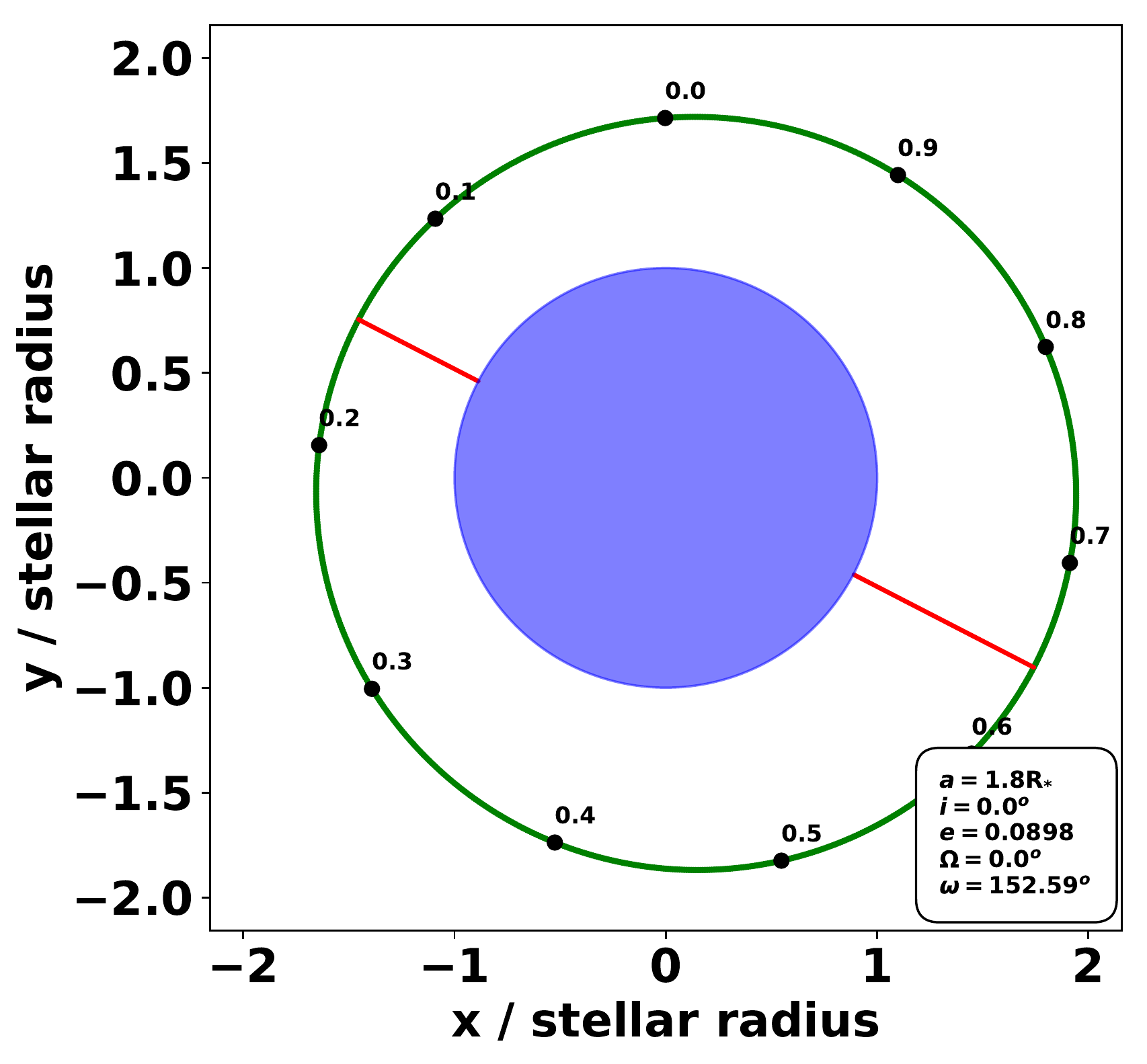}
    \caption{(top panel) Sketch of the Vela~X-1 system as seen from Earth based on the ephemeris of \citet{Kreykenbohm:2008}, using an intermediate assumed inclination within the known constraints (see Section~\ref{sec:system:mass} and Table~\ref{tab:masses}). Points mark the position of the neutron star at given orbital phases. (bottom panel) Top view of the system.}
    \label{fig:sketches}
\end{figure}

\object{Vela X-1} (4U\,0900-40) is an eclipsing high-mass X-ray binary (HMXB) at a distance of about 2~kpc (Section~\ref{sec:system:distance}). The binary system consists of an accreting neutron star that emits X-ray pulses with a period of $\sim$283~s (Sect.~\ref{sec:obs:xray:pulse}) in a close, mildly eccentric orbit with a period of $\sim$8.964~days (Section~\ref{sec:system:orbit}, Table~\ref{tab:orbit}) around the supergiant \object{HD 77581}, also known as GP~Vel, which is most frequently described as  B0.5~Ia supergiant. Because the orbital separation of $\sim 1.7 R_{\star}$ (Section~\ref{sec:system:orbit}) is small, the accreting neutron star is deeply embedded in the dense stellar wind of the supergiant star, with a mass loss of about $10^{-6} M_\sun\, \mathrm{yr}^{-1}$ (Sect.~\ref{sec:system:wind}). Figure~\ref{fig:sketches} illustrates the system geometry. In this figure and in the remainder of this paper, the orbital phases quoted are expressed with respect to the mid-eclipse time $T_\mathrm{ecl}$ defined in \citet{Kreykenbohm:2008} as zero-point; the different ephemerides used in the literature are discussed in Section~\ref{sec:system:orbit}. The orbital eccentricity is high enough for the orbital phase 0.5 to be reached significantly after inferior conjunction of the neutron star. This detail is sometimes ignored when the system with its ``almost circular'' orbit is discussed.

The intrinsic X-ray luminosity of Vela X-1 is only moderate \citep[$\sim 4\times 10^{36} \mathrm{erg~s}^{-1}$][]{Nagase:86}, but because of its proximity, it is still one of the brightest persistent point sources in the X-ray sky. Together with a wide variety of interesting phenomenology, this has made Vela~X-1 a well-studied target ever since its detection. It has frequently been called an ``archetype'' of wind-accreting HMXB.

Because of its eclipsing nature, the extent of the donor star with respect to the orbital separation is well constrained, and so is the mass transfer mechanism. Because the orbit of the slowly spinning neutron star around its blue supergiant companion is almost circular ($e\sim 0.0898$), the stellar line-driven wind provides a continuous source of material to the compact object. The sources of stochastic variability of the intrinsic X-ray emission induced by, for instance, overdense small-scale regions in the wind (hereafter called clumps), can be isolated from the orbital modulation. The cooperative behavior of Vela X-1 from an observational and theoretical point of view has made it a privileged target of study for modelers, as further detailed in Sect.~\ref{sec:models}.

As an example of a high-mass neutron star, with a mass that has been determined to be clearly higher than the Chandrasekar limit (Sect.~\ref{sec:system:mass}), Vela~X-1 is also frequently used as an example for models of compact stars and equation-of-state calculations or studies of neutron star mass distributions \citep[e.g.,][]{Gangopadhya:2013,Ozel+Freire:2016,MafaTakisa:2017,Gedela:2019}. Over the years, however, the knowledge about many system parameters has evolved, and different authors have used different conventions to report their results. Even more importantly, quite different assumptions about essential parameters of the system have been used by different authors in the interpretation of their observations or in their modeling, for instance, the velocity profile of the stellar wind, the nature and shape of larger structures, clumps and their parameters, or the role of ionization.

In this article we revisit a large sample of observational findings. We treat the information in a consistent manner as far as possible, and we compare the observations with information obtained from recent detailed modeling, especially of the wind structure \citep{Sander:2018} and of the flow of matter around the neutron star \citep{El-Mellah:2018}. The remainder of this paper is structured as follows:
Section~\ref{sec:elements} summarizes the different elements of the Vela~X-1 system in order to describe the framework in which observational diagnostics and modeling studies take place.
Section~\ref{sec:obs} gives an overview of the observational diagnostics at different wavelengths and describes which parts of the system they explore.
Section~\ref{sec:models} describes semianalytical and numerical models of Vela~X-1 or close analogs, and each time we pay particular attention to how different physical ingredients are handled.
Section~\ref{sec:system} summarizes the knowledge about basic system properties, as found in the literature, with an emphasis on clearly describing the differences between the results obtained in different studies.
In Section~\ref{sec:future} we give our view on how further observations and modeling efforts may help us to answer the open questions and add to our knowledge of this prototype system.
Finally, Section~\ref{sec:summary} summarizes the essential points we raised throughout this overview.

\section{Elements of the Vela X-1 system}
\label{sec:elements}

The rich observational phenomenology of an HMXB system such as Vela~X-1 is the result of a complex interplay of its various elements, where the relevant physics cover more than six orders of magnitude in size, as indicated in Figure~\ref{fig:radii}. In the following we describe these elements from larger to smaller size scales. For the sake of clarity, we introduce the main physical concepts relevant to this and similar systems when we discuss observations
and modeling efforts in later sections. For a more complete discussion of stellar wind and HMXB physics see, for example,  \citet{Martinez-Nunez:2017}. 

\subsection{Supergiant and its stellar wind}
\label{sec:elements:starwind}

HD~77581 looses significant mass through a line-driven wind whose launching mechanism, first identified by \cite{Lucy+Solomon:1970} and \cite{CAK:75}, relies on the resonant line-absorption of UV photons by partly ionized metal ions. The amount of mass lost and the intrinsic acceleration by the radiative field of the supergiant evidently strongly depend on parameters such as temperature, luminosity, radius, mass, and effective surface gravity, which are in themselves not perfectly well known; see  Sections~\ref{sec:system:mass} and~\ref{sec:system:wind}. One major factor of uncertainty is that the effective temperature of the supergiant star (about 25,000\,K, as found in most studies; see Table~\ref{tab:stellarpar}), lies in a sharp transition called the bistability jump between slow winds on the cool side and fast winds on the hot side \citep{Vink:1999}. 

In a spherically symmetric configuration, theoretical arguments and observational insights support a velocity profile for line-driven winds that follows a so-called $\beta$-law \citep{PulsVinkNajarro:2008},
\begin{equation}
\label{eq:beta-law}
    v(r)=v_{\infty}\left(1-\frac{R_{\star}}{r}\right)^{\beta}
,\end{equation}
with $R_{\star}$ the stellar photosphere radius, $v_{\infty}$ the terminal wind speed, and $\beta$ a positive exponent: its value determines how quickly the wind reaches $v_{\infty}$, with a more gradual acceleration for a higher value of $\beta$. 
In the case of Vela X-1, quite different values for $v_{\infty}$ ($<$400 to $\sim$1700~km s$^{-1}$) and $\beta$ (0.8 to $\sim2.2$) have been reported by different authors, and the usefulness of a $\beta$-law as a description has been questioned (Section~\ref{sec:system:wind}).

In line-driven winds, internal shocks are prone to develop because of the line-deshadowing instability \citep{Lucy+White:1980,Owocki+Rybicki:1984}. The local density departs from the smooth wind value, with an overall density ratio of $\sim$100 between the most and least dense regions at a given distance from the donor star. In the case of Vela~X-1, current work (see Section~\ref{sec:models:wind} for details and references) suggests that at one stellar radius above the photosphere, most of the wind mass is contained in clumps. This wind clumping has often not been taken into account for discussions of the overall wind structure in the system. The internal shocks in line-driven winds that are responsible for the clump formation also produce X-rays, although the X-ray luminosity due to this process is low ($\sim$10$^{33}$erg s$^{-1}$).
The sizes of clumps in stellar winds and in HMXBs have been estimated with widely varying ranges in the past \citep[see the corresponding discussion in][]{Martinez-Nunez:2017}. Recent simulations for hot and massive stars \citep{DriessenSundqvistKee:2019} suggest that most of the wind mass is contained in clumps that are expected to appear as coherent structures of mass 10$^{17-18}$\,g and of a size of about 1\% of the stellar radius. This is about $10^{10}$\,cm for HD~77581.

The presence of the orbiting neutron star strongly affects the outflowing stellar wind through its gravity, the X-ray emission, and the orbital movement. \citet{Bessell:75} first highlighted the anisotropic distribution of material in the orbital plane induced by  tidal forces, in agreement with the observational indications for a trailing wake that was reported in the early papers on Vela X-1 \citep[see references in][]{Conti:78}. As further described in Section~\ref{sec:models:wind}, simulations of the Vela~X-1 system (or close proxies) find a wind focused toward the neutron star, forming an unsteady bow shock that leads to an accretion wake. This wake trails the neutron star throughout its orbit \citep[e.g,][]{Blondin:91,ManousakisWalterBlondin:2012}.

The line-driven acceleration of the stellar wind is affected by the intense X-ray emissions that are emitted from the immediate vicinity of the neutron star. This modifies the ionization structure of the wind in which it is embedded and thus the ability of the stellar radiative field to accelerate the wind through resonant line-absorption \citep{HatchettMcCray:77}. An additional ionizing contribution, which is expected to be very low for Vela~X-1, however, may come from the X-rays that are produced in shocks in the wind. 
The local degree of ionization in an optically thin wind in thermal balance can be evaluated using the ionization parameter \citep{Tarter:69},
\begin{equation}
\label{eq:ionization}
    \xi=\frac{L_\text{X}}{n r_{\text{X}}^2}
,\end{equation}
where $L_\text{X}$ is the X-ray luminosity, $n$ the local atomic number density of the gas, and $r_{\text{X}}$ the distance to the X-ray point source. Above a certain critical ionization parameter $\xi_\text{crit}$, most of the elements responsible for the wind acceleration (e.g., C, N, O, and Fe) are expected to be fully ionized and the wind acceleration is essentially quenched. Depending on the shape of the irradiating X-ray spectrum and on the details of the resonant line-absorption mechanism, the value of $\xi_\text{crit}$ is expected to range between 10$^2$ and 10$^4$erg$\cdot$cm$\cdot$s$^{-1}$. Shocks between the radiatively driven wind and stagnant gas around the compact object can in principle lead to a trailing accretion wake \citep{Fransson+Fabian:1980}, but for Vela~X-1 and similar systems, this is expected to be a lesser contribution to the gas density \citep{Blondin:90}. 
It is important to note, however, that the ``on/off switch'' for wind acceleration set by the critical ionization parameter is a simplifying assumption, as is further explained in the models that we present in section\,\ref{sec:ion_struct}. Moreover, wind clumping weakens the effect of X-ray ionization from the accreting compact object through the increased recombination inside the clumps \citep{OskinovaFeldmeierKretschmar:2012}. 

Another large-scale feature that is common around massive hot stars are corotating interaction regions \citep{Mullan1984}. None has been reported in Vela~X-1, however. 

\subsection{Mass transfer to the neutron star}
\label{sec:elements:mass_transfer}

A fraction of the mass lost by HD~77581 is captured and accreted by the neutron star and fuels the intense X-ray emission. While the system is generally labeled a ``wind accretor'', the exact mechanisms of mass transfer in the Vela~X-1 system are not fully determined and have again become a topic of more intense research; see Section~\ref{sec:models:mass_transf}. 
Uncertainties remain on the orbital inclination, and to a lesser extent, on the mass ratio (Section~\ref{sec:system:mass}), therefore it can currently not be excluded in principle that HD~77581 would fill its Roche lobe and thus lose mass through Roche-Lobe overflow (RLOF).
However, the moderate X-ray luminosity of Vela X-1, the long and
erratically varying spin period (Section~\ref{sec:obs:xray:pulse}), the
absence of signatures for a permanent accretion disk, and the stable
orbital period \citep{Falanga:2015} make an RLOF scenario very unlikely (see Section\,\ref{sec:models:mass_transf}).

\begin{figure}[htb]
\centerline{\includegraphics[width=1.0\linewidth]{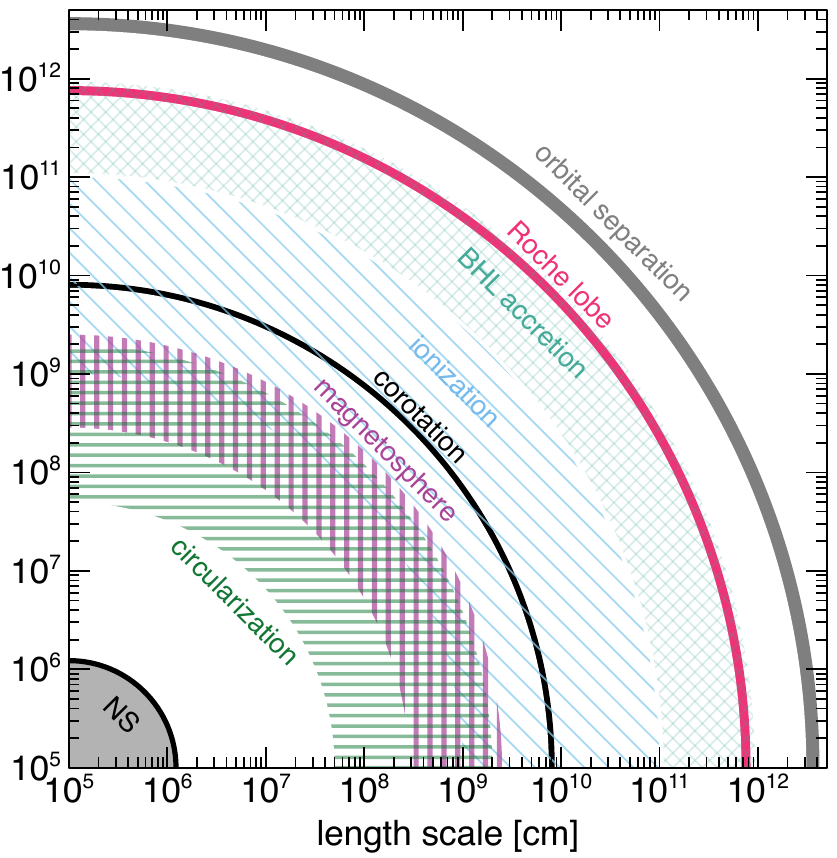}}  
\caption{Overview of characteristic length scales in the Vela~X-1 system. Some of them are relatively well determined, on the order of 10\%, and can be considered effectively fixed over the timescale of known observations. Other scales, shown with different types of hatching or shading, strongly depend on different parameters, as explained in the text. They may vary on short timescales of sometimes over an order of a magnitude or more, and a variation in one may drive changes in another. The ranges shown in this figure for these parameters are only to be taken as indicative. For comparison, the orbital separation corresponds to approximately 1/4~AU.}
\label{fig:radii}
\end{figure}

It is therefore generally assumed that the main mass transfer in Vela X-1 occurs through gravitational capture from the stellar wind.
The length scale at which the wind is beamed toward the accretor by its gravitational field (on the order of the Roche-lobe radius of the neutron star) is much larger than the extent of the neutron star itself or its magnetosphere (see Figure\,\ref{fig:radii}). In the fast-wind limit, the so-called Bondi-Hoyle-Littleton (BHL) formalism \citep{HoyleLyttleton:1939,BondiHoyle:1944,Edgar:2004} provides a robust framework. In this description, the fraction of the wind with an impact parameter lower than the accretion radius $R_\text{acc}$ will be captured,
\begin{equation}
\label{eq:R_acc}
    R_\text{acc}=2GM_\text{NS}/v_\text{rel}^2    
,\end{equation}
where $M_\text{NS}$ is the mass of the neutron star and $v_\text{rel}$ is the relative speed of the wind with respect to the neutron star. In this case, the mass accretion rate is approximately given by
\begin{equation}
\label{eq:BHL_mdot}
    \dot{M}_\text{acc}=\frac{4\pi \rho\left(GM_\text{NS}\right)^2}{v_\text{rel}^3}
,\end{equation}
where $\rho$ is the wind density at the orbital separation. 
Subsequently, empirical refinements were proposed to correct $v_\text{rel}$ for slower winds \citep{DavidsonOstriker:73}. Still, the mass accretion rate is very sensitive to $v_\text{rel}$ , and to a lesser extent, to $\rho,$ and a wide range of X-ray luminosities can be derived from smaller changes in these parameters, which are only relatively loosely constrained from observations and modeling (Sections~\ref{sec:models:wind} and~\ref{sec:system:wind}). In addition, the geometry of the problem itself might differ from the planar picture of the BHL framework because orbital effects are strong enough in Vela X-1 to significantly bend the flow to a point that it could form a disk-like structure before it reaches the magnetosphere.

\subsection{Accretion close to and within the magnetosphere}
\label{sec:elements:magnetosphere}

The mass transfer rate is an upper limit on the mass accretion rate at the neutron star surface that eventually produces the observed X-ray luminosity. A fraction of the captured material could either stall in intermediate regions or might in principle be ejected, although there are no reports on outflows from close to the X-ray source in Vela~X-1 so far (but see Section~\ref{sec:obs:radio}). When the flow reaches the magnetosphere, it is highly ionized by the intense X-ray flux that is emitted from the immediate vicinity of the neutron star surface. It behaves as a plasma, and its dynamic is entirely determined by the neutron star magnetic field, which channels the material all the way down to the neutron star magnetic poles \citep{DavidsonOstriker:73}.

For a qualitative understanding of the different possible accretion regimes within the magnetosphere, the following length scales play a role. First, the corotation radius which is the distance at which the neutron star magnetosphere, which rotates with the neutron star spin period, has the same angular velocity as the local Keplerian velocity. The mass, size, and spin period of the neutron star are fairly constant over timescales shorter than years, therefore the corotation radius can be considered as fixed. Second, the magnetospheric radius\footnote{For the sake of simplicity, we ignore the distinction between magnetospheric and Alfv\'{e}n radii.} which is the distance to the neutron star below which the magnetic field dominates the dynamics of the inflowing plasma. A rigorous evaluation of this radius depends on the geometry of the inflow, but it is typically estimated by comparing the magnetic pressure to the ram pressure such that it increases with decreasing mass accretion rates. Other factors such as mass, radius, and magnetic field strength are again taken to be essentially fixed. Third, the circularization radius which is the radius of the Keplerian orbit that has the same specific angular momentum as the orbit of the accreted flow. Semianalytic and numerical estimates exist to evaluate the upper limit of the specific angular momentum of the inflow, but they vary by more than an order of magnitude between each other \citep{IllarionovSunyaev:75,Shapiro:76,Wang:81,Livio:86}.

%
%
%

Different estimates exist for the angular momentum accretion rate, but the upper limit commonly reads
\begin{equation}
    l=\frac{1}{2}\Omega R_\textrm{acc}^2
,\end{equation}
where $\Omega=2\pi/P$ is the angular orbital speed and $P$ is the orbital period. We can then deduce the circularization radius,
\begin{equation}
    R_\textrm{circ}=\frac{1+q}{4} \left(\frac{R_\text{acc}}{a}\right)^4 a
,\end{equation}
where $q$ is the ratio of the mass of the donor star to the mass of the neutron star, and $a$ is the orbital separation. With realistic values for the orbital separation, the mass ratio, and the accretion radius, we obtain circularization radii in Vela X-1 that range between $\sim 5\times 10^{7}$ cm and $\sim 2\times 10^{9}$ cm for wind speeds at the orbital separation that range between 600 and 300 km$/$s. 

In the case of pure wind mass transfer, the flow carries a negligible amount of angular momentum, and the circularization radius is much smaller than the other two length scales (fast-wind case). A bow shock forms ahead of the neutron star with an opening angle smaller for higher Mach numbers of the inflow \citep{ElMellah:15}. The flow is deflected by the shock, and provided its impact parameter is smaller than the accretion radius, it returns to the neutron star, essentially from the back hemisphere \citep[for simulations of BHL accretion onto a magnetic dipole, see][]{Toropina:12,Aaron:14}. If the mass accretion rate is high enough, the flow efficiently cools down downstream of the shock through bremsstrahlung or Compton processes, and the material freefalls at supersonic speeds onto the neutron star magnetosphere \citep{Burnard:83}. This regime is highly unsteady and is referred to as the direct accretion regime. Otherwise, the shock is adiabatic and the flow forms a quasi-spherical shell of hot material that subsonically settles onto the neutron star magnetosphere \citep{Davies:81,Shakura:2012}. This regime has ramifications for different models depending on the intensity of the plasma-magnetosphere coupling \citep[referred to as strong, moderate, and weak coupling in][]{Shakura:2012,Shakura:18}. In general, the plasma penetrates the magnetosphere through the interchange instability, which is the magnetohydrodynamics counterpart of the Rayleigh-Taylor instability \citep{Arons:76a}. For a more comprehensive description of the different accretion regimes onto a neutron star magnetosphere in wind-fed HMXBs in general, see \citet{Bozzo:08} and \citet{Martinez-Nunez:2017}.

\begin{figure}[htb]
\centerline{\includegraphics[width=0.95\linewidth]{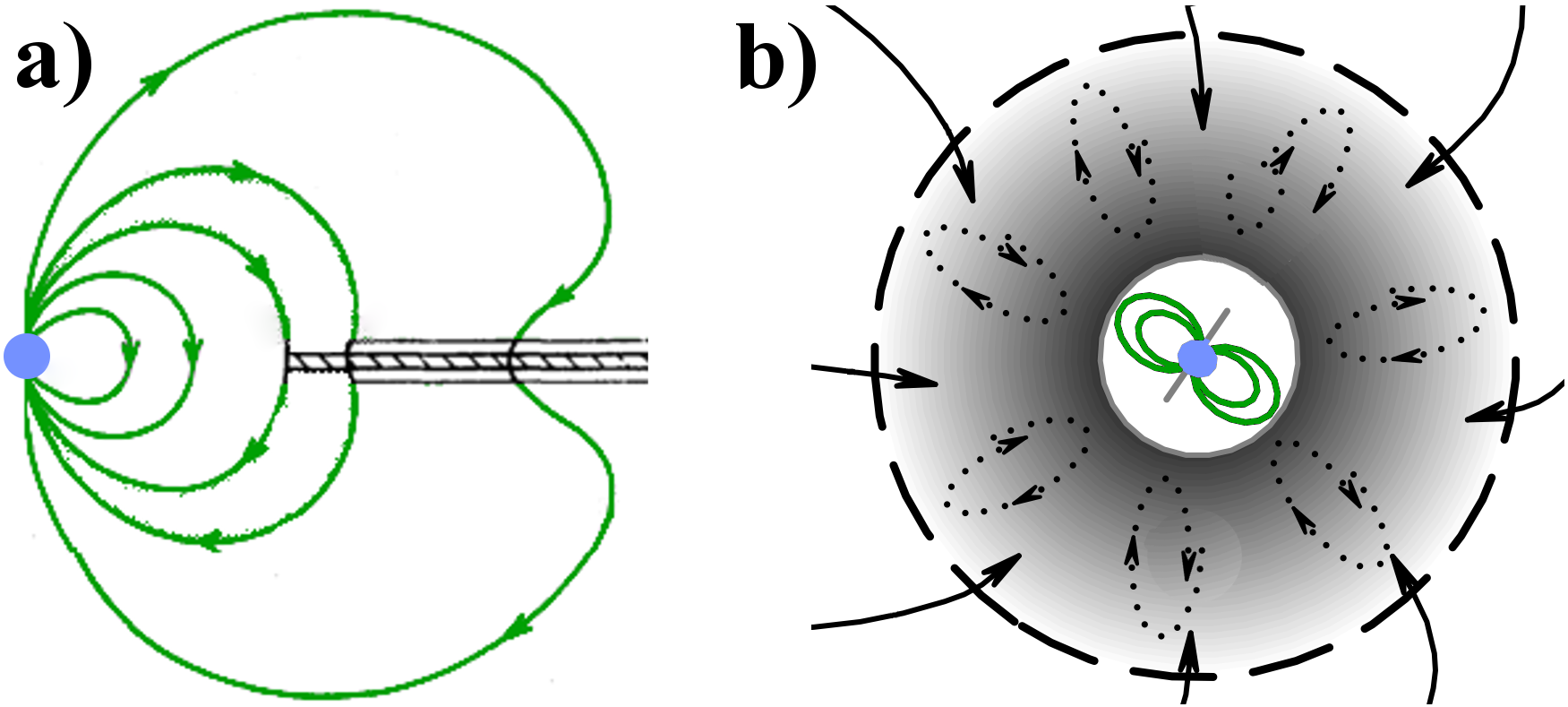}}
\caption{Disk (a) and quasi-spherical (b) accretion geometries at the outer edge of the neutron star magnetosphere (represented in green as a dipole), adapted from \cite{GhoshLamb:78} and \cite{Shakura:2012}, respectively. In each of these ideal geometries, different accretion regimes are expected according to semianalytic derivations, depending on how the plasma couples to the magnetic field.}
\label{fig:accr_geom}
\end{figure}

For slower winds, the wind-RLOF mass transfer mechanism becomes more realistic. The bow shock extends up to a significant fraction of the Roche lobe of the neutron star ($\sim$10\%). As detailed in Section~\ref{sec:models:mass_transf}, in this regime, the accreted flow acquires and retains enough angular momentum through the shock so that the circularization radius is larger than the magnetospheric radius and a centrifugally maintained structure, called a wind-captured disk, can form within the shocked region. A disk like this would still be truncated at its inner edge by the magnetosphere, which is a few hundreds times larger than the neutron star radius, and thus such a disk would not radiate sufficiently at higher energies to be detected. Indirect evidence has emerged in recent years, however, for the existence of transient disks in wind-fed HMXBs in general \citep{Hu:17,Taam:18,Taani:19} and for in Vela X-1 in particular \citep{Liao:2020}.
In this configuration, the disk and the magnetosphere are coupled by magnetic reconnection, the Kelvin-Helmholtz instability, and Bohm diffusion \citep{Ghosh:79}. The ratio of the corotation radius and the magnetospheric radius then controls accretion and the transfer of angular momentum between the neutron star and the accreted matter. In the most simple cases, plasma that rotates faster at the magnetosphere than at the magnetic field lines that are anchored in the neutron star can dissipate kinetic energy, accrete, and spin the neutron star up, while at the other extreme, in the propeller regime, plasma that rotates more slowly than the magnetic field lines at the magnetosphere is expected to be expelled by the centrifugal force at the magnetocentrifugal barrier. This reduces the angular momentum of the neutron star \citep{IllarionovSunyaev:75}. Intermediate cases are discussed in \citet{Ghosh:79} and many publications based on this work \citep[see][for an overview and application to different sources]{Bozzo:2009}. Strong torques for spin-up and spin-down can also be transmitted in quasi-spherical accretion \citep{Shakura:2012}. A comparison of disk and quasi-spherical model predictions for various accreting pulsars is presented in \citet{Malacaria:20}. The two geometries are sketched in Figure\,\ref{fig:accr_geom}, and the implications for Vela~X-1 are further discussed in Section~\ref{sec:models:acc-ind_torq}.

\subsection{Accretion column and X-ray emission}
\label{sec:elements:column}

When the plasma has coupled to the magnetic field, it is funnelled to the magnetic poles of the neutron star where it forms a polar cap or accretion columns and emits most of the X-ray emission we observe \citep{DavidsonOstriker:73,Lamb:73,Shapiro:75,Arons:87}. No hysteresis is expected in the sense that a given mass accretion rate will correspond to a specific X-ray luminosity (averaged over the spin period): The energy acquired by the flow in its journey from the stellar photosphere to the neutron star surface has to be released, except when there are significant outflows from the immediate vicinity of the neutron star. This has not been observed at this point. When the spin and magnetic axis of the neutron star are not aligned, which is frequently the case, regular pulses can be observed at the rotation frequency of the neutron star \citep[see, e.g., the catalog of][where 66 of 114 HMXBs in the Galaxy have an identified pulse period]{Liu+vParadijs+vdHeuvel:2006}. The intensity and shape of the observable pulse profile, that is, the distinct pattern of the light curve folded with the pulse period, results from a complex interplay of multiple factors. The emission itself is expected to be nonuniform and is often highly beamed \citep{BaskoSunyaev:75,Mushtukov:2018b}. The emitted radiation is therefore strongly affected by relativistic light that in the curved space time bends close to the neutron star \citep[e.g.,][and references therein]{Meszaros+Riffert:88,Riffert+Meszaros:88,Kraus:95,Odaka:14,Sotani:18,Falkner:2018PhD}. Close to the neutron star surface, nondipolar components of the magnetic field might also have a significant effect on the dynamics of the plasma \citep{Petri:15,Petri:17,deLima:2020}. Even when the magnetic field is purely dipolar, there could be many regions on the neutron star surface to which the plasma is funnelled, and they do not necessarily correspond to the magnetic poles \citep{Romanova:04}.


The emitted X-ray spectrum of the column is dominated by Compton scattering of thermal seed photons. Cyclotron emission and scattering processes play an important role as well \citep[][and references therein]{Schwarm:2017a,Schwarm:2017b}. 
A substantial fraction of accreting X-ray pulsars show relatively broad absorption line features in their X-ray spectra \citep[see][for a recent review]{Staubert:2019}. These features result from resonant scattering of X-ray photons on electrons, whose energies perpendicular to the magnetic field are quantized into so-called Landau levels. This causes the the plasma to become optically thick for X-ray photons at these energies and triggers the formation of cyclotron resonant scattering features (CRSFs) or often simply cyclotron lines. The spacing of these lines is determined by the energy difference between adjacent Landau levels, leading to centroid line energies of
\begin{equation}
    \label{eq:Ecyc}
     E_{\mathrm{CRSF},n} = \frac{n}{1+z} \frac{\hbar e B}{m_e c} \approx \frac{n}{1+z} 11.6 \times B_{12}\,\mathrm{keV}
,\end{equation}
where $n$ is number of Landau levels involved in the scattering, $z$ is the gravitational redshift due to the neutron star mass, $e$ and $m_e$ are the electron charge and mass, $B$ is the magnetic field in the scattering region, $c$ is the speed of light, and $B_{12}$ is the magnetic field strength in units of $10^{12}$~Gauss. Determining cyclotron line energies thus gives a relatively direct measure of the magnetic field strength of the neutron star. CRSF studies for Vela~X-1 are discussed in Section~\ref{sec:obs:xray:cyclo}, and the implications for the neutron star magnetic field are summarized in Section~\ref{sec:system:B-field}.

Despite significant efforts and some progress, as further detailed in Sections~\ref{sec:models:column} and~\ref{sec:models:CRSF}, there is currently no self-consistent, physics-based description that can correctly describe the spectra of accreting X-ray pulsars under all circumstances.


\section{Observational diagnostics}\label{sec:obs}

In this section, we review the different methods which were used to constrain the parameters in Vela X-1. We treat the different wavelength domains separately.

\subsection{X-ray diagnostics} \label{sec:obs:xray}

\subsubsection{Continuum spectrum} \label{sec:obs:xray:spectrum}

The X-ray luminosity of \vel is dominated by the X-ray continuum produced in the accretion column. For a distance of 2\,kpc, the median luminosity is between 4--5$\times10^{36}$\,erg\,s$^{-1}$ in the 20--60\,keV energy band \citep{Nagase:86, Fuerst:2010}. This approximately translates into a bolometric luminosity of about $1\times10^{37}$\,erg\,s$^{-1}$. However, the \vel luminosity varies strongly \citep[e.g.,][]{Kreykenbohm:2008} and follows a log-normal distribution \citep{Fuerst:2010}. Figure~\ref{fig:fluxhistos} (top) shows the distribution of the observed hard flux, which is a good tracer of the intrinsic luminosity. Even on the day-long timescales sampled by \textsl{Swift}/BAT, large flares of 5--11$ \text{ times}$ the average luminosity are regularly observed. 

The X-ray emission of Vela~X-1 outside the eclipse can be phenomenologically described, as in many other accreting X-ray pulsars, by a power-law shape with a rollover at higher energies, beyond 15--30~keV \citep[e.g.,][]{Nagase:86,Kreykenbohm:99,Orlandini:98-VelaX1,Odaka:2013}. However, the exact shape of this rollover is unknown. In the simplest approach, it can be modeled with an exponential cutoff alone, but observations with a high signal-to-noise ratio at energies beyond 30\,keV often require a more complex shape of the rollover, and different phenomenological models have been used in the literature \citep[e.g.,][]{Orlandini:98-VelaX1,Odaka:2013}.
The continuum is modified at lower energies by variable amounts of absorption (see Section\,\ref{sec:obs:xray:flux}), a significant iron fluorescence line with an occasional iron edge \citep{Nagase:86} and a frequent thermal soft excess below $\sim$3~keV \citep{Sato:86PASJ,Pan:94,Haberl:94a}, as well as further fluorescence lines from elements such as silicon, magnesium, and neon \citep[e.g.,][]{Sako:99, Goldstein:2004}.
At higher energies, the continuum is modified by CRSFs between 25~keV and 30~keV and at $\sim$55~keV, as further detailed in Sect.~\ref{sec:obs:xray:cyclo}. 

The overall continuum slope and the CRSFs vary strongly with the phase of the X-ray pulse \citep[e.g.,][]{Kreykenbohm:2002,LaBarbera:2003}, but within the scope of this paper, we do not discuss this further.

During the eclipse, the residual X-ray emission shows no pulsations and is dominated by line emission in addition to a flat continuum \cite{Nagase:94,Schulz:2002}.

In general, the X-ray flux observed from Vela~X-1 varies significantly in a largely random pattern around the mean value when the clearly visible X-ray eclipse is excluded. These variations are driven by variations in the intrinsic X-ray emission and by strong variations in the absorbing material between observers and the neutron star, especially for instruments in the soft X-ray band.

\begin{figure*}[htb]
\centerline{\includegraphics[width=0.86\linewidth]{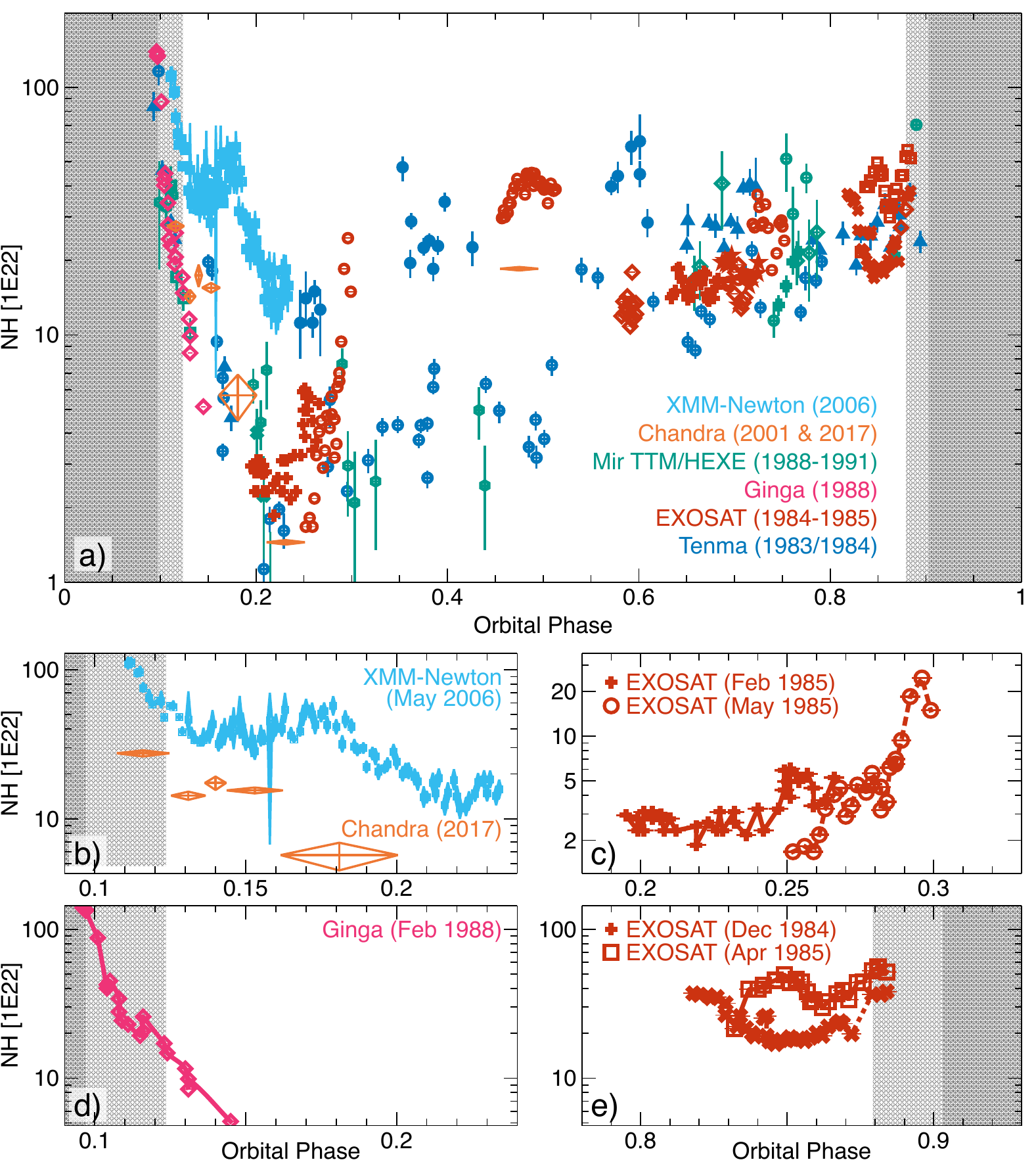}}
\caption{Variations in derived absorbing column density ($N_\mathrm{H}$ / cm$^2$) throughout the binary orbit of Vela X-1 as determined by various observations and different satellites. See the text for caveats when absolute values in $N_\mathrm{H}$ are compared. The shaded gray areas mark the eclipse (dark gray) and the ingress and egress (light gray) as determined by \citet{Sato:86PASJ}. Binary phase values have been corrected for the differences in phase-zero definition (using $T_\mathrm{ecl}$) and orbital period relative to \citet{Kreykenbohm:2008}. The uncertainty in this conversion is 1--4$\times 10^{-3}$ in phase, i.e., usually within the symbol size. Care has been taken to separate measurements taken during different binary orbits by the same instrument and team. Where possible, this is marked by different symbols. The data have been taken from \citet[Fig.~5 (\textit{Tenma})]{Sato:86PASJ}, \citet[Fig.~2 (\textit{EXOSAT})]{HaberlWhite:90}, \citet[Fig.~2 (\textit{Ginga})]{Lewis:92}, \citet[Fig.~3 (\textit{TTM/HEXE})]{Pan:94}, \citet[Tab.~2 (\textit{Chandra} 2001)]{Watanabe:2006},  \citet[Fig.~7 (\textit{XMM-Newton})]{Martinez-Nunez:2014} and \citet[Tab.~1 (\textit{Chandra} 2017)]{Liao:2020}. Panel a) gives an overview of the overall distribution, demonstrating the large scatter of results at intermediate orbital phases and that the larger structures driving $N_\mathrm{H}$ are not stable from one orbit to the next \citep[see also][]{Malacaria:2016}. Panels b) to e) show subsets of the data in more detail to better visualize short-term variations. In all these panels, the X-axis covers the same range, and the Y-axis covers the same factor between minimum and maximum.}
\label{fig:nh_orbit}
\end{figure*}

\subsubsection{Eclipse timing} \label{sec:obs:xray:eclipse}
Despite the sparse observations of
the time, it was soon realized in the early years after the detection that Vela~X-1 was a variable source   \citep[and references therein]{Giacconi:72-UhuruCat}. Based on long observations with the \textsl{OSO-7} UCSD X-ray telescope, \citet{Ulmer:72} found evidence for periodic flux variations with phases of zero flux at a period of $8.7\pm0.2$~d and proposed that the X-ray source might be a member of an eclipsing binary system. This result was confirmed, and the periodicity was refined to $8.95\pm0.02$~d by \citet{Forman:73}, based on \textsl{Uhuru} data. In subsequent years, the eclipse timing and derived orbital parameters were refined by various authors, for example, by \citet{Ogelman:77}, using 20 days of \textsl{COS-B} data, or by \citet{WatsonGriffiths:77}, using \textsl{Ariel~V} data covering 17 binary cycles, see Table~\ref{tab:orbit}. Most recently, \citet{Falanga:2015} combined eclipse times derived from \textsl{INTEGRAL} observations with the earliest data to investigate a possible evolution of the orbital parameters (Sect.~\ref{sec:system:orbit}).

\subsubsection{Variations in the X-ray absorption} \label{sec:obs:xray:abs}
Early observations of Vela~X-1 \citep{Eadie:75,WatsonGriffiths:77,Becker:78} already found indications for significant and variable absorption in the system, together with brightness variations. Because of instrumental limitations, the brightness variations could not always be clearly ascribed to variations in absorbing material or in the intrinsic X-ray flux. 
Later observations with improved X-ray continuum spectra allowed better measurements of the actual absorption and found strong variability on many timescales.  In \textit{Tenma} data, \citet{Ohashi:84} found large variations of the absorbing column density $N_\mathrm{H}$ that ranged from 2 to $50\times 10^{22}$~cm$^{-2}$ on a timescale of about an hour.

Over the years, very many observations have among other properties followed the temporal evolution of the absorbing column density in the Vela~X-1 system; see Figure~\ref{fig:nh_orbit} and Table~\ref{tab:observations_xray}. 
When they are presented in the same figure as a function of the orbital phase, different sets of $N_\mathrm{H}$ measures can be misleading because they often do not belong to the same orbit and because the column density at a given orbital phase is not stable from one orbit to the next. The overall behavior can be summarized by the following statements. First, from orbital phase $\phi_\mathrm{orb}\sim0.1$ (eclipse egress) to $\sim0.25$, $N_\mathrm{H}$ generally decreases strongly, often by more than one order of magnitude. This is usually explained by suggesting that the X-ray source shines through the extended atmosphere of its companion. Occasionally, shorter deviations from this trend have been found \citep{Nagase:86,Martinez-Nunez:2014}. These are most probably caused by absorbing material closer to the X-ray source. Second, between $\phi_\mathrm{orb}\sim0.25$ and $\sim0.6,$ strong variations of $N_\mathrm{H}$ on timescales of hours to days are observed. At any given phase, high and low absorbing column densities have been found in different binary orbits, sometimes using the same instrument and analysis method \citep[e.g.,][]{Nagase:86}. This also applies to $\phi_\mathrm{orb}\sim0.5$, where \citet{HaberlWhite:90} and \citet{Watanabe:2006} found high $N_\mathrm{H}$ values, which the latter modeled as broad cold cloud, but where \citet{Nagase:86} found an intermediate minimum, as shown in Figure~\ref{fig:nh_orbit}. In a systematic study of MAXI data, \citet{Malacaria:2016} found that $\sim 15$\% of all orbital light curves showed a dip around $\phi_\mathrm{orb}\sim 0.5$, which was best explained by Thomson scattering in an extended and ionized accretion wake. Third, for late orbital phases ($\phi_\mathrm{orb}>0.6$) up to the eclipse ingress, the observed absorbing column densities always remain high, but are still significantly variable by a factor of a few.

The following caveats apply when results from different publications on the absolute values of $N_\mathrm{H}$ are applied. First, spectral modeling results are usually somewhat degenerate for the parameters defining the spectral slope (e.g., a power-law index) versus the amount of absorbing material. The impact of this depends on the energy range that is covered, on the spectral response of each detector, and on the assumptions about permitted models in the fit procedure. For historical missions, it is frequently impossible to repeat the data reduction and analysis, even if time and knowledge were available for lack of suitable data in archives. Second and similarly, the inclusion or exclusion of an assumed soft excess can have a strong effect on the derived $N_\mathrm{H}$. Third, while many publications have used a single, unique absorbing column to obtain their results, in some particularly detailed studies, multiple absorption components were required to fit all of the available spectra \citep{HaberlWhite:90,Martinez-Nunez:2014}. In this case, one absorbing component may have significantly higher $N_\mathrm{H}$ values, but this only affects a fraction of the total observed radiation. Fourth, as detailed in \citet[][Sect.~4.4.1,]{Martinez-Nunez:2017} various different absorption models have been published in the literature and used over the years by different authors. These models rely on different assumptions for the interstellar medium cross-sections and abundances of the different elements, which can lead to differences of more than 20\% in the obtained $N_\mathrm{H}$ .

%
%

\begin{figure}
    \centering
    \includegraphics[width=1.0\linewidth]{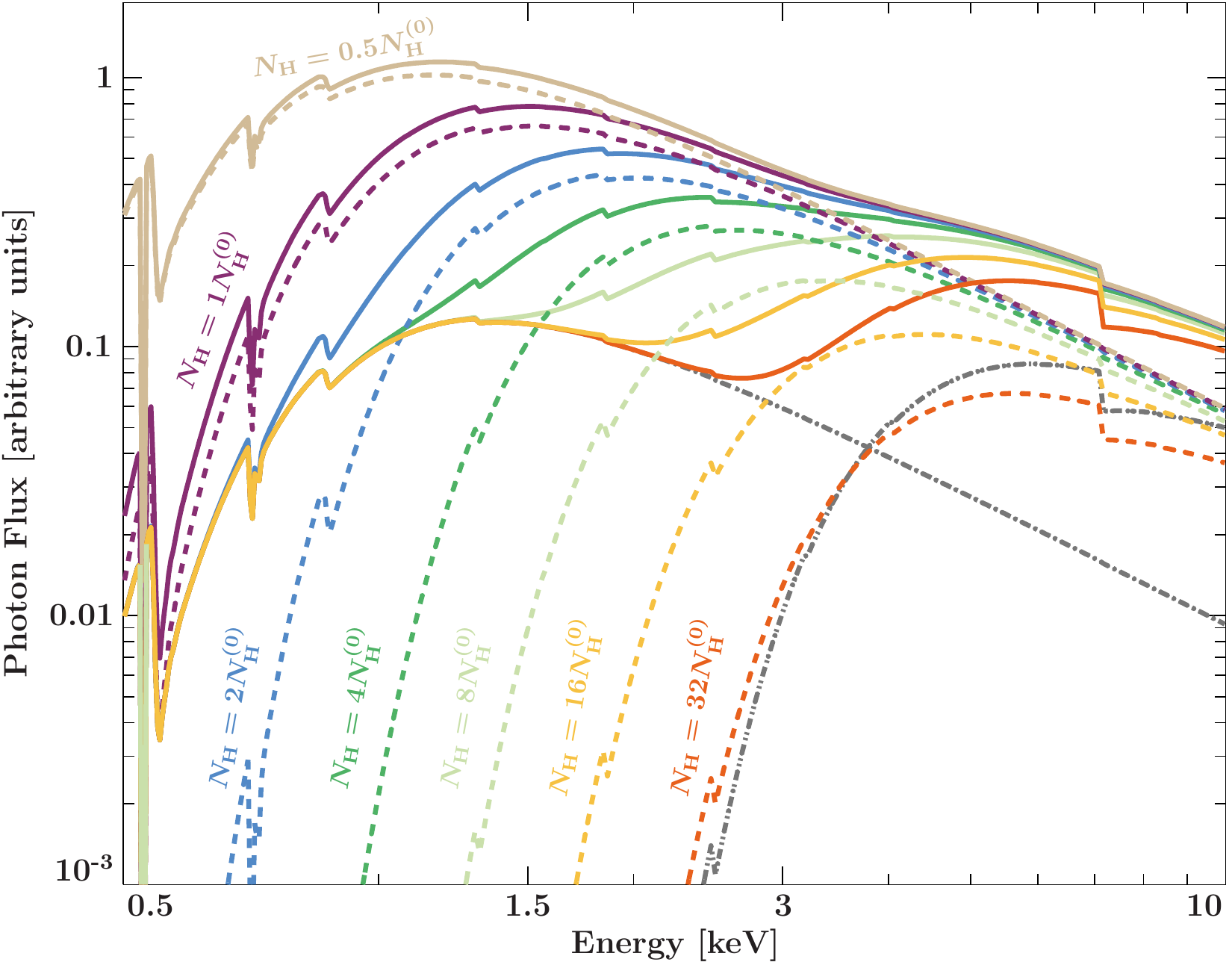}
    \caption{Effect of variable absorption on the overall shape of the spectrum of \vel in the 0.5--10\,keV range. Following \citet{Martinez-Nunez:2014}, we assume a spectral model consisting of three power-law components with the same slope, but different normalizations, each absorbed by a different absorption component. In accordance with trends seen by \cite{Martinez-Nunez:2014}, we then vary the total amount of absorption that the second of these power-law components experiences. The dot-dashed gray line shows the first power-law component. The dashed lines show the second power-law component, and different colors represent different absorption strengths, as indicated in the plot. The dot-dot-dashed line shows the third power-law component. The sum of the three components, i.e., the total spectrum, is shown as solid lines, and colors again represent different absorption levels for the second component.
    \label{fig:absvar}}
\end{figure}

We show the effect of variable absorption on the spectrum of \vel in Fig.~\ref{fig:absvar}. In particular, we demonstrate that even a combination of relatively simple continuum models can lead to complex variability in spectral shape. To do so, we used the model that was introduced to describe the 0.5--10\,keV spectrum of \vel by \citep{Martinez-Nunez:2014}, namely a sum of three power-law components that have different normalizations, but the same spectral index. Each of the power laws is absorbed by a distinct absorption component. A possible physical interpretation of this model is a complex partial coverer, possibly with an additional soft excess contribution \citep[e.g.,][]{Grinberg:20}. \citet{Martinez-Nunez:2014} found that the main driver for the variable absorbed spectral shape of \vel during a flaring episode was the absorption of the second, most prominent power-law component. Fig.~\ref{fig:absvar} is calculated for typical values of the power-law index and relative power-law component contributions. It clearly shows how changes in absorption in this component can lead to strong nonlinear variability in the overall spectral shape.

\citet{Grinberg:2017} have explored a different temporal and spatial scale. In order to test the effect of clumps in the wind on short-term variations in the absorbing column density, they compared data from a \textsl{Chandra} observation of Vela~X-1 at orbital phases $\sim$0.21 to $\sim$0.25 with a toy model that consisted of a constant X-ray source and a clumpy wind, based on the simulations by \citet{Sundqvist+Owocki+Puls:2018}. They found that undisturbed inhomogeneities in the model wind could not account for the observed enhanced absorption \citep[see also][for an extended model]{ElMellah2020_NH_paper}, indicating that more complex factors such as shock-clump interaction, the possibility of transient disk structures, or the effect of the compact object and its X-ray emission on the wind itself must be included. 

\subsubsection{X-ray fluorescence lines and radiative recombination continua}\label{sec:obs:xray:lines}

Narrow spectral features such as lines, edges, and radiative recombination continua (RRCs) are imprinted onto the X-ray continuum by the material surrounding the neutron star. They can in turn be used to constrain the properties of the surrounding (wind) plasma, such as ionization and density structure.

Especially the iron lines in the spectral region of ~6--7~keV, the so-called Fe complex, constitute a fundamental tool for probing the physical conditions of the material in the close vicinity of the X-ray source \citep{George:1991}. In the spectra of accreting X-ray binaries, these lines are prominent because of their intrinsic X-ray brightness and ubiquitous stellar material. In \vel, iron lines were first reported by \citet{Becker:78} and were studied by several teams since then, including studies with \textsl{Chandra} and \textsl{XMM-Newton} \citep[see, e.g.,][]{ Goldstein:2004,Watanabe:2006,Torrejon:2010,Gimenez-Garcia:2016,Grinberg:2017}. Fe~K$\alpha$ and K$\beta$ have been detected with multiple instruments and are narrow and usually unresolved, even with the highest currently available resolution using the \textsl{Chandra} transmission gratings \citep[e.g.,][]{Watanabe:2006,Goldstein:2004,Grinberg:2017,Amato2021}. The line energy is consistent with either neutral or lowly ionized ions, below $\sim$\ion{Fe}{xii}. Three reprocessing sites have been suggested based on observational trends, such as deviations from curve of growth for spherical geometry. All three likely contribute to some extent: the stellar wind as a whole, the companion atmosphere, and cold material in the vicinity of the neutron star, for example, in the accretion wake \citep[e.g.,][]{Sato:86PASJ,Watanabe:2006}.

A plethora of other narrow features, in particular, multiple fluorescent lines from different instruments, have already been detected  with ASCA \citep{Sako:99} and were resolved in detail   using grating instruments on board \textsl{Chandra} and \textsl{XMM-Newton}. Detections include prominent lines in both emission and absorption, edges, and RRCs of sulfur, silicon, magnesium, neon, and oxygen at different ionization levels \citep{Schulz:2002,Goldstein:2004,Watanabe:2006,Grinberg:2017,Lomaeva2020,Amato2021}. The strength of different features can change with orbital phase \citep{Goldstein:2004,Amato2021}, with the observed amount of absorption \citep{Grinberg:2017}, and with changing irradiation through the neutron star, for example, during a flare \citep{Lomaeva2020}. While the total normalization of the spectrum changes dramatically between eclipse and $\phi_\mathrm{orb} \approx 0.5$, striking similarities between the relative strength of fluorescence lines have been seen in two \textsl{Chandra}/HETG observations at these orbital phases covering the same orbit \citep{Sako:2003, Watanabe:2006}. This behavior allows us to rule out not only a smooth wind model, but also several simple clump distributions, such as a population of identical clumps \citep{Sako:2003}. 

Multiple modeling approaches have been employed to describe the observed spectra. \citet{Sako:99} determined the differential emission measure distribution at $\phi_\mathrm{orb} \approx 0$  from modeling contributions of individual ions and from comparison to theoretical emission measure distribution for a neutron star immersed in the stellar wind that are parameterized by wind parameters (mass loss and velocity) alone. They can obtain a good description of the observed spectrum and constrain the mass loss. However, their model predicts a lower ionization at $\phi_\mathrm{orb} \approx 0.5$ that has subsequently not been observed; the discrepancy is due to the assumption of spherical symmetry, which is clearly not present in \vel \citep{Sako:2003}. \citet{Lomaeva2020} and \citet{Amato2021} attempted to describe \textsl{XMM-Newton}/pn and \textsl{Chandra}/HETG high-resolution spectra, respectively, using photoionization models and comparing different photoionization models, namely XSTAR, CLOUDY, and SPEX/pion. While the fits are satisfactory, the residuals clearly point toward the existence of more than one component, in agreement with the assumption of a multiphase, clumpy medium. Their results clarify that a more sophisticated treatment of the system is necessary. \citet{Watanabe:2006} performed a quantitative modeling and spectral analysis using \textsl{Chandra} HETG during three orbital phase ranges within a single binary orbit in 2001: $\phi_\mathrm{orb} \approx 0.25$, $\phi_\mathrm{orb} \approx 0.5$, and $\phi_\mathrm{orb} \approx 0$ (during eclipse). Their aim was to constrain the ionization parameters and the geometrical distribution of the material in the X-ray irradiated stellar wind. They computed the ionization structure of a wind whose density and temperature distribution is given by a simple empirical model. With a Monte Carlo simulation of X-ray photons propagating through the wind, they computed synthetic X-ray spectra and established a self-consistent radiative-hydrodynamics picture of the system, but assuming a smooth and not a clumpy wind. For a more thorough discussion of simulations of the plasma properties in the \vel system of these and other authors, we refer to Sec.~\ref{sec:ion_struct}.



\begin{figure}
    \centering
    \includegraphics[width=1.0\linewidth]{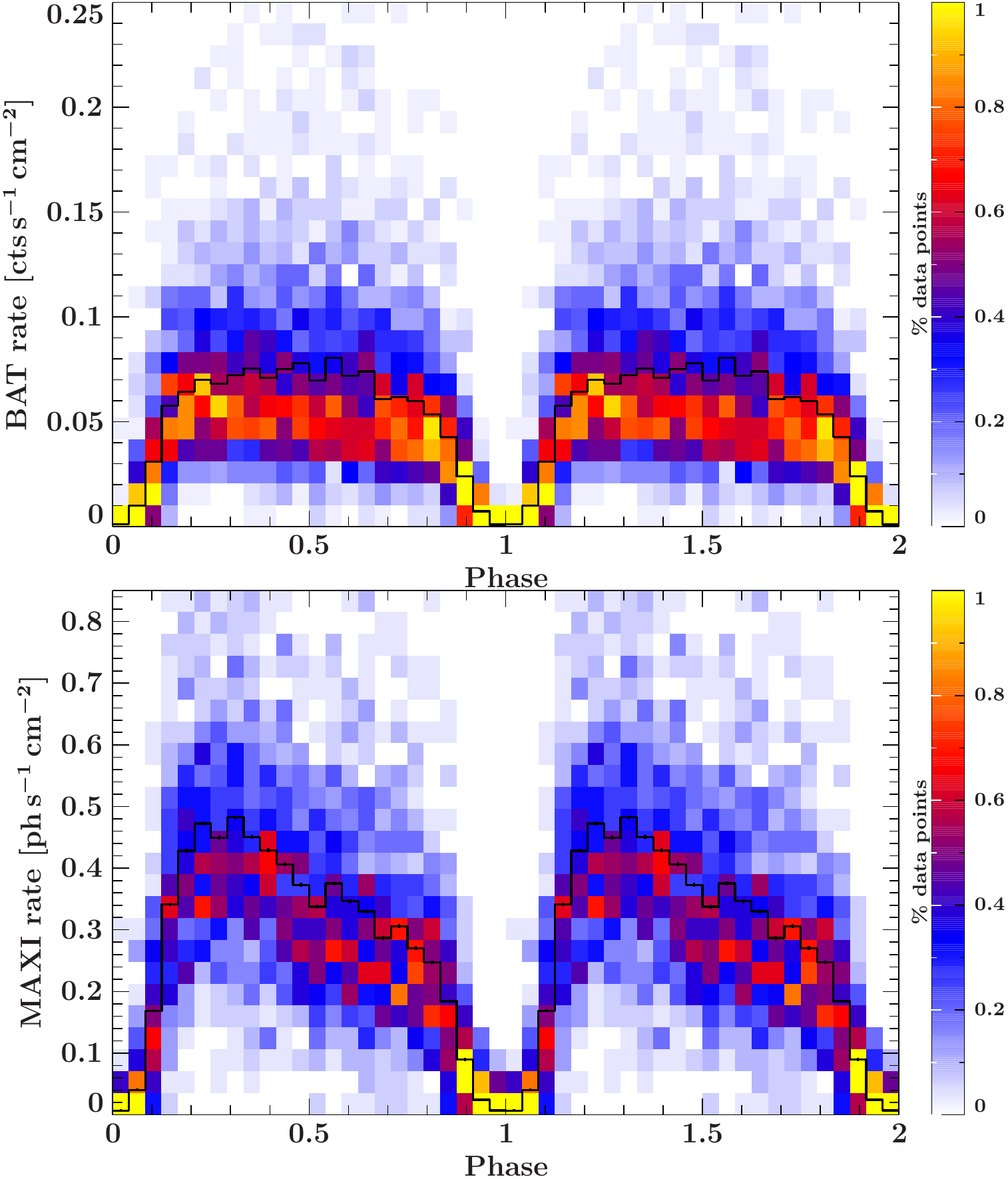}
    \caption{Brightness distribution of Vela~X-1 throughout its orbit as measured with \textit{Swift}/BAT (15--50\,keV, top panel) and MAXI (2--20\,keV, bottom panel). In comparison, 1~Crab corresponds to values $\sim$0.22 for \textit{Swift}/BAT and $\sim$4 for MAXI. We used the full history for each instrument, binned to a resolution of 1\,d. The mean orbital flux profile is superimposed in black. The probability for a measurement to fall into the respective histogram bin is color-coded according to the color scales on the right. The profile is repeated once for clarity. For details, see text.
    \label{fig:fluxhistos}}
\end{figure}

\subsubsection{Overall flux and continuum variations} \label{sec:obs:xray:flux}
In addition to the effect of absorption variations, any longer observation of \vel has found significant variations of the source flux on many different timescales, hours or tens of minutes \citep[see, e.g.,][]{HaberlWhite:90,Kreykenbohm:2008}, down to changes from one X-ray pulse to the next \citep[e.g.,][]{Inoue:84,Boerner:87}. There are many examples of flares, with timescales ranging from a single pulse to several hours \citep[e.g.,][]{WatsonGriffiths:77,Boerner:87,Odaka:2013,Martinez-Nunez:2014}. 

Off-states outside eclipse with no or very little detectable X-ray emission have been reported by various authors \citep{Inoue:84,Lapshov:92,Kreykenbohm:99,Kreykenbohm:2008,Doroshenko:2011,Sidoli:2015}. These states have been explained by different combinations of possible low-density zones or accretion that is choked or gated by magnetospheric effects (see Section\,\ref{sec:models:variability}). One caveat for a comparison of different publications is that the effective sensitivities of the instruments used and the definitions of ``off'' vary significantly.

\citet{Fuerst:2010} used data taken in the hard X-ray band (20--60~keV) with INTEGRAL for a 
systematic study of the brightness distribution in an energy range that is not affected by photoabsorption
below $\sim3\times10^{23}\,\mathrm{cm}^{-2}$. They found a log-normal distribution of the X-ray
brightness and that the power spectral density of the light curve follows a red-noise power law. They suggested that a mixture of a clumpy wind, shocks, and turbulence can explain the measured distribution. 

The strong variability of Vela~X-1 can be seen in the orbital-phase-resolved histograms shown in Fig.~\ref{fig:fluxhistos}. While the average orbital profile is stable and well defined, large variations around that mean value are also present. This is indicative of large flares and low states. Notably, the hard \textsl{Swift}/BAT profile is much more symmetric between eclipses than the softer one as measured by MAXI, which indicates a strong rise in absorption (on average) at later orbital phases.
The difference between the mean value and the median that is apparent through the color-coding is expected for a log-normal flux distribution. 

\begin{table*}
\caption{Overview of off-states or low-flux observations of Vela X-1 reported in the literature.}
\label{tab:offstates}
\begin{center}
\begin{tabular}{lrcrr}
\hline\hline
References   & \multicolumn{1}{c}{Date} & Instrument &Energy range&Off-state luminosity  \\ 
\hline
\citet{KallmanWhite:82}, \citet{Hayakawa:84} &  9 May 1979 & \textit{Einstein} & 2-10~keV & \emph{not quantified}  \\
\citet{Inoue:84}, \citet{Hayakawa:84}        & 12 Mar 1983 & \textit{Tenma} & 3-9~keV & \emph{$<$10\% of preceding} \\
\citet{Lapshov:92}                           &  9 Jan 1991 & \textit{Granat}/WATCH & 8-15 keV & \emph{not quantified}  \\
\citet{Kreykenbohm:99}                       & 23 Feb 1996 & \textit{RXTE}/PCA & 3-30~keV & \emph{$<$15\% of normal}   \\
\citet{Kretschmar:CGRO99}                    & 22 Jan 1998 & \textit{RXTE}/PCA & 2-60~keV & \emph{not quantified}  \\
\citet{Kreykenbohm:2008}                     &  8 Dec 2003 & \textit{INTEGRAL}/ISGRI &  20-40~keV & \emph{not quantified}  \\
\citet{Doroshenko:2011}                      & 17--18 Jun 2008 & \textit{Suzaku}/XIS & 0.4--70~keV & $\sim 2.4\times10^{35}$~erg~s$^{-1}$  \\
\citet{Sidoli:2015}                          &  2002--2012 & \textit{INTEGRAL}ISGRI &  22-50~keV & $\lesssim 3\times10^{35}$~erg~s$^{-1}$    \\
\hline
\end{tabular}
\end{center}
\end{table*} 

Hand in hand with the flux variations, a hardening of the spectrum has been observed in \vel, that is, there is an anticorrelation between the photon index $\Gamma$ and the observed flux \citep{Fuerst:2014}. This behavior agrees with the  expected subcritical accretion regime of \vel and is also seen in sources of similar luminosity \citep{klochkov:2011}.

\subsubsection{X-ray polarization} \label{sec:obs:xray:polarization}

For lack of sensitive X-ray polarimeters in space, no observations of X-ray polarization parameters in Vela~X-1 or similar systems from X-ray satellites have so far been made. A notable recent exception was the balloon-borne hard X-ray calorimeter, X-Calibur, which observed Vela~X-1. Because the flight duration was much shorter than anticipated, however, no constraining data could be obtained \citep{Abarr:2020}.

Clear polarization signatures are expected from the physics of the radiation transfer in the emission region as well as in the system as a whole, however. Photon-scattering opacities within the accretion column will depend strongly on energy, on the direction of propagation, and and on polarization, as discussed by \citet{Meszaros:1980} and in various subsequent publications. Thus, a significant variation of polarization parameters is expected as function of pulse phase and energy for the X-ray emission originating at the neutron star.

Recently, \citet{Caiazzo+Heyl:2021a}  have  presented  a  detailed  calculation  of  the  intrinsic  polarization of the X-ray point source for accreting pulsars in X-ray binaries. They took the structure and dynamics of the accretion region into account and included relativistic beaming, gravitational lensing, and quantum electrodynamics in their treatment of the propagation of the radiation toward the observer.

Complementary to this, \citet{Kallman:2015}  computed scattering of the emitted X-rays within the dense winds of HMXBs. They showed that it may lead to a variation in polarization with orbital phase, energy, and system geometry, especially if large-scale structures are present in the wind. They described how they obtained the three Stokes parameters of linearly polarized light from this scattering. However, their calculations assumed an unpolarized, isotropic X-ray emitter, which is likely not the case for accreting X-ray pulsars. They found fractional polarization values of up to $\sim$10\% for a spherically symmetrical wind and $>$20\% in their more realistic hydrodynamical model with a focused wind.




\subsubsection{Cyclotron resonance scattering features} \label{sec:obs:xray:cyclo}

\citet{Kendziorra:1992FXRA} first reported CRSFs (Section~\ref{sec:elements:column}) in \vel at 54~keV and possibly 27~keV, based on pulse-phase-resolved spectra obtained from the High-Energy X-ray Experiment (HEXE) on board the Mir space station. This made \vel one of the early identified sources with such line features \citep{Staubert:2019}. Evidence for these features was also given by \citet{MakishimaMihara:92}. 
While the detection of line features was not challenged, it was debated in the literature for at least a decade whether the feature between 25~keV and 30~keV was the fundamental cyclotron line \citep[e.g.,][]{Kretschmar:97,Kreykenbohm:99} or rather the harmonic \citep{Orlandini:98-VelaX1} because the lower-energy feature was not always apparent, especially in phase-averaged spectra. \citet{Kreykenbohm:2002} found in a deep pulse-phase-resolved analysis of spectra taken with \textit{RXTE} a fundamental line feature at $23.3^{+1.3}_{-0.6}$~keV during the pulse maxima that was much less significant during the pulse minima. All line parameters showed clear pulse-phase dependence.

\begin{figure}
    \centering
    \includegraphics[width=1.0\linewidth]{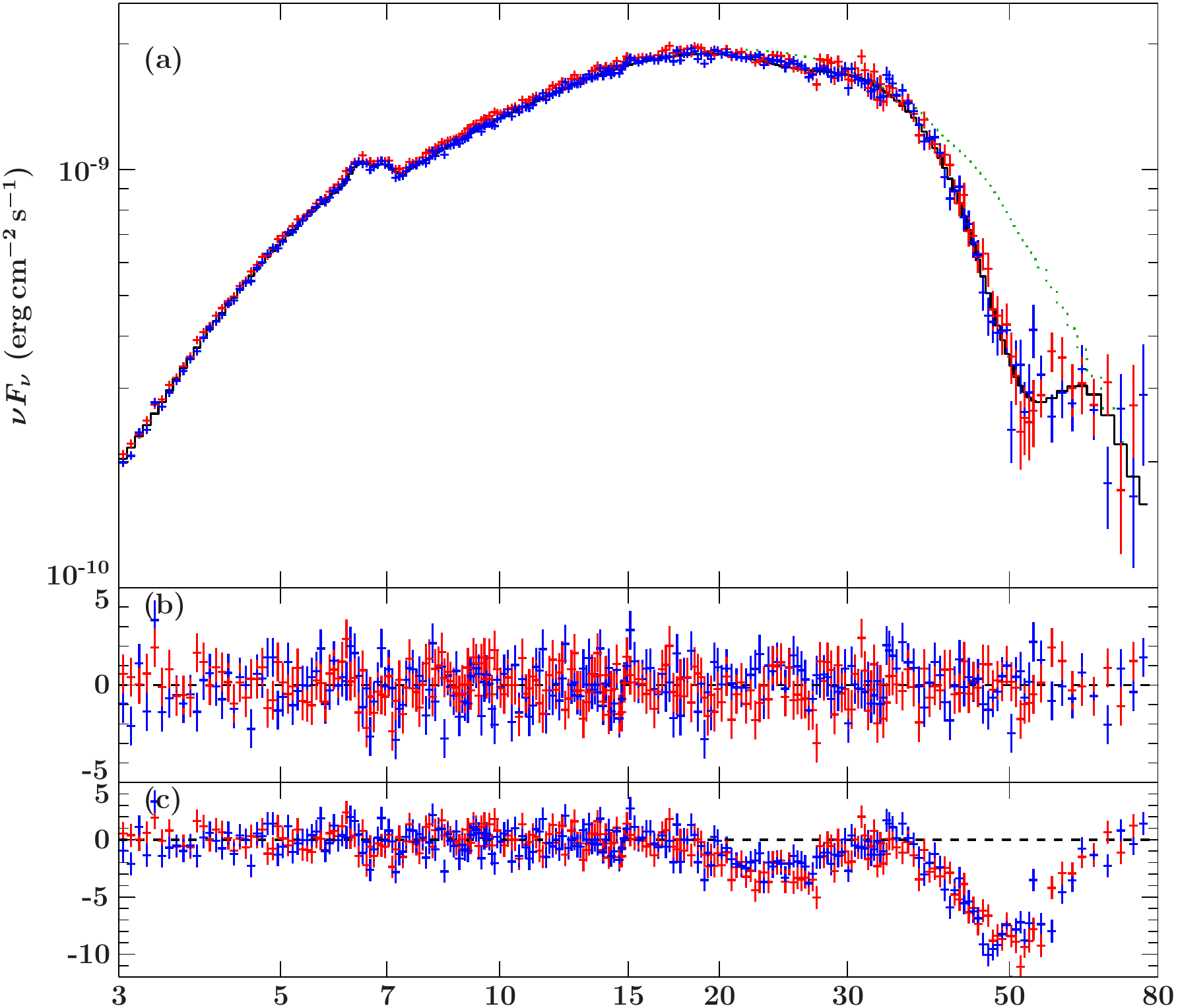}
    \caption{\textit{a)} X-ray spectrum of \vel as measured with \textsl{NuSTAR} (ObsID 30002007003). Data from the two instruments of \textsl{NuSTAR}, FPMA and FPMB, are shown in red and blue, respectively. The best-fit model as presented by \citet{Fuerst:2014} is shown in black. The dashed green lines indicates the continuum model without the CRSF components. \textit{b)} Residuals in terms of $\chi^2$ between the data and the best-fit model. \textit{c)} Residuals between the data and the model without the CRSF components. The harmonic line around 55\,keV is much more prominent than the fundamental line around 25\,keV.
    \label{fig:spectrum}}
\end{figure}

Observations of \vel with \textit{NuSTAR} \citep{Fuerst:2014} clearly detected the fundamental line at $\sim$25~keV together with the more prominent harmonic at $\sim$55~keV. Fig.~\ref{fig:spectrum} shows one of the spectra presented by \citet{Fuerst:2014}, which is a typical out-of-eclipse \vel spectrum. The data presented here were extracted with NUSTARDAS v2.0.0 and CALDB v20201217. The model is based on the best-fit model presented by \citet{Fuerst:2014}, which uses an absorbed Fermi-Dirac cutoff continuum spectrum, modified by two multiplicative lines with Gaussian optical depth profiles to describe the CRSFs. In addition, two Gaussian emission lines are used to model the Fe K$\alpha$ and K$\beta$ lines, and a Gaussian absorption line was added to describe the 10\,keV feature \citep{LaBarbera:2003}. 
In time-resolved spectroscopy, \citet{Fuerst:2014} found that the strengths of the two CRSFs  appeared to be anticorrelated, which could be explained by photon spawning \citep{Schoenherr:2007}. 

Earlier studies found no significant variation of the CRSF energy with flux. \citet{Fuerst:2014} found a positive correlation of the harmonic (higher-energy) line with flux, as expected for a source in the subcritical accretion regime as defined by \citet{Becker:2012}. 
\citet{LaParola:2016} undertook a long-term study of \textit{Swift}/BAT data of Vela~X-1. While the fundamental line was not evident in their 15--150~keV spectra, the first harmonic of the CRSF was clearly detected between $\sim$53 and $\sim$58 keV, and with an apparent positive correlation of the line energy with luminosity, flattening above $L_\textrm{1--150~keV}\approx 5\times 10^{36}$~erg s$^{-1}$.
Recently, \citet{Ji:2019} reported
a secular decrease of the energy of the harmonic line, again based on \textit{Swift}/BAT data  . If confirmed, this would indicate a slow change in the magnetic field configuration close to the polar caps.

One important caveat when results from different studies are compared is that different choices of mostly phenomenological spectral models for the continuum and for the cyclotron line parameters themselves lead to fit parameters that are intrinsically not directly comparable. Even the line centroids determining $E_{\text{CRSF},n}$ (Eq.~\ref{eq:Ecyc}) can be systematically different between different line models by a few keV for the same fitted spectrum, as explained in detail in \citet[Sect.~3]{Staubert:2019}.

\subsubsection{Pulse period variations}\label{sec:obs:xray:pulse}

\begin{figure}[hbt]
    \centering
    \includegraphics[width=1.0\linewidth]{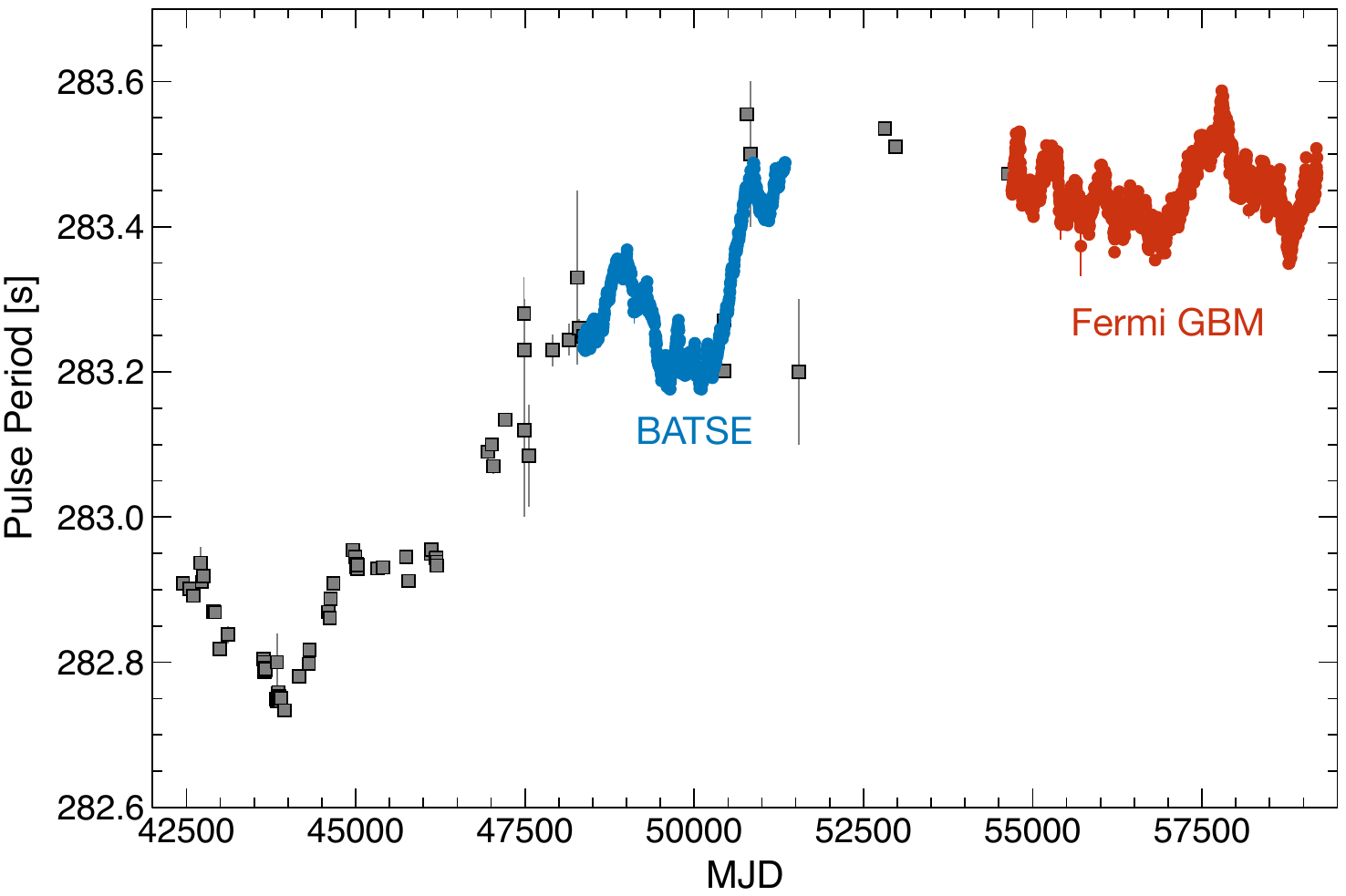}
    \caption{Evolution of the pulse period of Vela X-1 as observed by diverse instruments (gray squares, see Table~\ref{tab:pulse_periods} for details), by \textit{CGRO}/BATSE (blue circles) or by \textit{Fermi}/GBM (red circles). 
    BATSE and GBM period determinations are derived over two and three intervals of the orbit, respectively.
    Note the scale. The longest reported pulse period differs by $\sim$0.3\% from the shortest known pulse.
    \label{fig:pulseperiods}}
\end{figure}

The long-term pulse period evolution of Vela~X-1 is usually described as a random walk. In an in-depth study of \textsl{\textit{HEAO-1}} data, \citet{Boynton:84} found that the power density spectrum covering timescales from 0.25 to 2600 days can be described by assuming white noise in angular acceleration, or equivalently, a random walk in pulse frequency.
Short-term angular accelerations were found to be as large as $\dot{\Omega}/\Omega = (5.8\pm 1.4)\times 10^{-3} \textrm{yr}^{-1}$. According to their analysis, apparent changes in secular trends and frequency variations on shorter timescales of at least days are all consistent with the same random noise process in angular acceleration. This result was reinforced and refined in subsequent papers \citep{Boynton:86,Deeter:89} and also by a similar study based on the \textsl{Hakucho} and \textsl{Tenma satellites} \citep{Deeter:87b}. \citet{deKoolAnzer:93} extended these studies by combining data from \citet{Nagase:89} and \citet{RaubenheimerOegelman:90}, again confirming that the pulse period behavior is very well fit by a random walk.
\citet{Boynton:84} also noted that the short-term angular accelerations reported in their study were much larger than expected for accretion from a uniform wind, assuming the wind speeds given in \citet{Dupree:80}, and that the flow may form a ring or disk around the neutron star.

\subsubsection{Pulse profiles} \label{sec:obs:xray:pulseprofiles}
\begin{figure}
    \centering
    \includegraphics{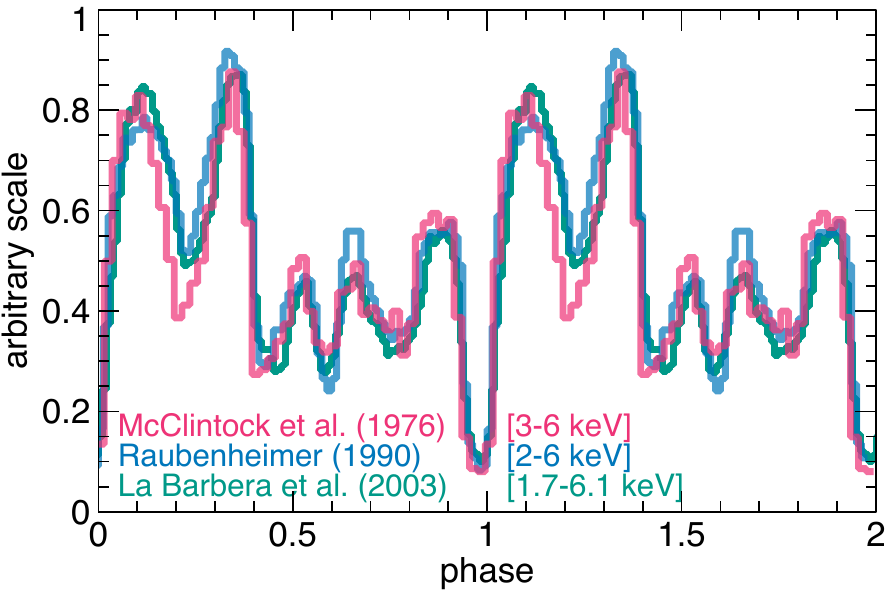}
    \caption{Comparison of the complex low-energy pulse profile in the roughly 2 to 6~keV energy range from observations taken decades apart with three different X-ray missions, \textit{SAS3, EXOSAT,} and \textit{BeppoSAX}, respectively. The data points have been taken from \citet{McClintock:76,Raubenheimer:90}; and \citet{LaBarbera:2003}, but shifted in phase and scaled arbitrarily for a visual match.}
    \label{fig:pp_comparison}
\end{figure}

The first observations discovering the periodic pulsations of Vela X-1 with the \textsl{SAS-3} X-ray observatory have found a complex pulse profile structure with five visible peaks at energies below 6~keV and two broad peaks at the higher energies above $\sim$10 keV, which are both somewhat asymmetric, with a sharper trailing than leading edge \citep{RappaportMcClintock:75,McClintock:76}. The overall structure of the pulse profile has been confirmed in many subsequent studies of the system \citep[e.g.,][]{Nagase:83,Raubenheimer:90,Kreykenbohm:99,LaBarbera:2003}, across significant variations in X-ray brightness. It can be considered a fingerprint of the X-ray pulsar, like in many other systems. In other words, while some variations in the observed profile are visible when different observations are compared, especially in the peak fluxes of different pulse components, the overall shape and thus the underlying emission geometry is evidently very stable. This is also demonstrated in Figure~\ref{fig:pp_comparison}. The main visible differences at the lower energies can be caused by the sometimes very strong absorption and scattering (Section~\ref{sec:obs:xray:abs}), which can smear out especially the low-energy profile \citep{Nagase:83}. For detailed pulse profiles across a wide energy band, see especially \citet[Fig.~1]{Raubenheimer:90} and \citet[Figs.~1--4]{LaBarbera:2003}.

When different observations are compared, we recall that like in many other accreting X-ray pulsars, there are clear variations between individual pulses, as demonstrated in the first detailed study of the hard X-ray pulsations \citep{Staubert:80} and later studies \citep{Kendziorra:89,Kretschmar:2014PMB}. These variations may affect the mean pulse profile of observations that cover only a limited number of pulse period intervals. 


In earlier publications, pulsations were found to cease during off-states \citep{Inoue:84,Kreykenbohm:99,Kreykenbohm:2008}. This behavior might be interpreted as the onset of the propeller regime (Section~\ref{sec:elements:magnetosphere}). In contrast, \citet{Doroshenko:2011} analyzing three off-states observed with \textsl{Suzaku}, did observe remaining pulsations, but with significant changes in the pulse profile. They interpreted this behavior as gated accretion, with some remaining accretion.

A very peculiar observation was reported by \citet{Kretschmar:CGRO99}: For a duration of several hours, the source flux diminished, but was still significantly above the background level, while pulsations ceased practically completely. After this low-state phase, there was a high state with strong, flaring pulsations, but a very similar spectrum to the preceding nonpulsating state. At the time, an explanation by a very thick clump in the wind was put forward, but in the light of newer results on wind properties and realistic clump sizes (see Sect.~\ref{sec:elements}), this seems very unlikely. 

\citet{Liao:2020} reported another case of a low-flux state with ceasing pulsations. Based on spectral evidence, this was interpreted as formation of a (temporary) accretion disk in this case. We also note the explanation put forward by \citet{King+Lasota:2020} for suppressed pulsations in ultraluminous X-ray sources by scattering in a thick beaming funnel. This role would be played in Vela~X-1 by the transient disk-like structure.

\subsection{Optical, UV, and IR diagnostics}\label{sec:obs:uvopt}

\subsubsection{Photometric light curve} \label{sec:obs:uvopt:photometry}

The study of photometric data, and in particular, the shape and amplitude of the optical light curve, is a tool that beyond determining the photometric period can be used through comparison with models to infer information on system parameters such as the size of the components relative to the orbit or the mass ratio. For a review, see \citet{Wilson:1994BinaryLC}; a visualization of a model light curve for Vela X-1 is included in \citet{Wilson:1994IAPP}. Several studies of this type have been performed in the 1970s and 1980s. All of them reported variability on the photometric light curve from one cycle to another. 
At later times, system parameters were instead obtained from X-ray studies, which allowed a more accurate determination of the system parameters.

The first photometric study of HD 77581 was carried out by \citet{Vidal:73} in the V band over a time span of 16 days. 
They observed four minima in the light curve that were consistent with the reported period of $8.95 \pm 0.02$~days by \citet{Forman:73}. They reported a small amplitude of the light curve of $\sim 0.14$ mag and found some evidence of variability with the period, particularly at the primary minimum. This behavior was attributed by \citet{Vidal:73} to the important role played by tidal effects.

The same periodicity was confirmed by \citet{JonesLiller:73} using UBV photometric observations of the companion over 27 consecutive nights, including two maxima and two minima.
They reported an unexpected nonrepeatability of the optical light curve from cycle to cycle with color-free erratic changes. They explained this as due to complex changes in the upper layers of the star that are comparable to the gravity-darkening effect. 

\citet{Petro+Hiltner:1974} combined the observations of \citet{Vidal:73} and \citet{JonesLiller:73} with three new campaigns to obtain a more precise determination of the photometric period of 8.972(1) days. They noted systematic deviations between the data and the best-fitting circular-orbit light curve, which might be explained by an elliptical orbit.

\citet{Vidal:74} presented a more extensive set of observations in V band with 126 observations in total, including those reported in \citet{Vidal:73}, which were obtained during the period of 22 January to 19 June 1973. They reported that the most interesting feature of the light curve was the high scatter of the data throughout the whole period and claimed that this variability was intrinsic to the source. They observed that the rising and descending branches of the light curve during primary minimum were different and interpreted the shape of the light curve as due to different aspects of a tidally distorted companion. 

\cite{Zuiderwijk:77b} presented a four-color photometric study of HD 77581 in the Strömgren \textit{uvby} system. They observed the source on 46 nights in three campaigns in 1975. The magnitudes are listed in tabular form in \citet{Zuiderwijk:77a}. 
These observations also showed changes in the light curve from one orbit to another. They found a regular wave-like variation in the color index c1 and possibly in b-y in phase with the light curves and twice noted the disappearance of a maximum during their observations, as also reported by \citet{JonesLiller:73}. The observed variability in the light curve and the colors were interpreted with a model of a tidally distorted rotating primary.

\citet{vanGenderen:1981} carried out a VBLUW photometric study in 1976 and 1977. 
They obtained a photometric period of $8.9615\pm0.0025$~days, which is closer to the contemporary measured period of the X-ray light curve. They reported the peculiarities of the mean light curve, which had a total amplitude of $\sim$ 0.10 mag, with a difference in height between the maximum and the minimum of $\sim$ 0.04 mag. The minimum seemed to be delayed by $\sim$0.1 in phase with respect to the center of the X-ray eclipse. The high peaked maximum varied in height and possibly in phase with a timescale of $\text{about}$ two years. They concluded that the observed intrinsic variability of the light curve was correlated with the appearance and disappearance of hotter or cooler areas in the stellar atmosphere.

\citet{KhruzinaCherepashchuk82} performed an analysis of all published photometry data and revealed a regular long period of 93.3 days of the optical light curve  (B band). This period was later confirmed by \citet{Cherepashchuk82} using independent new UBV photometric data obtained at Siding Spring Observatory in February-June 1980. \citet{KhruzinaCherepashchuk82}  and \citet{Cherepashchuk82} associated this long-term period with forced precession of the rotation axis of the optical companion. This induces changes in the eclipsing gas streams and in the accretion structure. While super-orbital periods have been detected for other HMXBs \citep{Corbet+Krimm:2013}, no such periodicity has been reported for \vel.

\begin{figure}
    \centering
    \includegraphics{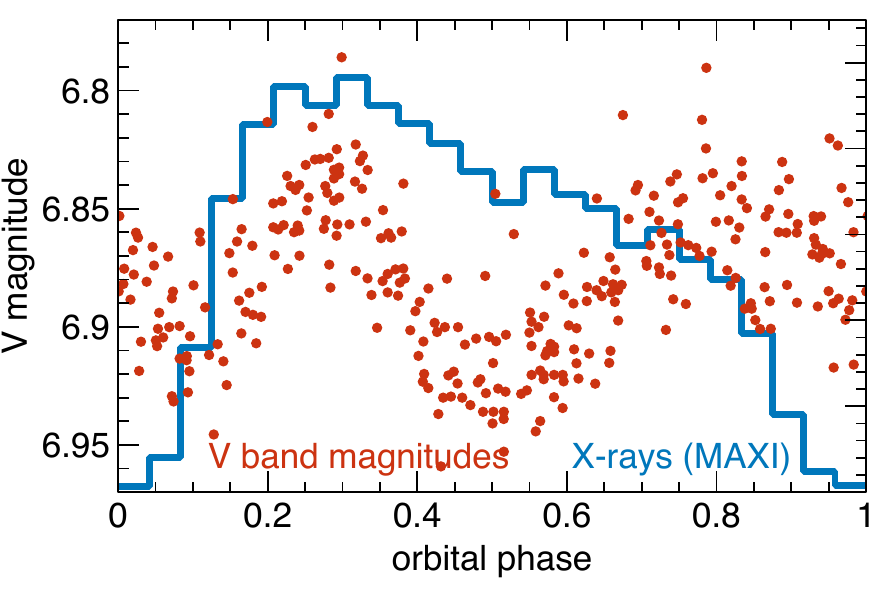}
    \caption{V-band magnitudes of HD~77581 as compiled by \citet[Fig.~8]{Tjemkes:1986}, demonstrating the intrinsic scatter, which is much larger than measurement uncertainties. For visual comparison, we plot the mean orbital profile in X-rays as observed by the MAXI instrument from Figure~\ref{fig:fluxhistos} in the background.}.
    \label{fig:OptXray}
\end{figure}

\citet{Tjemkes:1986} developed a simple geometric model to analyze the optical light curves of several X-ray binaries and compared their predictions with real data. In the case of HD 77581, they used all published data in addition to data obtained in 1979 and 1980 
to compute an average visual light curve. Ellipsoidal light variations were observed in the light curve, which is indicative of a donor star close to filling its Roche lobe. The light curve had two brightness maxima and two minima, and a photometric period of 8.965 $\pm$ 0.001 was reported. However, the light curve presented some particularities: 1) the two maxima are different in shape and brightness level, 2) the minimum is somewhat asymmetric and is shifted with respect to the mid-time of the X-ray eclipse by 0.05, and 3) the scatter around the average light curve is $\sim$ 0.02, which is much larger than the accuracy of the photometric data. These variations occurred on timescales from hours to many days. They carried out a search of the long-term period reported by \citet{KhruzinaCherepashchuk82} and \citet{Cherepashchuk82} and detected some indications in favor of a minimum variance near 93.2 d, but no significant evidence to support the existence of this super-orbital period. 
Overall, \citet{Tjemkes:1986} found serious discrepancies between their model and the data for the minima and concluded that the model did not describe the region near the inner Lagrangian point well. They suggested that basic model assumptions, in particular the instantaneous adjustment of the primary shape to the equipotential surface, are invalid for an eccentric orbit.

\citet{Wilson+Terrell:1994AIPC} fit optical light curves, optical velocities, and X-ray pulse arrival times simultaneously to reproduce the observed X-ray eclipse duration. They assumed a binary star model on equipotentials with an eccentric orbit and adjusted the varying tidal effect. The companion star rotated at a constant angular rate.  They concluded that the optical star rotates subsynchronously and presents phase shifts that the authors interpreted as tidal lags.  Both of these properties have major implications for understanding the scenario of the system evolution. According to \citet{Wilson+Terrell:1994AIPC}, the system could be on the verge of a common evolution scenario.

\subsubsection{Radial velocity curve} \label{sec:obs:uvopt:radial}


In an eclipsing X-ray binary system, the procedure for determining the neutron star mass is based on the radial velocity (RV, hereafter) curve of the companion star, as determined from shifts in the positions of various observed spectral lines, knowing the orbits of the two components, and knowing or assuming the inclination of the system. This procedure is valid under the assumption that RV curve variations truly represent the orbital motion of the star. In the case of HD~77581, this assumption is highly questionable for two reasons: the prominent line profile variability \citep{vanParadijs:77b,vanKerkwijk:95, Barziv:2001,Quaintrell:2003}, and the asymmetries over the orbital cycle of the RV curve \citep{vanParadijs:77b,vanKerkwijk:95, Barziv:2001}.

The first RV study of HD~77581 was performed by \citet{Hiltner:1972} using a Coudé spectrograph with a total of 85 spectrograms. They found variability on the source, and the velocity curve was not sufficiently defined to warrant a detailed analysis. They concluded that the system was a spectroscopic binary with a provisional period of 7d and a minimum mass of the companion near 1.4~$M_\sun$. 

In 1974, several publications \citep{Wickramasinghe:74,MikkelsenWallerstein:74,Zuiderwijk:74,Hutchings:74} carried out an analysis of the RV curve. No consistent orbital elements could be derived, however, leading to estimates of the mass of the compact object that were too large to be consistent with those of a neutron star.

\cite{vanParadijs:76} performed the first relatively accurate determination of masses for both partners using RV measurements. They used 26 Coudé spectrograms to determine the total mass of the system as well as the masses of each component. In contrast to previous work, they excluded the lines of hydrogen because they are the most affected by gas motions in the system. Using He I lines and lines of heavier ions, they derived mean values of the RV for all the lines and for the He I and heavier-ion lines independently. They obtained a semiamplitude of the RV curve of $K_\textrm{opt}= 20 \pm 1\,\textrm{km s$^{-1}$}$  and derived a total mass of the system of $\sim 21.6 M_\sun$, with a mass of the compact object of $\sim 1.6 M _\sun$, and $\sim 20 M_\sun$ for the companion (see Table~\ref{tab:masses}).

\cite{vanParadijs:77b} carried out a more detailed optical RV study to improve the accuracy of previous works and to investigate the possible presence of the distortion effect predicted in \cite{vanParadijs:77a}. They found short-term autocorrelated nonorbital variations, but because so many factors were involved, they were unable to estimate the amplitude of the individual nonorbital contributions. They derived a more accurate $K_\textrm{opt} = 21.75 \pm 1.15\,\textrm{km s$^{-1}$}$ than in \citet{vanParadijs:76} and claimed that the mass of the primary was rather low for its spectral type and luminosity class (see Table~\ref{tab:masses}), a similar type of undermassiveness as was found for SK~160 in SMC~X-1.

Nearly 20 years later, \citet{vanKerkwijk:95} performed a RV curve analysis using optical and IUE data. Since the previous mass determination, the statistical and systematic accuracy of stellar spectroscopy had improved significantly following the introduction of CCDs detectors.  Thus, they expected to obtain a smaller error on the RV determination. However, this was hampered by the large deviations from a pure Keplerian RV curve. As did \citet{Hiltner:1972,Zuiderwijk:74, vanParadijs:77b}, they reported velocity differences with respect to the orbital fit dominated by random excursions that they found to be autocorrelated within one night, but not from night to night. They found substantial and very similar changes in the shape of the profiles of all the lines. These changes in the lines plus the velocity variability were interpreted as due to large-scale motion of the surface induced by tidal forces due to the presence of the neutron star. This is the same underlying physical mechanism that \citet{Tjemkes:1986} proposed to explain the irregular variations of the optical light curve. \citet{vanKerkwijk:95} derived a $K_\textrm{opt}$ = $18.0-28.2\,\textrm{km s$^{-1}$}$ (95$\%$ confidence range) and concluded that the accuracy of the RV study is limited by three factors: (i) the velocity excursions, (ii) possible effects induced by the tidal deformation, and (iii) the possible presence of systematic positive deviations in velocity close to the time of minimum velocity. To derive more accurate constraints on the orbital parameters, they suggested that the short- and long-term behavior of the system should be studied to better understand the tidal interaction.

For a better understanding of the radial-velocity excursions, \cite{Barziv:2001} performed an extensive long-term spectroscopic campaign for about nine months following one of the recommendations given by \citet{vanKerkwijk:95}. They expected that the velocity excursions would average, thus allowing a more accurate determination of the RV amplitude than previous works. They determined the RV from several lines using the same cross-correlation algorithm as \citet{vanKerkwijk:95}. As in previous analyses, they found strong deviations in the RVs from those expected for a pure Keplerian orbit, with root-mean-square amplitudes of $\sim 7\,\textrm{km s$^{-1}$}$ for strong lines of Si\,IV and N\,III near 4100\,\AA, and up to $\sim 20\,\textrm{km s$^{-1}$}$ for weaker lines of N\,II and Al\,III near 5700\,\AA. They found that systematic deviations depended on the orbital phase, with the largest deviations observed near inferior conjunction of the neutron star and near the phase of maximum approaching velocity. They also reported a so-called blue excursion in the H$\delta$ data and interpreted it as a photoionization wake. They inferred a radial-velocity amplitude $K_\textrm{opt} = 21.7 \pm 1.6\,\textrm{km s$^{-1}$}$ with an uncertainty not much smaller than was found in previous analyses, in which the effect of systematic phase-locked deviations were not taken into account.

To overcome the velocity excursions issue, \citet{Quaintrell:2003} carried out a comprehensive RV study with maximum phase coverage over two consecutive orbits of the system instead
of averaging the velocity excursions over many orbital periods. They found evidence for tidally induced nonradial oscillations from the power spectrum of the residuals to the RV curve fit. Moreover, when they allowed the zero phase of the RV curve to vary instead of constraining it to the X-ray ephemeris, the fit significantly improved and they obtained a semiamplitude of the RV curve of $K_\textrm{opt} = 21.2 \pm 0.7\,\textrm{km s$^{-1}$}$. This is the most accurate value to date (see Table~\ref{tab:masses} for a comparison of the determined $K_\textrm{opt}$). \citet{Quaintrell:2003} concluded that this apparent shift in the zero-phase may indicate an additional RV component at the orbital period, which could be another manifestation of the tidally induced nonradial oscillations and an additional source of uncertainty in RV studies.

\citet{Koenigsberger:2012} have explored the manner in which surface motions induced by the tidal coupling between the blue supergiant and the neutron star affect the RV curve \citep[for the theory of tides in general, see][]{Zahn:89}. They developed a 2D code that provides the time-dependent shape of the stellar surface and its surface velocity field for the general case of an elliptic orbit and asynchronous rotation without considering the effects of a nonuniform temperature distribution over the stellar surface. They concluded that the tidal effects on the RV curve produce orbital phase-dependent variations in the profiles that lead to asymmetries, blue or red wings. Moreover, the line-profile variability induces a significant variation from a Keplerian RV curve that artificially enhances the semiamplitude. 
One point of note is that a prominent feature appears in their synthetic RV curves: A blue dip occurs shortly after the maximum. This is caused by the asymmetrical shape of the line profiles. This feature coincides with the blue excursion reported by \citet{Barziv:2001}, and it could be explained by a higher mass outflow after periastron passage, where the tidal effects are stronger.

\subsubsection{Quantitative spectroscopy} \label{sec:obs:uvopt:spectra}

The stellar and wind parameters of hot massive stars such as HD~77581 are frequently determined through quantitative spectroscopy, that is, by fitting synthetic spectral energy distributions and normalized spectra to observations, mainly optical and UV spectra. The best-fitting spectra and thus parameters are frequently found by eye in a visual comparison of models and data or by minimization on a grid of spectra. 

The main parameters derived in these studies are the measured color excess ($E_{B-V}$), the (effective) stellar temperature ($T_\star$) from the ionization equilibrium determined from line ratios, the surface gravity ($g_\star$), and the projected rotational velocity ($v_\mathrm{rot} \sin i$). Stellar radii and thus luminosities are then derived using the temperature, absolute magnitude (depending on the assumed or derived distance), and the bolometric correction. 
An overview of methods and diagnostics used for these studies and caveats to consider are given in \citet{Martinez-Nunez:2017}. 

Early studies in  the 1980s were undertaken before codes that describe hot stellar atmospheres were developed. These studies relied on an examination of specific lines and a comparison to similar stars. 
\citet{Dupree:80} carried out a simultaneous observation program in the X-ray, ultraviolet \textit{(IUE)}\footnote{The International Ultraviolet Explorer satellite} , and optical bands. They mainly used the comparison of P~Cygni lines with the atlas of \citet{Castor+Lamers:1979} and theoretical profiles provided by Olson (\textit{privat communication} of \citeauthor{Dupree:80}) to estimate the terminal velocity and mass-loss rate of the line-driven wind. 
\citet{Sadakane:85} analyzed \textit{IUE} spectra of HD~77581 by comparing individual lines with the corresponding lines from well-studied single B0-B1 supergiants. 
Other examples in which specific line features were used to estimate some stellar wind or line properties include \citet{Prinja:90} for terminal velocities of massive star winds or \citet{Zuiderwijk:95} for the rotation period and rotation velocity of HD~77581. 

From the the 1990s onward, nonlocal thermodynamic equilibrium (NLTE) hydrostatic codes began to be used to genereate synthetic spectra. In a seminal paper, \citet{Vanbeveren:93ADS} used a plane-parallel code by D.~Kunze for the comparison with optical spectra with high signal-to-noise ratio in the range 4175--4525~\AA\ to derive stellar parameters. Using the orbital parameters of \citet{RappaportJoss:83}, they determined a range of stellar parameters satisfying the orbital, X-ray eclipse, and stellar atmosphere analysis. 

The NLTE codes SYNSPEC\footnote{\url{http://tlusty.oca.eu/Synspec49/synspec.html}}  \citep{Hubeny:1985Synspec} and TLUSTY\footnote{\url{http://tlusty.oca.eu/}} \citep{HubenyLanz:1995} were used by \citet{Fraser:2010} to calculate model atmosphere grids. These were then compared with high-resolution ($R\sim 48000$) spectra obtained with FEROS\footnote{The Fiber-fed Extended Range Optical Spectrograph \url{https://www.eso.org/sci/facilities/lasilla/instruments/feros.html}} for very many massive stars, including HD~77581, in order to determine the atmospheric parameters (effective temperature, surface gravity, and microturbulent velocity), the surface nitrogen abundances, and the rotational and macroturbulent velocities.
These codes do not take the inhomogeneities of the wind into account, however, and are optimized for hot stars with no significant wind. 

A stellar wind can considerably alter the
spectral appearance and for example spoil the derived stellar parameters
such as $\log g$ if it is not taken into account. So-called unified model
atmospheres are necessary to consistently describe the outermost
layers of the star and their winds.  Model atmospheres like this are
inherently in NLTE and have considerably improved over the past decades through  higher computational power and the better understanding of
physical processes in stellar atmospheres \citep{Sander:2017b}, for
example, by properly accounting for the complex effects of the millions of
iron lines. While modern codes provide a more accurate determination of
the stellar and wind parameters, a proper comparison of observed
and model spectra can only be automatized to a certain degree \citep[see
also][]{Martinez-Nunez:2017}. In any case, spectral analysis studies
should be performed using as many lines as possible because a particular
single line can be affected by a variety of parameters, and it requires
detailed knowledge about which features are affected by which parameters.

\citet{Gimenez-Garcia:2016} carried out the first detailed study of HD~77581 using models with an expanding stellar atmosphere for line-driven winds generated by the Potsdam Wolf-Rayet (PoWR)\footnote{\url{http://www.astro.physik.uni-potsdam.de/~wrh/PoWR/}} model atmosphere code \citep{Graefener:2002,Hamann+Graefener:2003}. The PoWR code takes line-driven winds, wind clumping, and a plethora of ions into account. These were compared to the previously mentioned \textit{IUE} and FEROS spectra to derive stellar and wind parameters. The rotational velocity, the $\beta$ exponent for the wind velocity profile (Eq.~\ref{eq:beta-law}), and the macroturbulence were set based on insights from previous works and on spectral line shapes and depths. They also found indications of chemical evolution in the star, with a moderate overabundance of He and N, together with an underabundance of C and O. 

In recent years, the PoWR code has been extended with the option to consistently solve the hydrodynamic and statistical equations in 1D, together with the radiative transfer, in order to obtain a hydrodynamically consistent atmosphere stratification \citep{Sander:2017}. This was applied to the Vela~X-1 system in \citet{Sander:2018}, who in a simplified manner also included the effect of the X-ray radiation (see also Section~\ref{sec:ion_struct}).

Tables~\ref{tab:stellarpar} and \ref{tab:wind} in section~\ref{sec:system:wind} give an overview of the parameters that were derived mainly from quantitative spectroscopy. In some cases, these results were also used to infer information about other properties of the system (see Sections~\ref{sec:system:distance} or~\ref{sec:system:mass} and table~\ref{tab:masses}).

\subsubsection{UV resonance lines}
\label{sec:obs:uvopt:uv_resonance}

\citet{HatchettMcCray:77} predicted that an X-ray source in the stellar wind of an early star produces changes in the photospheric ultraviolet resonance lines of ions such as Si\,{\sc iv}, C\,{\sc iv}, Al\,{\sc iii,} or N\,{\sc v}. They also predicted that the lines vary along the orbit and that these variations can be used to determine the size of the stellar wind region that is affected by the X-ray ionization induced by the compact object. However, the amplitude of the observed orbital modulation also depends on the density and velocity of the stellar wind. 

\citet{Dupree:80} presented the first UV resonance line study of the system using IUE observations. They detected very prominent P Cygni profiles in the resonance lines of Si\,{\sc iv} (1393.75$\AA$ and 1402.77$\AA$) and C\,{\sc iv} (1548.19$\AA$ and 1550.76$\AA$). They observed variability in the profiles over the orbital phase with more dramatic changes in the Si\,{\sc iv} than in the C\,{\sc iv} profiles.
They reported that when the compact object is in front of the mass donor, around orbital phase 0.5, the P Cygni absorption is ionized to a higher state than at eclipse, and the emission component is increased. They concluded that the observed variability and the variation in the edge velocity could be qualitatively explained by the Hatchett-McCray effect (hereafter, the HM effect).

\citet{Sadakane:85} detected P Cygni profiles of the C\,{\sc iv} and Si\,{\sc iv} resonance lines in the same way as \citet{Dupree:80}, and in addition, they reported the Al\,{\sc iii} resonance doublet at 1854.72$\AA$ and 1862.79$\AA$. They found similar profiles and variations on the C\,{\sc iv} and Si\,{\sc iv} as \citet{Dupree:80}. In the case of Al\,{\sc iii}, they found a remarkable difference in the profiles before and after mid-eclipse. They explained this behavior as due to the complex structure of the stellar wind with two different components, the high- and low-velocity components. The low-velocity component seemed to be present only from mid-eclipse and during the second half of the orbital cycle. They claimed that it was formed near the secondary and that its origin might be a trailing wake behind the compact object. On the other hand, the high-velocity component was observed throuhgout the orbit and was interpreted as originating from a cooler expanding region.

\citet{Kaper:93} carried out a study to investigate the dynamical structure of the stellar wind in the system using IUE data. They also observed orbital variations in the blueshifted absorption part in the Si\,{\sc iv} and C\,{\sc iv} resonance lines. When the compact object was in front of the optical companion, the Si\,{\sc iv} absorption was weaker not only at high velocities ($\sim$ -700 to -1300 km s$^{-1}$) as reported by \citet{Dupree:80}, but also at low velocities ($\sim$ -300 to -500 km s$^{-1}$). In the case of C\,{\sc iv}, they found variations at high velocities, whereas at low velocity, the variations were not significant. Like \citet{Sadakane:85}, they also found variations in the Al\,{\sc iii} resonance, but in this case, also at low and intermediate velocities. They interpreted the larger Al\,{\sc iii} variations as due to ionization effects.  In addition, they reported N\,{\sc v} resonance lines at 1238 and 1242 $\AA$ for the first time, with variations in the absorption part that are consistent with the changes found in the Si\,{\sc iv} and C\,{\sc iv} profiles and consistent with the variability predicted by the HM effect. Regarding the emission of the P Cygni lines, \citet{Kaper:93} found stronger emission in the  Si\,{\sc iv}, C\,{\sc iv,} and Al\,{\sc iii}  profiles around the mid-eclipse phase when the X-ray source is in the line of sight. They estimated that these orbital variations of the redshifted emission extended to about 1000 km s$^{-1}$ for the  Si\,{\sc iv} and C\,{\sc iv} lines. The N\,{\sc v} resonance doublet showed different variability at the short- and in long-wavelength sides. In the former, the strongest emission was observed during eclipse, and in the latter, the red emission peak is remarkably stable throughout the orbit. In conclusion, \citet{Kaper:93} reported an orbital modulation of the P Cygni profiles not only in the high-velocity part of the profiles, but also at intermediate and low velocities. This shows the occurrence of the HM effect for the full range of velocities.  Their observations could not be explained with a monotonic stellar wind combined with partial ionization. They concluded that the resonance lines can be better understood if in some regions, the wind decelerates, in agreement with the predictions from hydrodynamical calculation of radiation driven winds performed by \citet{Owocki:88}.

\citet{vanLoon:2001} presented the first quantitative analysis of the UV spectral variability in five HMXBs including HD 77581, for which they used high-resolution spectroscopy obtained with the IUE satellite. Their synthetic profiles were obtained using a modified version of the Sobolev exact integration method of \citet{Lamers:87}, including a nonmonotonic wind structure, turbulence, and an extended X-ray photoionized zone. The line profiles of Si\,{\sc iv}, C\,{\sc iv,} and Al\,{\sc iii} and their orbital modulation were rather well reproduced by their model, with an additional absorption component due to a photoionization wake that was previously indicated by \citet{Kaper:94} using strong optical lines such as the hydrogen Balmer lines and He lines. In the case of the N\,{\sc v} resonance line, their model was unable to properly reproduce the profile, and it seemed that this line was more affected by the photoionization wake than the other lines. From these data, they were able to derive the extent of the region in which line-acceleration is quenched because the energy levels that are most strongly responsible for it are depopulated by the X-ray photoionizing feedback.

\subsubsection{Infrared diagnostics}
\label{sec:obs:uvopt:ir}

Observations of Vela~X-1 beyond the near-infrared region of the spectrum have been reported only rarely. \citet{Smith+Beall+Swain:1990} listed HD~77581 with significant \textsl{IRAS} (Infrared Astronomical Satellite) fluxes in their collection of 81 X-ray binary counterparts, but did not discuss this further in their study, which focused on infrared emission from accretion disks. \citet{Friedemann:1996} noted that the \textit{IRAS} fluxes reveal cold dust and that the binary may be surrounded by a fossil dust shell.

As detailed in Sect.~\ref{sec:system:VelaOB1}, \vel is associated with a wind bow shock in the interstellar medium. This structure was first discovered in a H$\alpha$ image \citep{Kaper:97}, but has been studied in the infrared based on \textsl{IRAS} \citep{Huthoff+Kaper:2002}, \textit{Spitzer} \citep{Gvaramadze:2011c}, and \textit{WISE} \citep{MaizApellaniz:2018} observations.

\citet{Choquet:2014}  obtained infrared interferometric observations of \vel with the VLTI in the K band (2.2$\mu$m) in 2010 and in the H band (1.6$\mu$m) 
in 2012. The second observations covered different orbital phases (0.89, 0.11, 0.33, and 0.55), but with no significant variation in the interferometric observables. In the K band, a structure of  $8\pm3 R_\star$ was resolved, while two years later, a centro-symmetric structure of radius $2.0^{+0.7}_{-1.2} R_\star$ was found in the H band. The authors offered three different possible explanations for the difference in derived diameters: (1) A strong temperature gradient in the supergiant wind, where hot material at 1720~K is more compact than material at 1350~K; (2) a diffuse gaseous shell observed in 2010, which had diffused two years later; or (3) the structure observed in the H band was the stellar photosphere and not the supergiant wind. 


%

The chemical evolution noted for HD~77581 could indicate that shock chemistry may be dominant in the surrounding interstellar medium. For other regions of the sky, it has been shown that relevant dynamical and physical parameters of the shocked gas can be determined by analysing its far-infrared (far-IR) emission lines \citep[e.g.]{Lerate:2008}. In the same vein, the study of far-infrared emission lines could provide a better insight into the wind clumps and structure.

The region of \vel has been briefly observed by \textit{Herschel} in 2012, but no associated publications are found in the Herschel Science Archive.

\subsection{Radio observations}\label{sec:obs:radio}

On 15 May, 2018, van den Eijnden et al. (in prep.) observed Vela X-1 with the Australia Telescope Compact Array. This four-hour observation yielded the first radio detection of the source at flux densities of $92 \pm 11$ and $122 \pm 10$ $\mu$Jy at $5.5$ and $9$ GHz, respectively. These measured flux densities imply a spectral index measurement of $\alpha = 0.56 \pm 0.36$, where $S_\nu \propto \nu^{\alpha}$. The observation was performed completely during the eclipse of the neutron star, at orbital phases of 0.92--0.94. 


With the currently known radio properties of Vela X-1, it is challenging to determine the dominant radio emission process. While in LMXBs radio emission has been established as a signature of synchrotron-emitting jets (except for the handful of LMXBs hosting a strongly magnetized neutron star), the massive donor and its wind in HMXBs can also contribute to the observed radio emission. The properties of Vela X-1 fit both scenarios: The spectral index is consistent with that of a steady, compact jet, as seen typically in hard states of persistent low-mass X-ray binaries \citep[$\alpha = 0$--$0.7$;][]{Fender+Belloni+Gallo:2004}. The radio luminosity of Vela X-1, interpolated to be $\sim 2\times10^{27}$ erg s$^{-1}$ at 6 GHz, is also similar to the radio emission of other jets launched by strongly magnetized (e.g. $B>10^{12}$ G) accreting neutron stars \citep{vdEijnden:2018GX1+4,vdEijnden:2018HerX-1,vdEijnden:2018Nature,vdEijnden:2019SwiftJ0243}.
Similarly, the spectral index fits the predicted spectral index for thermal radio emission from an ionized stellar wind 
\citep[$\alpha = 0.6$;][]{Wright+Barlow:1975,Guedel:2002ARAA}. The flux densities also fit within their uncertainties the predicted brightness of such a wind following the formalism in \citet{Wright+Barlow:1975}, assuming a mass-loss rate of $10^{-6}$ $M_{\odot}$ yr$^{-1}$ and a terminal wind velocity of $\sim 700$ km s$^{-1}$ \citep{Grinberg:2017}.

Both scenarios could have significant implications: First, strongly magnetized neutron stars were long thought to be incapable of launching jets \citep[e.g.,][]{Fender+Hendry:2000,Migliari+Fender:2006,Massi+KaufmanBernado:2008}, until the recent detection of such a jet showed otherwise \citep{vdEijnden:2018Nature}. If Vela X-1 also launches a jet, it would add to the still small but growing sample of strongly magnetized neutron stars launching jets (van den Eijnden et al., in prep.). Alternatively, if a wind is observed in radio, it would only be the second HMXB with a radio wind detection (after GX 301-2; Pestalozzi et al. 2009). Based on its high inclination and clear orbital variation of the wind absorption, Vela X-1 would provide a unique and powerful testbed in which to explore the properties of stellar winds in HMXBs in radio, while at the same time, radio observations would provide a new and complementary view of Vela X-1 itself. 

At the time of writing, however, the first challenge is to probe the origin of the detected radio emission: jet, wind, or a combination of both? Detailed radio and X-ray monitoring over several orbital phases will be vital for this purpose, revealing the variability of the radio properties and allowing a comparison between accretion, X-ray, wind, and radio properties along the orbit.

\section{Models of Vela X-1}
\label{sec:models}

Like most wind-fed HMXBs, Vela X-1 is a system that can only be appreciated with multiphysics and multiscale models (see Figure\,\ref{fig:radii}). Many efforts have been made to capture the predominant mechanisms at each scale. In simplified geometries assuming an isotropic stellar wind (1D models) or focusing on the orbital plane (2D models), some semianalytic results were derived. However, since the late 1980s, numerical simulations have been the privileged tool for capturing the whole 3D complexity of this archetypical system and provide suitable models for interpreting the observations.



\subsection{Stellar wind}
\label{sec:models:wind}

\subsubsection{Hydrodynamic structure}
\label{sec:models:hd_structure}


The donor star is not isolated in Vela X-1. The presence of the orbiting neutron star breaks the spherical symmetry of the problem, and the wind cannot be fully described in a 1D framework. If the wind speed were much higher than the orbital speed, the wind would only be affected by the gravitational field of the neutron star in a region far smaller than the orbital separation (see the discussion of the accretion radius in Section\,\ref{sec:models:accretion}). Except in the thin wake of the accretor, which is axisymmetric with respect to the axis joining the two bodies, the 1D representation would hold. On this implicit assumption \cite{Watanabe:2006} relied when they empirically introduced a cone-shaped dense and cold cloud in the wake of the neutron star to explain the excess of absorption at inferior conjunction. In the same spirit, \cite{Fryxell1987} performed wind tunnel simulations: They ran 2D axisymmetric hydrodynamic simulations on a spherical grid of a planar adiabatic flow coming from infinity with a supersonic relative speed that was deflected by the gravitational field of an accretor. This configuration corresponds to the problem that was first addressed by \cite{HoyleLyttleton:1939} and \cite{BondiHoyle:1944} in ballistic terms, the BHL configuration. \cite{Fryxell1987} found a good agreement with the BHL predictions and that an axisymmetric bow shock forms around the accretor, with an opening angle depending on the Mach number of the flow. Provided the orbital effects are negligible, that is, when the wind speed is much higher than the orbital speed, the flow in a wind-fed HMXB can be divided into two parts: A purely radial wind flowing away from the donor star, and in the immediate vicinity and in the wake of the compact object, a flow that is well described by the BHL solution or by its hydrodynamics counterpart. 


Nevertheless, in Vela X-1, a more realistic wind speed at the orbital separation of about or even lower than the orbital speed, such as the one used by \cite{Watanabe:2006} and \cite{Sander:2018} (see Section\,\ref{sec:ion_struct}), indicates an anisotropic wind that surrounds the donor star. The outflow is strongly shaped by the orbital motion, as is corroborated by the systematic asymmetries observed in the absorbing column density profile\footnote{An eccentric orbit in an isotropic wind would also result in an asymmetric $N_\textrm{H}$ profile, although of more limited contrast than observed.} between orbital phases 0-0.5 and 0.5-1 (see Secion\ref{sec:obs:xray:abs}). In this case, a fully 3D model is required to describe the hydrodynamic structure of the wind and interpret the evolution of the observations with the orbital phase in a high-inclination HMXB such as Vela X-1. \cite{Bessell:75} first highlighted the anisotropic distribution of material in the orbital plane induced by the tidal forces. It was later confirmed by 2D hydrodynamic simulations in the equatorial plane of a spherical mesh by \cite{Blondin:90,Blondin:91}. They characterized the gravitational and radiative impacts of the neutron star on the wind: They showed that the nonsteady and overdense envelope of the bow shock that is formed upstream trails the neutron star throughout its orbit. This structure is referred to as the accretion wake. \cite{Blondin:91} further showed that in simulations of Vela X-1, the wind was highly focused into a steady tidal stream between the star and the neutron star. This feature is clearly visible in the 2D simulations by \cite{ManousakisWalterBlondin:2012}. On the other hand, wind tunnel simulations centered on the neutron star but with a nonplanar inflow were previously performed on 2D cylindrical $(r,\phi)$ grids \citep{TaamFryxell:88,FryxellTaam:88,TaamFryxell:89}. The aim was to parameterize the shearing in velocity and/or mass density of the inflow that is caused by the orbital effects. They identified an oscillation of the accretion wake around the main axis of the cylinder, which they called the flip-flop instability \citep[see also][for the seminal simulations in which this effect was determined]{Matsuda:87,Matsuda1991}. First thought to be responsible for the formation of transient disks in wind-fed HMXBs, this feature was later found to be a numerical artifact due to the unrealistic geometry of the mesh and to the unphysically large size of the sink particle standing for the accretor \citep{Ishii1993,Ruffert1999,Blondin2012}. In addition to the tidal stream and the accretion wake, a spiral-shaped density enhancement is to be expected at the side of the star opposite to the neutron star due to the tidal forces \citep{ElMellah2018a}. With the 3D framework they worked with, these authors also solved the dynamics of the wind off the orbital plane and found a significant flattening of the wind in the orbital plane when the ratio of the wind speed to the orbital speed was similar to what is expected for Vela X-1. 


In Vela X-1, the microstructure of the wind is still poorly constrained, although \cite{Gimenez-Garcia:2016} measured a density contrast in the wind that is consistent with theoretical predictions. Observational diagnostics that depend quadratically on density (e.g., H$\alpha$ line and thermal radio emission) lead to overestimated mass-loss rates when clumping is not accounted for \citep{Sundqvist+Owocki+Puls:2011}. Conversely, the mass-loss rates deduced from diagnostics that depend linearly on density (e.g., UV resonance lines) can be underestimated \citep{Sundqvist:18}. Therefore the current uncertainties on the clumping factor prevent us from drawing a definitive conclusion concerning the stellar mass-loss rate. In Vela X-1, the size of the clumps derived from simulations and from the two main observational diagnostics at our disposal, absorption variability and X-ray flares, differs. While the first is thought to be due to unaccreted clumps passing the line of sight \citep[see, e.g., the observations and model by][respectively]{Grinberg:2017,ElMellah2020_NH_paper}, the second have been proposed to be due to the serendipitous capture of a clump by the neutron star \citep{Fuerst:2010,Martinez-Nunez:2014}. However, hydrodynamics simulations by \cite{El-Mellah:2018} showed that the bow shock around the accretor significantly lowers the variability compared to a purely ballistic model of the clump capture. It suggested that although intrinsic X-ray flares might be triggered by clump capture, the amount of mass involved in a flare was probably much more important than the amount of mass contained in a single clump. A better understanding of the whole accretion process, from the orbital scale all the way down the accretion columns where most of the X-rays we observe originate, is necessary to determine the connection between the flares and the clumpiness of the wind.

\subsubsection{Ionization structure and inhibition of wind acceleration}
\label{sec:ion_struct}

It is commonly accepted that the X-ray irradiation from the neutron star significantly affects the wind velocity profile, but it has proven very challenging to quantify the extent of this effect on the dynamics of the flow. In a seminal paper, \cite{Fransson+Fabian:1980} described how the photoionization of the wind by the accretor in high-mass X-ray binaries could lead to the formation of a shock between the accelerating wind and the stalling photoionized plasma. \cite{Blondin:90} quantified this effect with 2D numerical simulations. They modeled the effect of the X-ray ionizing feedback on the hydrodynamic structure of the wind. Assuming that wind acceleration was unaltered below $\xi_\text{crit}=10^{2.5}$erg cm s$^{-1}$ and fully inhibited above, they showed that the wind speed in the direction of the neutron star was lowered by a factor of $\sim$2. Because of the stalling of the line-driven wind that is caused by X-ray photoionization, in some simulations, the wind formed a trailing spiral density enhancement between the neutron star and the star. \cite{Kaper:94} relied on this photoionization wake to interpret an absorption component that they were unable to attribute to the accretion wake or to the tidal stream. 


Although wind acceleration is inhibited in the limit case of a high X-ray illumination, for intermediate fluxes, it is not a step function nor even a monotonous dependence of the X-ray luminosity. For instance, as already noted by\citet{MacGregor:82}, for intermediate X-ray photoionizing fluxes, the wind would overall be more efficient at absorbing UV photons, and the line-driven acceleration would even become higher. In Vela X-1, a few studies initiated by \cite{Sako:99}, went beyond the on/off switch assumption in different ways. In a preliminary approach, \cite{Watanabe:2006} assumed simplified descriptions of the line acceleration, the radiative transfer, and the ionization structure to compute 1D velocity profiles. They found a wind speed at the orbital separation in Vela X-1 of 180km$\cdot$s$^{-1}$ for $L_\text{X}=$3.5$\times$10$^{36}$erg$\cdot$s$^{-1}$ instead of 570km$\cdot$s$^{-1}$ without X-ray ionizing feedback. However, they did not iterate the radiative transfer and computation of the ionization structure and only accounted for a very limited number of line transitions.


More advanced treatments have emerged in the past decade to determine the effect X-rays have on the wind acceleration. In these methods, realistic stellar atmosphere models accounting for the inherent NLTE conditions and the moving outer layers were irradiated by an external X-ray source in order to study the imprint of the different elements and ions on the line acceleration. \cite{Krticka:2012} derived wind solutions with the \texttt{METUJE} code \citep{KrtickaKubat:2017} for different axis in the orbital plane and iterated the radiative transfer and statistical equilibrium equations to obtain a consistent wind ionization structure. In a follow-up study, \citet{KrtickaKubatKrtickova:2018} accounted for optically thin wind clumping and obtained similar and very low values of the critical ionization parameter, above which line-acceleration was significantly inhibited, ranging between 5 and 25 erg$\cdot$cm$\cdot$s$^{-1}$. In both cases, they computed wind velocities that were severely lower in the direction of the neutron star in Vela X-1, with a speed that hardly peaked at 100km$\cdot$s$^{-1}$. Following a spectral analysis by \cite{Gimenez-Garcia:2016} that yielded updated parameters for the donor star and wind in Vela X-1, \cite{Sander:2018} performed an analysis with the dynamically consistent branch of the \texttt{PoWR} code \citep{Sander:2017} to study the effect of X-ray irradiation on individual driving contributions and spectral lines. They showed that with the updated parameters, an X-ray ionizing source even as high as 10$^{37}$erg$/$s does not necessarily inhibit wind acceleration, but might even enhance it upstream the neutron star. While both \texttt{METUJE} and \texttt{PoWR} perform calculations in 1D, their results demonstrate that the wind speed could be so low that the outflow significantly departs from a spherically symmetric wind. However, the computational cost of these methods precludes any 3D direct computation in the next years.

Finally, it is worth mentioning that a preliminary global model of the radiation-driven photoionized wind was sketched and applied to Vela X-1 in \cite{Mauche:2007} and \cite{Mauche:2008AIPC}. The authors combined different numerical tools. First, 2D and 3D hydrodynamical simulations were performed with FLASH \citep{Fryxell2000}. Second, the wind ionization structure was computed with the photoionization code XSTAR in order to derive the local heating and cooling rate, along with the ionization fractions of the different ions \citep{Kallman2001}. Third, the rich HULLAC database provided the emission models for the X-ray photoionized plasma \citep{BarShalom2001}. Finally, a Monte Carlo radiative transport computation enabled them to derive detailed synthetic X-ray spectra.

\subsection{Accreted flow}
\label{sec:models:accretion}

In Vela X-1, the ionization state and thus the whole hydrodynamic structure of the flow depends on the intensity of the X-ray emission from the neutron star, fed by the wind it illuminates. It emphasizes the need addressed by various authors for simultaneously modeling the orbital scale and the accretion process itself in order to achieve a fully consistent model that can connect the X-ray luminosity to the mass accretion rate. The accretion of stellar material by the neutron star proceeds in different steps. A mass transfer mechanism must first ensure that stellar material is brought into the Roche lobe of the neutron star if the flow has a negligible amount of specific kinetic energy, or is within the accretion radius given by equation\,\eqref{eq:R_acc} otherwise. These two characteristic length scales far exceed the other relevant scales of accretion, therefore we can first treat the mass transfer alone (Section\,\ref{sec:models:mass_transf}). Then, accretion can proceed onto the magnetosphere of the neutron star and eventually carry the plasma all the way down the X-ray emitting regions near the magnetic poles (Section\,\ref{sec:models:acc-ind_torq}).

\subsubsection{Mass transfer mechanisms}
\label{sec:models:mass_transf}

The donor star in Vela X-1 is unlikely to undergo RLOF. In X-ray binaries, mass transfer through RLOF usually leads to higher X-ray luminosities than in Vela X-1. Moreover, RLOF leads to the formation of a permanent accretion disk, which has not been observed in Vela X-1. The torques by such a disk would also spin the neutron star up, as is the case in Be X-ray binaries, but this contradicts the measured long neutron star spin period. An additional strong argument against mass transfer through RLOF comes from the stable orbital period \citep{Falanga:2015} because no outflow from the vicinity of the neutron star is observed and conservative mass transfer from a high-mass star to a low-mass accretor would lead to a quickly shrinking orbit \citep{Quast2019,ElMellah2020}. Instead, accretion of the stellar wind is to be preferred as the dominant mass transfer mechanism.

Wind accretion as described by the BHL formalism might be a misleading alternative, however. Equation\,\eqref{eq:BHL_mdot} has been widely used to interpret the intrinsic X-ray luminosity of Vela X-1 and its variations, but fundamental caveats must be acknowledged, especially because the wind speed at the orbital separation is not high compared to the orbital speed. It is plausible that a hybrid mass transfer mechanism is taking place in Vela X-1 \citep[called wind-RLOF and introduced by][]{Mohamed+Podsiadlowski:2007}, where the wind is significantly disrupted and the geometry of the problem no longer obeys that of the planar BHL (see Section\,\ref{sec:elements:starwind}). For instance, the wind is compressed in the orbital plane by the motion of the two bodies, so that the density in equation\,\ref{eq:BHL_mdot} cannot be deduced from the continuity equation in spherical geometry. An approach to obtain more accurate mass transfer rates was introduced by \cite{El-MellahCasse:2017}, who assumed a donor star in synchronous rotation with the orbit and applied their model to Vela X-1: Because the wind is highly supersonic, ballistic streamlines reveal how beamed the flow is toward the neutron star, and density-enhanced (depleted) regions are identified where the streamlines diverge slower (faster) than in the spherically geometric case. The streamlines that intercept within the Roche lobe of the neutron star indicate the fraction of the stellar wind that will take part in the accretion process. They also enable us to compute the net specific angular momentum of the flow, which significantly departs from zero: An even more important reason to avoid the use of equation\,\eqref{eq:BHL_mdot} in Vela X-1 is indeed that the flow might carry enough angular momentum to circularize before it reaches the neutron star magnetosphere, in contrast to the intrinsic assumption of cancelling angular momenta in the BHL picture. 

\begin{figure}
\centerline{\includegraphics[width=1.0\linewidth]{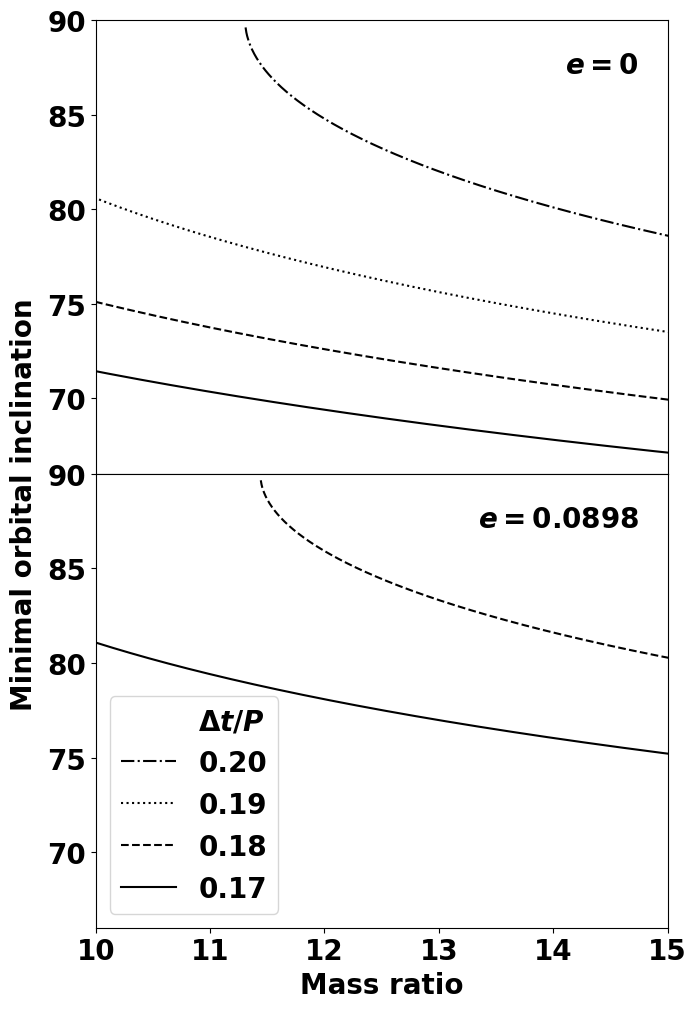}}
\caption{Minimum inclination angle of the orbit for avoiding RLOF as a function of the mass ratio and for different durations of the eclipse relative to the orbital period. In the upper panel, a circular orbit is assumed, and in the lower panel, we account for an eccentricity of 0.0898 and enforce no RLOF even at periastron.}
\label{fig:imin}
\end{figure}

The fact that the donor star does not fill its Roche lobe sets a stringent constraint on the orbital inclination angle $i$. If the eclipse duration relative to the orbital period is $\Delta t/P$, the ratio of the stellar radius $R$ to the orbital separation $a$ reads
\begin{equation}
\label{eq:eclipse_duration}
\frac{\Delta t}{P}=\frac{1}{\pi}\arcsin\left(\frac{\sqrt{R^2-a^2\cos^2 i}}{a\sin i}\right)
,\end{equation}
where for now, we neglect the eccentricity of the orbit and the tidal deformations of the star. The condition of absence of RLOF yields a lower limit on the inclination angle,
\begin{equation}
    i>i_{\text{min}}=\arcsin\left[\frac{\sqrt{1-\mathcal{E}^2(q)}}{\cos\left(\pi{\Delta t}/{P}\right)}\right]
,\end{equation}
where $\mathcal{E}$ is the estimate of the ratio of the stellar Roche-lobe radius to the orbital separation provided by \cite{Eggleton:83} and function of the mass ratio alone. For a given range of $q$ and $\Delta t/P$, we can thus deduce a range of minimum orbital inclination angles below which the donor star would overflow its Roche lobe. In the upper panel in Figure\,\ref{fig:imin}, we plot the minimum orbital inclination angle to avoid RLOF as a function of the mass ratio for realistic eclipse durations. At low mass ratios and for long eclipse durations ($q<12$ and $\Delta t/P>0.2$), the donor star would fill its Roche lobe even if the system were seen edge-on. Currently, the upper limit set on the eclipse duration is closer to $\Delta t/P=0.19$ \citep{Kreykenbohm:2008}, which would indicate a lower limit for the orbital inclination of $\sim$74$^{\circ}$. This conclusion must be tempered, however, because the orbit is not circular. Moreover, the proximity of the donor star to filling its Roche lobe means that the star should not be assumed to be spherical, and only an effective stellar radius can be defined.

\begin{figure}
\centerline{\includegraphics[width=1.0\linewidth]{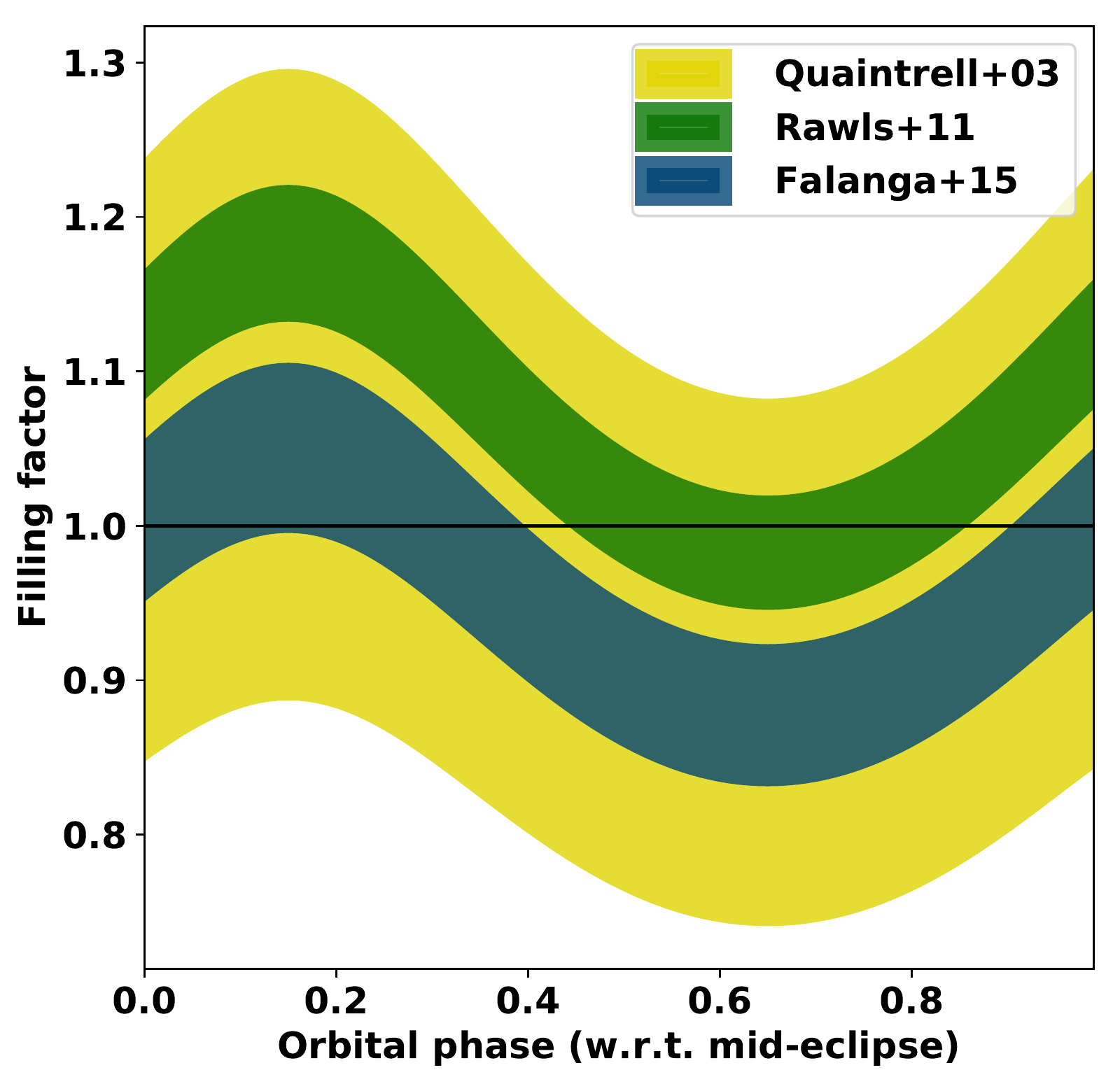}}
\caption{Stellar filling factor, i.e., the ratio of the stellar radius to the Roche-lobe radius, as a function of the orbital phase (0 is mid-eclipse). Each colored stripe is obtained for a different set of extreme values observationally derived by \cite{Quaintrell:2003}, \cite{Rawls:2011}, and \cite{Falanga:2015}.}
\label{fig:filling_factors}
\end{figure}

Furthermore, the eccentricity of the orbit in Vela X-1 leads to a filling factor that varies with the orbital phase: Because the orbit is eccentric, the orbital separation varies, and so does the filling factor. When we enforce the more severe constraint to avoid RLOF even at periastron, we obtain the minimum inclination angles displayed in the lower panel in Figure\,\ref{fig:imin}, which are significantly higher. Two conclusions can be drawn from this result. Either the system presents an orbital inclination and a mass ratio that lie at the upper edge of the intervals derived from observational diagnostics not based on the assumption of the absence of RLOF, or there are regular episodes of intermittent RLOF at periastron. In Figure\,\ref{fig:filling_factors} we show the evolution of the Roche-lobe filling factor of the donor star as a function of the orbital phase for three different sets of system parameters derived from observations by \cite{Quaintrell:2003}, \cite{Rawls:2011}, and \cite{Falanga:2015}. At a given orbital phase, the uncertainty ranges are upper limits, obtained by considering the extreme values obtained without accounting for the correlations between parameters. The striking result is that observationally derived parameters tend to give filling factors above 1. When we take the absence of RLOF for granted, it means that we can set stronger constrains on the system parameters. With an eccentricity of 0.0898, however, the filling factor necessarily varies by $\sim$15\% throughout the orbit. This major variation implies a mass transfer mechanism that might be modulated throughout the orbit, with a short phase of mass transfer through RLOF at periastron.

\subsubsection{Magnetospheric accretion and induced torques}
\label{sec:models:acc-ind_torq}

Following its formation, a pulsar spins down first due to its autonomous magnetodipole radiation \citep{Pacini:68,Gunn:69}. After a few thousand years, it spins down at a higher pace due to the torques associated with the repelling of infalling stellar material that is ejected during the propeller phase \citep{Francischelli:02}. After this, the magnetospheric and corotation radii are expected to be similar because the torques applied by the accreted flow onto the neutron star magnetosphere tend to spin it up (spin it down) if the corotation radius is larger (smaller) than the magnetospheric radius \citep[see][for a study of the long-term evolution toward spin equilibrium that covers these three phases]{Ho:20}. Because this equilibrium phase is reached within only a few 10 kyr, a first attempt to estimate the neutron star magnetic field from its spin period and the mass accretion rate can be made assuming that the magnetospheric radius and the corotation radius match. The corotation radius of the neutron star in Vela X-1 can be accurately deduced from the mass of the neutron star and from its spin period,
\begin{equation}
    R_{co}=\left(\frac{GMP^2}{4\pi^2}\right)^{1/3}\sim 8.1\times 10^9 \text{cm} \left(\frac{M}{2\,M_{\odot}}\right)^{1/3} \left(\frac{P}{283\,\text{s}}\right)^{2/3}
.\end{equation}
In addition to being stable over timescales shorter than years, the corotation radius is known with a 10\% precision given its low dependence on the mass and spin period of the neutron star. On the other hand, the expression that determines the magnetospheric radius depends on the geometry of the accretion flow (planar, disk, spherical), but an order-of-magnitude estimate can be obtained from \citep{Lamb:73,Wang:96}
\begin{equation} 
\begin{split}
\label{eq:R_mag}
R_{mag}\sim 4.9\times 10^8 \text{cm} & \left(\frac{\dot{M}_{acc}}{2\times10^{-10} \text{M}_{\odot}\cdot\text{yr}^{-1}}\right)^{-2/7} \left(\frac{M}{2\,M_{\odot}}\right)^{-1/7} ...\\
    & ... \left(\frac{B}{10^{12}\,\text{G}}\right)^{4/7} \left(\frac{R_{NS}}{12\,\text{km}}\right)^{12/7}
\end{split}
,\end{equation}
where the inertia moment of the neutron star is computed assuming a uniform sphere (i.e., $2MR^2/5$), where an accretion rate of $2\times10^{-10} \text{M}_{\odot} \text{yr}^{-1}$ corresponds to an accretion luminosity of 4$\times10^{36}$erg s$^{-1}$ for a conversion factor of 35\%, and where we used the definition of a magnetic dipole $\mu$ as $BR^3/2$.

At equilibrium, the magnetic field should be such that these two radii are equal. With numerical values coherent with those deduced for Vela~X-1, we obtain an estimate of the magnetic field of $B \sim 2.2\times10^{14}\text{G}$. This value is at odds with what was measured relying on the CRSFs diagnostics (see section\,\ref{sec:obs:xray:cyclo}). \cite{Fuerst:2014} measured a cyclotron line energy corresponding to a magnetic field $B \sim 2.6\times10^{12}\text{G}$, which is two orders of magnitude smaller than the value derived assuming spin equilibrium. The serious mismatch between the magnetic field values independently deduced from the corotation radius and from CRSFs in Vela X-1 would mean that the neutron star has been spun down to a spin period much higher than the equilibrium spin period set by the torques predicted by \cite{Ghosh:79b} and corresponding to the estimated mass accretion rate. This is probably the strongest theoretical point that can be made against the presence of a permanent disk beyond the neutron star magnetosphere in Vela X-1. The magnetic field required to account for the observed neutron star spin variations is also higher than the field measured by the CRSFs diagnostics by two orders of magnitude \citep[see][for a detailed discussion]{Staubert:2019}.

Alternatively, this overbraking could be indicative of a phase in the past in which stellar material was supplied to the neutron star at a much lower rate than today, and with a high amount of specific angular momentum, which is suggestive of a slow wind \citep{Wang:20,Ho:20}. The system would have been in the propeller regime, and the neutron star would have spun down until the mass accretion rate suddenly increased by so much that the magnetospheric radius became smaller than the corotation radius, as is currently observed if the magnetic field measured through CRSFs is accurate. We must acknowledge, however, that this scenario requires a fine-tuning of the increase in mass accretion rate, of the initial magnetic field, and of the magnetic field decay, which is unlikely. In what follows, we rely on the magnetic field value obtained by direct measures through CRSFs, that is, $B \sim 2.6\times10^{12}\text{G}$.


The first conclusion to draw from this value of the magnetic field is that in Vela X-1, the neutron star is never expected to be in the propeller regime. In equation\,\eqref{eq:R_mag}, the X-ray luminosity that is required for the magnetospheric radius to be as large as the corotation radius is about 10$^{32}$erg s$^{-1}$. This threshold is too low to be compatible even with the lowest observed X-ray flux values \citep{Liao:2020}. For the propeller effect to be triggered, a drop in mass accretion rate by 3 to 4 orders of magnitude compared to its median value would be needed \citep{Bozzo:08,Doroshenko:2011}. For a realistic wind microstructure, the inter-clump environment is therefore insufficiently empty to cause such a drop \citep{Sundqvist:18}, notwithstanding the lowered variability due to the mixing of the wind material at the hydrodynamics shock \citep{El-Mellah:2018}.


In spite of the absence of the propeller accretion regime in Vela~X-1, we do observe month- to year-long episodes of spin-up and spin-down. Each phase occurs at a steady spin period rate: $\sim$7$\times$10$^{-4}$s day$^{-1}$ to $\sim\times$10$^{-3}$s day$^{-1}$ for spin-up and $\sim$10$^{-4}$s day$^{-1}$ for spin-down \citep{Malacaria:20}. They are separated by sudden torque inversions, and the net spinning up over the past decades is compatible with zero. In neutron-star-hosting HMXBs in general and in Vela X-1 in particular, the spinning up or down of neutron stars has long been attributed to the coupling between the accreted material and the magnetic field of the neutron star. As described in Section\,\ref{sec:elements:magnetosphere}, the modalities of the coupling differ between the quasi-spherical and disk geometries, which both lead to different accretion-induced torques onto the neutron star magnetosphere. For disk accretion and given the parameters of Vela X-1, the torque applied to the neutron star computed by \cite{Ghosh:79b} spins it up and produces a spin period derivative of approximately \citep{Ho:2014,Malacaria:20}
\begin{equation} 
\begin{split}
| \dot{P}_\text{disk,+} | \sim  & 3.5 \text{s} \text{yr}^{-1} \left(\frac{P}{283\,\text{s}}\right)^{2} \left(\frac{M}{2\,M_{\odot}}\right)^{-4/7} ... \\
    & ... \left(\frac{\dot{M}_{acc}}{2\times10^{-10} \text{M}_{\odot} \text{yr}^{-1}}\right)^{6/7} \left(\frac{B}{2.6\times 10^{12}\,G}\right)^{2/7} \left(\frac{R_\text{NS}}{12\,\text{km}}\right)^{-8/7} 
\end{split}
.\end{equation}

For a quasi-spherical geometry, the evolution of the neutron star spin period depends on the accretion regime. The transition from the direct accretion regime to the subsonic propeller regime occurs when the mass accretion rate decreases below $4\times 10^{36}$\,erg s$^{-1}$, which corresponds to the median X-ray luminosity of Vela~X-1 \citep{Shakura:2012}. Then, both regimes should occur in this system. \cite{Shakura:2012} set the theoretical framework for computing the accretion-induced torques applied by a quasi-spherical flow onto the magnetosphere. \cite{Shakura:18} interpreted the absence of net spinning-up or -down of the neutron star over the past decades as an indication that the system must be in spin equilibrium. Based on this assumption and on the observed spinning-up and -down rates in Vela X-1, they estimated the value of the dimensionless parameters of their model ,which encapsulate the physical mechanisms at stake at the outer rim of the magnetosphere. With the parameters of Vela X-1, it yields the following spinning-up and -down torques for Vela X-1: 
\begin{equation}
\begin{cases}
  &|\dot{P}_\text{sph,+}|\sim 0.1 \text{s} \text{yr}^{-1} \left(\displaystyle\frac{v_r}{700\,\text{km s$^{-1}$}}\right)^{-4} \left(\displaystyle\frac{R_\text{sh}}{R_\text{acc}}\right)^{2} \\
  &|\dot{P}_\text{sph,$-$}|\sim 1   \text{s} \text{yr}^{-1} 
\end{cases}
,\end{equation}
respectively, where $R_\text{sh}$ is the distance from the bow shock to the neutron star, which typically is on the order of $R_{\text{acc}}/5$ \citep{ElMellah:15}, $\dot{P}_{sph,-}$ weakly depends on the system parameters, and only the two strongest dependences are shown for the spin-up rate. These values are to be compared to the spin period derivatives measured in Vela X-1, which are about 0.25--0.5 s yr$^{-1}$ (spin up) and 0.05 s yr$^{-1}$ \citep[spin down,][]{Malacaria:20}. In contrast, the quasi-spherical subsonic settling model predicts higher spin-down than spin-up rates: While the predicted spin-up rate is on the order of the observed rate, the spin-down rate is overestimated by more than an order of magnitude. Furthermore, in the moderate coupling regime, Vela X-1 is expected to accrete. In this regime, a correlation of the value of the observed spin period derivatives and the X-ray luminosity is expected but has not been reported. These inconsistencies might be due both to the numerous unknowns of this model and to the strong dependence of the spin-up rate on the location of the bow shock with respect to the accretion radius, which like the relative wind speed is unknown to within at least a factor of 2. Finally, in the supersonic accretion regime of the quasi-spherical settling model, the neutron star is generally spun up by accretion-induced torques. Spinning down can only occur if the neutron star spin axis is upside down with respect to the orbital angular momentum axis, or if clumps carry enough angular momentum to momentarily cause an inversion of the sign of the accreted net angular momentum. 

\subsection{Emission from the accretion column or polar cap} \label{sec:models:column}

Most of the hard X-rays we observe are produced very close to the neutron star within the accretion column. In this column, the hot in-falling plasma can upscatter through Compton-interaction seed photons to very high energies that roughly resemble the observed continuum spectrum \citep{Davidson:73, DavidsonOstriker:73, MeszarosNagel:85a, Rebetzky:88}. However, because the problem is complex, these early models had to make many simplifying assumptions, such as reducing the problem to 1D or neglecting the accretion flow dynamics. When they escape the accretion column, the photons cool the plasma and prevent a full thermalization of the flow \citep{Arons:87}. The source of the seed photons is still debated: A  large part very likely is due to cyclotron radiation \citep{Nagel:81b,Bussard:85, Arons:87}, but thermal photons from the accretion mound on the neutron star surface will also play a role \citep{BeckerWolff:2005a}.

While this basic idea of Comptonized photons agrees roughly with the observed spectra, detailed modeling of the spectrum produced in the accretion column continues to be difficult even with modern computers. These difficulties mainly arise because for a correct description of the particles and their interaction within the column, a full quantum-mechanical as well as relativistic treatment is necessary. In particular, the electron cross-sections are strongly polarization dependent and are increased at the cyclotron energy  by orders of magnitude \citep[e.g.,][see also Sect.~\ref{sec:models:CRSF}]{DaughertyHardin:86}, which is computationally expensive to model. Additionally, the open question remains of where and how the in-falling plasma is decelerated \citep[e.g.,][]{BeckerWolff:2005a}.

In the past decade, \citet{BeckerWolff:2005a, BeckerWolff:2005b, BeckerWolff:2007, Becker:2012} have presented a more detailed model of the spectral formation in the accretion column, taking bulk (first-order Fermi acceleration) and thermal Comptonization into account and injecting seed photons from bremsstrahlung, cyclotron radiation, and thermal contribution. This model is able to reproduce the spectral shape of a few selected X-ray binaries, but requires a simplified description of the velocity in the accretion column and does not account for the variability of theses sources \citep{Wolff:2016}. It assumes a high luminosity, in which a radiation-dominated shock is formed above the neutron star surface and decelerates the in-falling plasma to subsonic speeds.

\citet{Farinelli:2012,Farinelli:2016} expanded on the model by \citet{BeckerWolff:2007}, allowing for different velocity profiles and a magnetic field gradient within the accretion column. Their model allowed them to report a satisfying description of the X-ray spectra of bright sources such as Her~X-1 and 4U\,0115+63.

\vel only rarely exhibits luminosities above $10^{37}$\,erg\,s$^{-1}$ . It is therefore unclear whether a radiation dominated shock forms. This intermediate-luminosity range is in particular hard to model because it is unclear which approximations can be made. \vel also shows a complex pulse profile, which might indicate that both fan and pencil beam components are present. The most recent update of the Becker \& Wolff models was presented by \citep{West:2017a, West:2017b}, but the model has not yet been applied to \vel.

The observed pulse profile strongly depends on the assumed emission profile of the accretion column and of the polar cap, and also on light-bending effects around the neutron star. For the emission geometry, two simple approximations are often made: a slab, or a column geometry. In the slab geometry, the accretion mainly escapes from the sides of the accretion column (and perpendicular to the magnetic field) close to the neutron star surface, creating a so-called fan-beam pattern (i.e., in a similar configuration as a polar cap).  In a column geometry, the emission escapes along the column and parallel to the magnetic field, forming a pencil beam, with a certain (narrow) opening angle \citep[]{ MeszarosNagel:85b,Meszaros+Riffert:88}. 

While these two different geometries predict significantly different pulse profiles \citep[]{MeszarosNagel:85b, WangWelter:81}, they prove difficult to compare with observations. The observed profiles are highly complex and strongly energy dependent (see Sect.~\ref{sec:obs:xray:pulseprofiles}), and this makes them far more complicated than simple models predict.

Over the years, various efforts were made to offer physical descriptions of the pulse profile shapes. Early studies were concerned with the physical production region of the X-rays and their emission geometry. They found that the models were able to match the data for \vel approximately \citep[e.g.,][]{WangWelter:81, Leahy:90, Sturner:94}. However, the omission of gravitational light-bending led to unrealistic estimates of the emission geometry \citep{Riffert:93, Leahy:95, Kraus:95}. 

More detailed work led to descriptions of profiles for single sources,  using a large number of assumptions \citep[][]{Kraus:95, Caballero:2011, Falkner:2018PhD}. Recently, models have been presented that allow the user to be very free in the choice of geometry and emission pattern, and they no longer require many of the previous assumptions \citep{Cappallo:2017, Falkner:2018PhD}. However, the sheer number of free parameters and large degeneracy between them has so far made it impossible to identify unique solutions. Some of the most relevant physical parameters are the amount of relativistic boosting of the in-falling plasma, the size and mass of the neutron star, and the configuration of the magnetic field. For \vel, no recent physical description of the pulse profile has been put forward at the time of this review.

\subsection{Cyclotron resonant scattering features} \label{sec:models:CRSF}

As explained in Section~\ref{sec:elements:column}, the CRSF features that are visible in broadband X-ray spectra of Vela~X-1 (see Sect.~\ref{sec:obs:xray:cyclo}) are caused by resonant scattering of X-ray photons at the corresponding energies on electrons with quantized motion perpendicular to the magnetic field.
Calculating the scattering cross-sections requires detailed fully relativistic QED-based calculations. Over the decades, such calculations have been undertaken by various authors and have included increasingly complex details \citep[e.g.,][]{HardingDaugherty:91,Sina:96PhD,Isenberg:98a,Isenberg:98b,ArayaHarding:99,Schoenherr:2007}. The most recent treatment based on elaborate Monte Carlo calculations is presented in \citet{Schwarm:2017a,Schwarm:2017b}.

The scattering cross-sections are very strongly peaked close to the Landau level energies, up to several orders of magnitude compared to Thomson scattering. This makes the plasma optically very thick for photons at these energies. Because the  the energy of the electrons is only quantized in the direction perpendicular to the magnetic field, however, the angle between the magnetic field and the photon becomes relevant. For large angles, the cross-sections become thermally broadened and shift to higher energies for the harmonic levels. 
An open question in CRSF research is the fact that model calculations \citep{ArayaHarding:99,Schwarm:2017a,Schwarm:2017b} tend to predict asymmetrical lines, frequently showing emission wings at energies below and above the central energy, while observed features tend to be broad, without a marked asymmetry, and without a clear sign of the predicted emission features.

In order for discrete CRSFs to be observable, the sample magnetic field has to be confined to a very narrow range, indicating a closely confined region within the accretion column, possibly a shock region in the column or close to the poles \citep[and references therein]{BeckerWolff:2007}. The often observed relatively broad and shallow CRSFs might arise because multiple line-forming regions contribute \citep{Nishimura:2008}. \citet{Poutanen:2013cyclo} proposed another explanation according to which, CRSFs are formed due to reflection of the downward-beamed radiation from the accretion column on the neutron star surface around the poles. 

The strong angular dependence of the line features predicted in theoretical calculations would in principle lead to another diagnostic for possible emission geometries from the pulse phase, and thus geometrically, resolved spectroscopy. This is limited by the fact that because of gravitational light-bending, we can expect that multiple zones of the accretion column or neutron star surface contribute at least most of the time \citep{Falkner:2018PhD}. We would therefore also need to have a good grasp on separating these  mixed contributions.

\subsection{Variability of the observed X-ray emission}
\label{sec:models:variability}

In variability studies, the two privileged simulation outputs to be compared to observations are occurrence diagrams and power spectral distributions. While the first show the fraction of time that a variable spends at a given value (e.g., the mass or angular momentum accretion rates, or the column density), the second retains the information and reveals coherence timescales and possible periodicities.

\subsubsection{Variable absorption along the line of sight}
\label{sec:abs_var}

The mass-loss rate from the donor star is high, therefore absorption along the line of sight from the neutron star to the observer is high but is also variable (Section\,\ref{sec:obs:xray:abs}). Because we observe Vela X-1 close to edge-on, we expect a periodic modulation of the column density as the neutron star orbits the blue supergiant. The average orbital $N_\textrm{H}$ profile grants access to the orbital parameters and to the stellar mass-loss rate, provided estimates are available for the wind speed and the stellar radius. The origin of the stochastic variability of the measured absorbing column density at a given orbital period is threefold. It can be due either to changes in the amount of material that is integrated along the \los  because of clumps in the stellar wind or because of a change in the flow structure near the neutron star magnetosphere, or modifications in the opacity of the medium can cause it through changes in the wind ionization structure.

Regardless of whether they participate in the accretion process, clumps intercepting the \los will produce a variability in column density on short timescales. \cite{OskinovaFeldmeierKretschmar:2012} designed a 2D model of massive ($3\times10^{21}$g) and intermediate-size clumps (2\% of the stellar radius at one stellar radius above the stellar photosphere) that propagate radially from the donor star. They computed the extinction coefficient as a function of time and showed how clumps that are compressed in the radial direction tend to produce a higher level of variability than their spherical counterparts \citep[see also][]{Oskinova:06}. \cite{Grinberg:2017} estimated the characteristic amplitude and timescale of absorption episodes with a simple model of clumps, to be compared to column densities derived from hardness monitoring with time-resolved spectroscopy at orbital phase $\phi\sim 0.25$. This motivated a comprehensive exploration of the $N_\textrm{H}$ variability induced by a clumpy wind: \cite{ElMellah2020_NH_paper} designed a 3D model of clumps flowing radially away from the star with different velocity profiles, orbital parameters, and clump properties. Vela X-1 was one of the two HMXBs to which the model was applied. The authors computed the median $N_\textrm{H}$ orbital profile over many orbital periods, along with the standard deviation and coherence timescale at each orbital phase. The timescale was found to be associated with the flyby time of the smallest clumps along the \los, which granted access to their spatial extent. Based on the porosity length, the authors also showed that the mass of the clumps could be constrained based on the standard deviation of the column density: Because they are rarer, higher-mass clumps produce a higher spread of the measured column density at a given orbital phase.

\cite{ManousakisWalterBlondin:2012} computed the structure of a smooth stellar wind in hydrodynamics numerical simulations in the orbital plane. Although their model was tailored to another high-mass X-ray binary and not to Vela X-1, they were able to reproduce the excess of column density at orbital phases 0.4-0.8 reported in \cite{Manousakis11} for IGR J17252-3616 and traced its origin back to an overdense accretion wake trailing the neutron star. In Vela X-1, \cite{Malacaria:2016} measured episodic enhanced absorption events near inferior conjunction. They performed a detailed spectral analysis that showed that the data could be attributed to an inhomogeneous accretion wake, in agreement with the idea proposed by \cite{Doroshenko:2013}. The eccentricity of the orbit also introduces a minor asymmetry between the evolution of the column density when the neutron star moves toward and away from Earth. 

Finally, intrinsic variations of the X-ray ionizing emission from the immediate vicinity of the neutron star (see Section\,\ref{sec:int_var}) alter the wind ionization structure. The highly ionized fraction of the wind, near the neutron star and in low-density regions, is transparent to X-rays and does not contribute to the absorbing column density. It induces $N_\textrm{H}$ variations even if the amount of material integrated along the LOS remains unchanged.

While off-states have usually been explained as due to a quenching of the accretion itself (see Section\,\ref{sec:int_var}), one model found that the lowered luminosity during off-states might be due to scattering of hard X-rays along the line of sight: \cite{Sidoli:2015} reported different off-state occurrence rates at orbital ingress and egress that they interpreted as an indication that off-states were due to X-rays that were being scattered in the photoionization wake identified by \cite{Blondin:90} in numerical simulations. 

\subsubsection{Variability in mass accretion rate}
\label{sec:int_var}

The dominant mechanisms responsible for the intrinsic variability in mass accretion rate in Vela X-1 have not yet been conclusively identified. Periodic changes are expected because the eccentricity of the neutron star orbit leads to non-negligible systematic differences in the mass density of the environment in which the neutron star is embedded: Notwithstanding wind acceleration and      nonisotropic effects, the density at periastron would be 40\% lower than the density at apastron. This number is a lower limit because the wind is still accelerating at the orbital separation and is more strongly beamed toward the neutron star at periastron, when the filling factor is the highest (see Section\,\ref{sec:models:mass_transf}). Concerning the stochastic variability of the mass accretion rate onto the neutron star surface, the very location of its origin varies from one model to the next.

The X-ray ionizing feedback has long been suspected to modulate the mass supply onto the neutron star in an unstable manner: An increase in X-ray luminosity leads to a slower wind because the line acceleration is inhibited, which increases the mass accretion rate (see Section\,\ref{sec:models:mass_transf}). The cycle closes when wind acceleration is so inhibited that stellar material no longer penetrates the Roche lobe of the neutron star. In an attempt to quantify this effect, \cite{HoArons:87a} designed a 1D model along the line that joins the two bodies with a prescribed smooth and isotropic wind. They equated the accretion X-ray luminosity associated with the BHL formula\,\eqref{eq:BHL_mdot} with the X-ray luminosity corresponding to the extent of the region around the neutron star in which the ionization parameter is above $\xi_\text{crit}=10^4$\,erg cm s$^{-1}$. Using the orbital parameters of Vela X-1, they obtained two solutions. In the first, the wind was hardly affected by the X-ray ionizing feedback and yielded an X-ray luminosity corresponding to the one observed in Vela X-1. This solution was later found by \cite{HoArons:87b} to be prone to fluctuations with realistic timescales \citep[see, e.g.,][]{Kreykenbohm:2008}. In the second solution derived by \cite{HoArons:87a}, the X-ray luminosity was $\text{about four}$  orders of magnitude higher, with a wind speed significantly lowered by line-acceleration quenching. In spite of episodic flares in which Vela X-1 reaches a few 10$^{37}$erg$\cdot$s$^{-1}$, the system has never been observed in such a high state, however. \cite{Karino:2014} performed a similar computation, but accounted for optical depth and a gradual effect of the X-rays on line acceleration. \cite{Karino:2014} retrieved a bimodal behavior, but in contrast to \cite{HoArons:87a}, Vela X-1 was lying on the branch corresponding to a high luminosity, associated with a slow wind. \cite{ManousakisWalter:2015a} revived the role of the X-ray ioninzing feedback that they found to be a possible origin of the observed off-states. In 2D hydrodynamic simulations in the orbital plane, they found that when wind acceleration was inhibited above a critical ionization parameter of $10^{2.5}$erg$\cdot$cm$\cdot$s$^{-1}$, a permanent X-ray luminosity of $4\times10^{36}$erg$\cdot$s$^{-1}$ could produce an oscillation of the position of the accretion front shock. This breathing mode induces a subsequent modulation of the mass accretion rate onto the neutron star, with $\sim$30-minutes-long off-states during which the accretion is quenched, separated by a characteristic duration that agrees reasonably well with the characteristic timescale of the X-ray emission from Vela X-1 reported in \citet{Kreykenbohm:2008}. Recently, \cite{Bozzo:2020arXiv} designed a detailed semianalytical model of a smooth spherically symmetric wind and showed that accounting for the eccentricity of the orbit in a simplified model of accretion leads to an average mass accretion rate almost three times higher near periastron than near apastron. Because the accretion wake intercepts the line of sight at orbital phases 0.4-0.8, precisely when the neutron star is near apoastron, it is difficult to separate between a decrease in intrinsic emission and an increase in absorption, even using hard X-ray observations. The authors also included the X-ray ionizing feedback and found that its effect was limited to the vicinity of the neutron star. This challenges the conclusions drawn by \cite{ManousakisWalter:2015a} about the origin of off-states, but agrees with \cite{Sander:2018}.



Because its structure is inherently clumpy, the line-driven wind from the donor star provides material to the neutron star at variable rates. In a seminal study, \cite{Ducci:09} designed a statistical model of clump capture in which they accounted for this component by assuming that each clump intersecting the disk of radius $R_\text{acc}$ around the neutron star is instantaneously captured, with a correction factor if the overlap is only partial. They found occurrence diagrams consistent with what is observed in Vela X-1. However, these results were challenged by 3D hydrodynamic simulations of the accretion of a clumpy wind.  \cite{El-Mellah:2018} showed that clumps were strongly compressed, distorted, and eventually mixed within the shocked region. The bow shock that forms around the neutron star significantly decreases the peak-to-peak variability compared to what is expected in the ballistic BHL framework, in which clumps are instantaneously accreted without accounting for their thermodynamical properties. These results indicate that the intrinsic variability of the X-ray emission cannot be considered a direct tracer of the wind microstructure and that other mechanisms are likely to dominate the intrinsic variability of the emission. We note that in simulations of accretion onto supermassive black holes mediated by a centrifugal gating mechanism, \cite{Gaspari:15} found similar properties for the variability of the mass accretion rate (e.g., a log-normal occurrence diagram, indicative of self-organized criticality).


Although realistic 3D clumps per se do not produce a variability high enough to account for the changes in intrinsic X-ray luminosity, they can play a role in modifying the conditions of accretion at the neutron star magnetosphere. Due to the proximity between the magnetospheric and circularization radii in Vela X-1 (see Figure\,\ref{fig:radii}), the accretion flow between the shock and the magnetosphere in Vela X-1 might transit between a disk-like and a spherical geometry, depending on the instantaneous and local wind properties. In each geometry, there are regimes of low and high mass accretion rates, depending on the stellar material supplied at the orbital scale (see Section\,\ref{sec:models:mass_transf}). Transitions between regimes and/or geometries might contribute to the overall variability. \cite{Bozzo:2016} investigated this scenario in a model of accretion of a clumpy wind accounting for the magnetic and centrifugal gating mechanisms. They showed that for orbital and neutron star spin periods close to the ones in Vela X-1, switches between regimes were occurring. Even if the authors warned the readers about the limitations of their model, which is uni-dimensional and does not solve the wind dynamics, their work highlighted the importance of considering the coupling between the plasma and the magnetosphere to fully appreciate the evolution of the intrinsic X-ray variability.

\subsection{Evolutionary scenario} \label{sec:models:evolution}

High-mass X-ray binaries are an important evolutionary stage on the way to form compact object binaries and thus an interesting benchmark to understand the origin of gravitational wave sources. With a neutron star being the compact object, Vela X-1 is a candidate for either a black hole-neutron star merger or a double neutron star merger, assuming the system remains bound after the current supergiant component has undergone core collapse. The identification of the evolutionary scenario for an individual system such as Vela X-1 is complicated. Given the high number of parameters in binary evolution and the different possible channels, a snapshot of the present status can thus only be used to rule out certain scenarios rather than pinpointing at a particular evolutionary track.

Neglecting any scenarios that could involve higher-order systems, the standard scenario to form an HXMB considers two massive stars with an uneven, but not too extreme mass ratio \citep[e.g.][]{TvdH2006}. The more massive primary star evolves faster and eventually fills its Roche Lobe, which will lead to mass transfer from the primary to the secondary \citep[e.g.][]{Paczynski:71,vdH1976}. Depending on the nuclear burning status of the primary, this is referred to as either case A (during central H-burning), case B (after central H burning, but before central He burning), or case C (after the start of central He burning) mass transfer. For typical  separations and assuming close mass ratios, this first mass transfer period in the system is generally assumed to be stable \citep[e.g.,][]{Podsiadlowski1992,Ritter:1999,deMink:2008b}. 
The mass transfer will stop after a large fraction of the hydrogen envelope has been removed from the primary. The secondary has now gained hydrogen-rich matter. Depending on the precise parameters, the secondary could now even evolve faster than the primary and reach core collapse first \citep[e.g.][]{Pols:1994}. However, given that the mass transfer might have happened in an already advanced stage of the primary, the primary as the first object to undergo core collapse is the more common outcome. If the system remains bound beyond this stage \citep{Blaauw:1961}, we are left with a compact object resulting from the primary and a secondary which has gotten additional material and -- depending on the mixing timescale of this ``new'' material into the deeper layers -- has been ``rejuvenated'' \citep[e.g.][]{Braun:1995}. Once the compact object is able to accrete matter from the secondary -- either by wind accretion or Roche Lobe overflow (RLOF) -- the system will appear as an HMXB.

In the case of Vela X-1, the compact object is a neutron star and the secondary has a current mass on the order of $20\,M_\odot$. Consequently, we would expect a primary more massive than that with suggestions going up to $60\,M_\odot$ \citep{Quast2019}. Given that enough mass needed to be removed to eventually yield a neutron star as the compact object, the latter number might be a bit high, but illustrates the severe uncertainty when trying to extrapolate the systems' backstory. Even taking into account the accompanying supernova during the formation of the neutron star, the core mass of the primary had to be low enough to avoid the formation of a black hole, which means that either the primary mass had to be low enough or the mass transfer was efficient enough to alter the evolution before a more massive core could be formed \citep[e.g.][]{Wellstein:1999}.

The near-circular orbit of Vela X-1 prompts the questions whether the first mass transfer in the system might have lead to a Common Envelope (CE) phase \citep{Paczynski:76}. Even though many uncertainties remain on the detailed physical processes involved in CE evolution, their occurrence is typically inferred from an evolutionary necessity \citep[e.g.][]{Ivanova:2013}. CE stages are thought to occur for unstable RLOF and help to shrink and circularize the orbit, at least if the system can eventually eject the envelope and avoid merging. However, CE stages are unlikely to occur for the first mass transfer stage in an HMXB \citep{TvdH2006}. The parameters for Vela X-1, including its ``runaway'' nature \citep[see also Figure~\ref{fig:OB1}]{Kaper:97}, indeed point more towards this standard scenario \citep{vdH2019}.

The remaining secondary in Vela X-1 is classified as ``overluminous'' \citep[e.g.][]{Wickramasinghe:75,Conti:78}. This term generally refers to that fact that the object is more luminous than expected for its mass, although the details differ in whether this refers to a simple comparison with single star evolution \citep[e.g.][]{Quast2019} or is also used for remaining discrepancies in the context of binary evolution scenarios \citep[e.g.][]{Vanbeveren:93ADS}. Regardless of terminology, it is important to account for the fact that the secondary has been rejuvenated and thus will usually have a different luminosity and temperature compared to a single star of the same spectral type. 

The recent atmosphere analysis of HD 77581, the donor of Vela X-1, yields $\log L/L_\odot \approx 5.5$ (see Table~\ref{tab:stellarpar}). Relying on stellar structure calculations by \citet{Graefener:2011}, a single star on the zero age main sequence with this luminosity would have a mass of approximate $47\,M_\odot$. This is twice as large as measured (cf.\ Sect.\,\ref{sec:system:mass}). The other extreme would be a He star with a thin hydrogen layer negligible in mass. In this case, the donor star in Vela X-1 would be about $15.5\,M_\odot$. The measured masses of about $20\,M_\odot$ are much closer to the second case and thus illustrate that the now donor star has probably evolved considerably before  gaining mass from the primary. \citet{Vanbeveren:93ADS} deduced that the high ratio between luminosity and mass ($L/M$-ratio) could only be explained if the central hydrogen abundance of the secondary was already down to a mass fraction of $0.1$ when it started accreting from the primary. Moreover, the star would need to have been fully convective during the accretion stage and its surface should now be enriched in nitrogen, but depleted in oxygen and carbon. The latter is indeed in line with the recent analysis by \citet{Gimenez-Garcia:2016}, although their determined surface hydrogen abundance of $X_\text{S} = 0.65$ is slightly higher than the range suggested by \citet{Vanbeveren:93ADS}. However, depending on the details of the accretion into the secondary, the resulting values of $X_\text{S}$ can vary quite a bit \citep{Hellings:1983,Braun:1995} with existing scenarios not excluding the measured abundance. In any case, the surface hydrogen abundance of the donor in Vela X-1 is below the solar value, thus giving rise to the label that the donor of Vela X-1 is ``He enriched''.

The high $L/M$-ratio and the still rather high surface hydrogen abundance both lead to a stronger stellar wind mass loss than assumed for an isolated star of the same spectral type as stellar wind mass is significantly boosted for higher values of the so-called ``Eddington parameter'' $\Gamma_\text{e} \propto q_\text{ion}\,L/M$ \citep[e.g.][]{GH2008,Vink2011,Sander2020,SV2020}. A higher rotation rate due to angular momentum transfer in the earlier stages of the evolution would add to this. Consequently, the wind mass-loss rate of the secondary is now higher than it ever was for the secondary in the system before the onset of mass transfer.

In the standard picture, the HMXB type with a supergiant donor was seen as a relatively short-lived stage \citep[e.g.][]{TvdH2006} due to the unstable nature of the mass transfer arising from the -- now -- high mass ratios between the donor and the compact object and due to the convective envelope of the donor. This claim has recently been challenged by \citet{Quast2019}, who argue that despite the shrinking of the orbit also the donor radius is decreasing as the He surface abundance increases due to the mass transfer. This scenario would significantly prolong the HMXB stage and provide a better explanation for the relatively large number of HMXBs hosting a blue supergiant donor star in our own Milky Way. Depending on the $L/M$-ratio \citep{SV2020}, the donor could eventually appear as Wolf-Rayet star \citep{Quast2019}. Eventually, a CE stage with the compact object seems to be unavoidable \citep{Hjellming1987,TvdH2006}, which would either lead to shrinking of the orbit or the eventual merger of the two stars after the formation of a so-called ``Thorne-\.Zytkow object''  \citep[TZO,][]{TZ1975,TZ1977}. Recently, \citet{Oskinova:2018} suggested that sgB[e] HMXBs could represent those HMXBs currently in or shortly after a CE stage.
A successful ejection of the CE would remove further mass from the system, leading to a He star even if the considerations by \citep{Quast2019} were not applicable. However, a system like Vela X-1 might not survive the CE phase as \citet{vdH2017} argue that all systems with mass ratios larger than approximately $3.5$ will lead to an inspiral of the compact object. In contrast, recent simulations suggest that the parameter space leading to successful CE ejection is more complex. Depending on the properties of the donor star envelope, CE evolution can be immediately followed by stable mass transfer \citep{Klencki:2020}.
So far, there are a lot of remaining uncertainties on the scenarios and detailed evolution calculations are missing, but given that the current mass ratio for Vela X-1 is about $10$, the system would be a prime candidate for this fate and thus never form a compact object binary. Yet, the observational confirmation of this scenario is hampered by the severe uncertainties for the mass loss of lower-mass He stars \citep[e.g.][]{Vink2017,Sander2020}. While traditional considerations often associated He stars with a WR-type spectral appearance, this is only true if the stars have a sufficiently high $L/M$-ratio \citep{Shenar2020,SV2020}. For the case of Vela X-1, this would mean that the current donor would have to lose about $5\,M_\odot$ -- while at least keeping the current luminosity -- before reaching a sufficient $L/M$-ratio for a WR-type mass-loss rate. Hence, the evolutionary fate of Vela X-1 is quite uncertain. However, the most expected outcome according to current considerations and known system parameters is a merger of two components in a CE stage \citep[e.g.,][]{Belczynski:2012}. The inspiral of the neutron star into the core of the now secondary will eventually lead to the formation of a black hole. This could give rise to a less common type of short gamma-ray bursts \citep{Fryer:2013}.

In addition to our knowledge of its evolutionary track, Vela X-1 was shown to move at high speed ($\sim$90~km s$^{-1}$) with respect to  its local environment. This is likely indicative of the natal kick produced by the asymmetric supernova explosion when the neutron star formed, probably a few million years ago (see Sect.\,\ref{sec:system:VelaOB1}). A bow shock associated with Vela X-1 was found by \cite{Kaper:97} based on its H$\alpha$ emission. Although its main axis is consistent with the relative motion of Vela X-1 through the interstellar medium, \cite{Gvaramadze:2018} identified nonsymmetric features in the vicinity of the shock, such as an elongated H$\alpha$-emitting structure. Based on magnetohydrodynamic simulations, they showed that the opening angle of the bow shock and its H$\alpha$ brightness could be reproduced by considering the interaction of a region of higher density in the interstellar medium with the stellar wind from the donor star in Vela X-1. Interestingly enough, they were able to reproduce the emission from the filamentary structure only by accounting for a helical magnetic field in the stellar wind that they tentatively attributed to the interaction between the stellar wind and the magnetosphere of the neutron star. 

\subsection{Taxonomy of wind models}
\label{sec:model:taxonomy}

Relevant models to describe the physics at stake in Vela X-1 have been published over the past decades. Most of them resorted to the numerical tool to solve the flow dynamics \citep[e.g.,][]{}, but a handful are essentially based on (semi-)analytical approaches to study the X-ray ionizing feedback \citep{HatchettMcCray:77}, the flow structure \citep{Foglizzo:97}, the clump capture rate \citep{Ducci:09}, the accretion mechanism \citep{Bozzo:2016}, the formation of wind-captured disks \citep{Karino:2019}, or the secular spin evolution of the neutron star \citep{Karino:20}, for instance. Democratization of access to computing facilities and exploratory studies led to an overall increase in the complexity of the models that were developed to provide more detailed explanations of the physical mechanisms in Vela X-1. As an unfortunate side effect, the numerous physical and, when applicable, numerical assumptions required to capture the physics and solve the underlying equations preclude any one-to-one comparison of results, even in the rare cases when studies focus on similar aspects of the system: The price to pay for more realistic models has been an accelerated division of the system description into a constellation of models that each address more specific aspects in more detail. When observational data are not enough to decide, the robustness of the provided scenarios does not lie in the strict reproducibility of the results, but instead in the repeated occurrence of qualitatively similar features from one simulation to the next and in the compatibility of an explanation with the models proposed by other teams.

As an illustration of this difficulty of comparing different frameworks although they all solve the equations of radiative-hydrodynamics in 1D, \cite{Watanabe:2006}, \cite{Krticka:2012}, and \cite{KrtickaKubat:2017}, \cite{KrtickaKubatKrtickova:2018} and \cite{Sander:2018} did not obtain the same wind ionization and velocity profile (see Section\,\ref{sec:ion_struct}). The models developed by the last two teams account for more physics than in the more simplified calculation by \cite{Watanabe:2006}, but they both chose to rely on different assumptions concerning the frame in which the radiative transfer is performed, the ions involved in the computation, the X-ray emission, and the stellar properties or the wind clumping, among others.

Similarly, the 2D and 3D simulations ran by \cite{Cechura:15}, which are primarily tailored to the properties of Cygnus X-1, another wind-fed high-mass X-ray binary, \cite{Blondin:90}, \cite{ManousakisWalterBlondin:2012}, \cite{ManousakisWalter:2015a}, \cite{ManousakisWalter:2015b} and \cite{El-Mellah:2019} did not handle the effect of the X-rays on wind acceleration in the same way. Because of their multidimensional aspect, none of these simulations can afford to treat radiative transfer in a way as realistic as the studies above. For instance, they cannot capture any spectral property, but instead have to focus on the overall photoionizing flux, which is taken to be the X-ray flux from the neutron star. In order to represent the wind acceleration quenching, they relied on the critical ionization parameter (see equation\,\eqref{eq:ionization}) beyond which the ions contributing most to the resonant line absorption are assumed to be so depleted that acceleration drops to zero. In this all-or-nothing approach, it can only be guessed which critical ionization parameter is realistic because the dominant ions are unknown and sharply depend on the stellar properties and chemical composition. Even in this case, the sudden quenching assumed by \cite{ManousakisWalter:2015b} or the even steady decrease of the acceleration assumed by \cite{Cechura:15} has been shown to be a far too simplified approximation \citep{Sander:2018}. Instead, \cite{El-Mellah:2019} relied on the 1D line acceleration profile computed by \cite{Sander:2018} for the donor star in Vela X-1, but assumed that the nonisotropic 3D features induced by the orbital motion in their simulations had no significant effect on this profile, but were purely a function of the distance to the donor star. This hypothesis remains to be confirmed. Below the critical ionization parameter (10$^{0.5}$ erg$\cdot$cm$\cdot$s$^{-1}$ in \citealt{Cechura:15}, 10$^{2.5}$ erg$\cdot$cm$\cdot$s$^{-1}$ in \citealt{ManousakisWalter:2015b}), line acceleration is computed differently in these studies, leading to different amounts of specific kinetic energy that was deposited in the wind by the stellar photon field when the flow reached the neutron star. 

The wind microstructure has also been treated in different ways: While some works assumed a smooth wind \citep{Watanabe:2006,El-MellahCasse:2017,Blondin:91}, others included wind clumpiness in their models or simulations. The clumps are either indirectly represented by their effect on the radiation field \citep{Sander:2018}, are parameterized as spheres of a given mass and radius \citep{OskinovaFeldmeierKretschmar:2012,ElMellah2020_NH_paper}, or are computed from 2D radiative-hydrodynamics simulations on pseudo-planar grids before being extended to 3D \citep{El-Mellah:2018}. Because of their small sizes, however, including more realistic clumps comes at the expense of a smaller simulation space in radiative-hydrodynamics simulations, or of simplified equations of motion of the flow.

A detailed taxonomy of the models of Vela X-1 in particular and of wind-fed neutron stars in supergiant X-ray binaries in general is beyond the scope of this paper. It would help to determine which model could prove fruitful in order to interpret a given set of observations, however.


\section{Properties of the Vela X-1 system} \label{sec:system}

In this section we compile the existing knowledge on the essential parameters of the Vela~X-1 system and how they have been obtained. We also indicate where results were strongly based on parameters from previous work. The section is broken down into subsections for different essential parameters, but we note that depending on the data and analysis methods that were used, various parameters have frequently been derived jointly or cannot be determined in an independent fashion from other parameters. A trivial example are the masses of the binary partners and inclination (Figure~\ref{fig:nsmass}).

\subsection{Distance from Earth and relation to the Vela OB1 association} \label{sec:system:distance}
\begin{figure}[thbp]
\centerline{\includegraphics[width=1.0\linewidth]{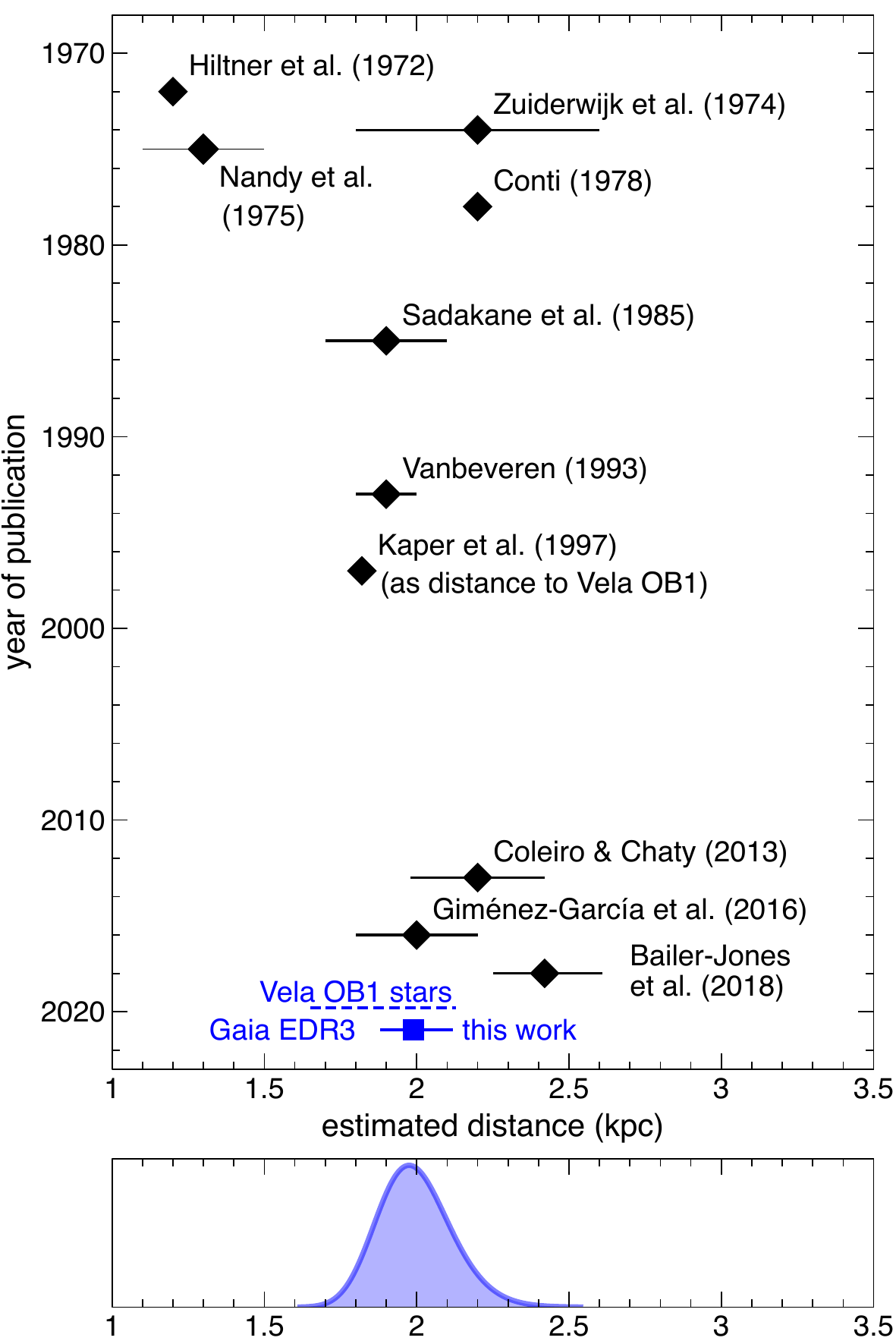}}
\caption{Upper panel: Evolution of distance estimates to the Vela~X-1 system over time derived with different approaches, as explained in the text. At the bottom we also show the distance ranges derived for stars in the Vela OB1 association and our new distance determination based on the new \textit{Gaia}-EDR3 data. 
Lower panel: Posterior distribution of our updated \textit{Gaia} distance estimate. 
}
\label{fig:distance}
\end{figure}

In order to derive the real luminosities of the supergiant and the X-ray source, it has been the goal for many different authors to determine the distance to Vela~X-1. Figure~\ref{fig:distance} illustrates the evolution
of different distance estimates over time.

\subsubsection{Evolution of the distance estimates}

The earliest distance estimate to the Vela~X-1 system, or more precisely, to HD~77581, was reported by \citet{Hiltner:1972}, who derived a distance of 1.2~kpc (distance modulus 10.5) based on an assumed absolute magnitude of $M_V=-6.0$ and a ratio of 3.0 for the total-to-selective absorption. This distance estimate was still used as a baseline in the 1980s, for instance, by \citet{Nagase:84ApJ}. Based on UV observations with the European satellite \textsl{Thor-Delta 1A (TD1)}, \citet{Nandy:75} arrived at a similar result of $1.3\pm0.2$~kpc.

In contrast, \citet{Wickramasinghe:74} derived a distance $>$2~kpc by comparing the reddening of HD~77581 with that of stars in its  vicinity. \citet{Zuiderwijk:74} used the equivalent width of interstellar Ca~\textsc{ii}~K, independent from the reddening diagnostics, to derive a distance of $2.2\pm0.4$~kpc.

Based on the spectral type \citep{Hiltner:1972} and duration of the eclipse \citep{Forman:73}, \citet{Conti:78}  derived a stellar radius of 35~$R_\odot$ and an absolute magnitude $M_V=-7.0$, which then led to an estimated distance of 2.2~kpc (no uncertainty given) based on the pho\-to\-met\-ry by \citet{JonesLiller:73}.

\citet{Sadakane:85} used ultraviolet spectra obtained with \textsl{IUE} to determine an effective temperature of $25\,000\pm 1000$~K. They combined this temperature with a radius of 31~$R_\odot$ \citep{RappaportJoss:83} to arrive at an absolute visual magnitude $M_V=-6.67$ and a derived distance of $1.9\pm0.2$~kpc.

From a spectroscopic analysis using NLTE models (see section~\ref{sec:obs:uvopt:spectra}), \cite{Vanbeveren:93ADS} derived a distance of 1.8--2~kpc.

\citet{Kaper:97} noted the possible connection to  the Vela~OB1 association (see below) and introduced the distance of  1.82~kpc to the Vela~OB1 \citep{Humphreys:78} as a distance value. This was then used by several subsequent X-ray studies without considering the implied uncertainties.

\citet{ColeiroChaty:2013} used a spectral energy distribution (SED) fitting procedure in order to derive the distance and absorption of a sample of HMXBs. For Vela~X-1, they derived a value of 2.2$\pm$0.2~kpc. 

This approach was further refined by \citet{Gimenez-Garcia:2016}. They combined data from IR (2MASS) to UV (IUE) for the SED with a detailed analysis of spectral lines using NLTE models created with the PoWR code (see section~\ref{sec:obs:uvopt:spectra}) in order to derive the extinction, luminosity, and distance to the source, arriving at a distance of 2.0$\pm$0.2~kpc.

In the huge catalog of stellar distances based on \textit{Gaia} Data Release~2 \citep[\textit{Gaia}~DR2,][]{Gaia:2018_DR2} parallaxes, published by \citet{Bailer-Jones:2018}, the distance to Vela X-1 is derived as 2.42$-$0.16$+$0.19~kpc from a Bayesian analysis, with a basic parallax of 0.38$\pm$0.03~$\mu$as.  Figure~\ref{fig:distance} shows that this distance is farther away than most previous estimates and is formally consistent with only very few estimates.

The recently published third {\it Gaia} data release (EDR3, \citealt{Gaia:2020}) gives us the opportunity to obtain an improved distance to Vela~X-1. {\it Gaia}~EDR3 lists a parallax $\varpi$~=~496.2$\pm$15.2~\muas. To derive a distance, we first applied a parallax zero-point of $-$13.5~\muas\ using the \citet{Lindegren:2020b} recipe for the relevant magnitude, color, and ecliptic latitude. For this specific star, this zero-point does not differ much from the median quasar value of $-$17~\muas,\ but we note that a star as bright as Vela~X-1 ($G_{\rm EDR3} = 6.7351$) is at the limit of the analysis in subsection~4.4 of \citet{Lindegren:2020b}. Next, we considered transforming from the internal parallax uncertainty of 15.2~$\muas$ to an external uncertainty. Following \citet{Lindegren:2018b}, astrometric uncertainties should be transformed from their internal ($\sigma_i$) to their external ($\sigma_e$) uncertainties using

\begin{equation}
 \sigma_e = \sqrt{k^2\sigma_i^2 + \sigma_s^2},    
\end{equation}

\noindent where $k$ is a value to be determined (and likely a function of at least magnitude), and $\sigma_s$ is the square root of the limit when the angular covariance reaches zero. As of the time of this writing, the value of $k$ has only been estimated for stars significantly fainter than Vela~X-1 \citep{Fabricius:2020,Maizetal21c}. We used a value of $k=1$, which may be a slight underestimation, but we compensate for this by using a very conservative value of $26$~\muas\ for $\sigma_s$, which is derived from (faint) quasars by \citet{Lindegren:2020a} and is likely to be an overestimation because the checkered pattern seen in the EDR3 data for the LMC has a smaller amplitude \citep{Lindegren:2020a}. In this way, we arrive at a corrected parallax of $\varpi$~=~509.7$\pm$30.1~\muas, which may be revised in the near future when the systematic uncertainties in {\it Gaia}~EDR3 are better characterized. Using this parallax as an input and the prior for runaway OB stars from \citet{MaizApellaniz:2001,MaizApellaniz:2005} with the updated parameters from \citet{MaizApellaniz:2008}, we arrive at a distance of $1.99^{+0.13}_{-0.11}$~kpc for Vela~X-1. We note from the analysis of \citet{Pantetal21} with {\it Gaia}~DR2 data (which should also apply to the EDR3 case) that when a reasonable prior is used for the stellar spatial distribution in the Galaxy, distances for stars with precise {\it Gaia} parallaxes are robust, that is, they are nearly independent of the specific prior, especially for objects within a few kiloparsec.

\begin{figure}[hbt]
    \centering
    \includegraphics[width=\linewidth]{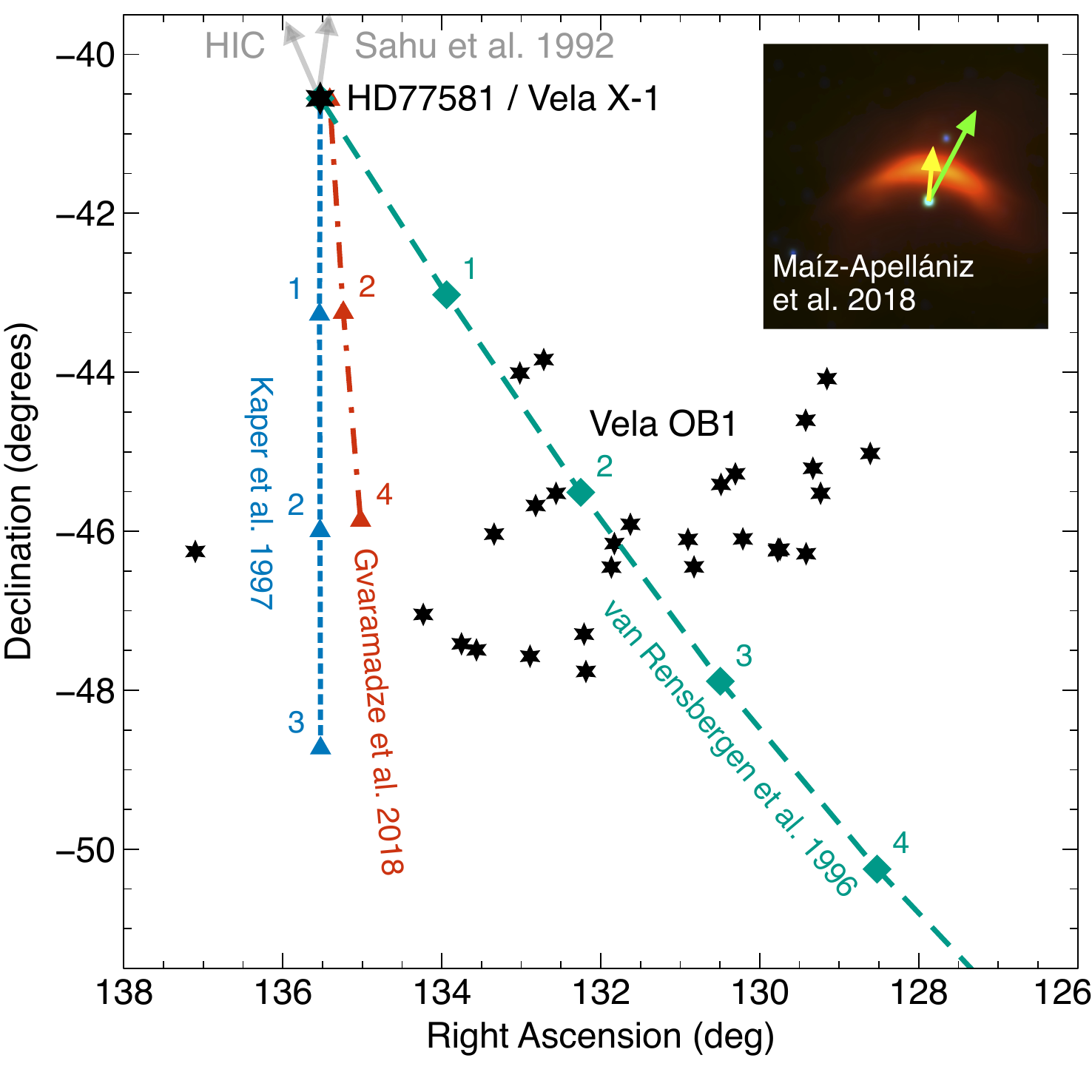}
    \caption{Comparison of the proposed tracks for the Vela~X-1 system relative to the Vela~OB1 association following Figure~7 of \citet{vanRensbergen:1996}, Figure~3 of \citet{Kaper:97}, and Figure~7 of \citet{Gvaramadze:2018}. The stars shown as belonging to the association (see Section~\ref{sec:system:VelaOB1}) are listed in Table~\ref{tab:Vela-OB1}. The numbers along the tracks indicate where the system would have been one, two, or three million years ago, based on the assumptions in these publications. The direction of proper motion, as listed in the Hipparcos Input Catalog (HIC) \citep{Turon:92HIC} and as obtained by \citet{Sahu:92PhD}, is shown by gray vectors. The inset is adapted from Figure 6 of \citet{MaizApellaniz:2018} and shows a $14' \times 14'$ WISE RGB mosaic. Green and yellow arrows indicate the raw and corrected proper motions, respectively (see the reference for details). The vectors shown in the inset are not to scale with those of the main drawing.}
    \label{fig:OB1}
\end{figure}

\subsubsection{Possible origin in the Vela~OB1 association}
\label{sec:system:VelaOB1}

In a wider study of six stars and four OB associations, \citet{vanRensbergen:1996} found that Vela~X-1 presumably originated in the Vela~OB1 association\footnote{\citet{Wright:2020} lists Vela~OB1 under ``OB associations for which very little is known''.}  \citep{Humphreys:78,Melnik+Dambis:2020a} of massive stars. They found that HD~77581 was a runaway OB star, probably expelled from the Vela~OB1 association by the natal kick that occurred during the supernova explosion associated with the formation of the neutron star. 

The connection with Vela~OB1 was again reported by \citet{Kaper:97}, who discovered a symmetric wind bow shock about 0.9 arcmin north of HD~77581 in a narrow-band H$\alpha$ image. This bow shock is created by the balance between the ram pressure of the stellar wind and the ambient interstellar medium, not to be confused with the much smaller bow shock that formed between the blue supergiant and the neutron star by the motion of the latter in the stellar wind of the former (Sect.~\ref{sec:elements:magnetosphere}). Based on the symmetry axis of the bow shock, they proposed a different direction of proper motion than was assumed by \citet{vanRensbergen:1996}. They also noted that the distance of 1.82~kpc to Vela~OB1 given by \citet{Humphreys:78} agreed with the distance to Vela~X-1 reported by \citet{Sadakane:85} and the X-ray luminosity, thus popularizing the use of this distance.

In a follow-up study, \citet{Huthoff+Kaper:2002} searched for bow shocks around other HMXBs or single OB runaway stars, but found no other example in the first group and only a minority in the second. They concluded that the formation of a bow show strongly depends on the temperature and density of the ambient medium and that apparently the majority of OB runaway stars is moving through hot bubbles in the Galactic plane.

In their search for Galactic runaway stars based on the \textit{Gaia} Data Release~1, \citet{MaizApellaniz:2018} found the bow shock around HD~77581 (GP~Vel in that publication) also in images obtained by WISE\footnote{\href{https://www.nasa.gov/mission_pages/WISE/main/}{Wide-field Infrared Survey Explorer}.} and demonstrated that the proper motion taken from the Tycho-\textit{Gaia} Astrometric Solution \citep[TGAS,][]{Michalik:2015TGAS} aligned very well with the symmetry axis of the shock, provided it was corrected for the mean proper motion of OB stars in that Galactic direction (see inset in Figure~\ref{fig:OB1}). 

Motivated by the results of \citet{Kaper:97}, \citet{Gvaramadze:2018} searched for an H$\alpha$ counterpart in the Southern H-alpha Sky Survey Atlas \citep[SHASSA,][]{Gaustad:2001SHASSA}. They found extended filamentary structures downstream of the bow shock within a wider area and presented the results of optical spectroscopy of the bow shock, comparing the observational data with 3D magnetohydrodynamic simulations. Recalculating the space velocity of the system (Tab.~\ref{tab:space_velocities}), \citet{Gvaramadze:2018} confirmed the possible association with the Vela~OB1 association, but warned that this connection might be spurious if the original binary system was ejected from its parent star cluster before the supernova explosion. Even the lower estimate for the actual space velocity of \vel places the system in the top few percentiles of the velocity distribution of runaway stars derived from recent numerical calculations of massive binary evolution \citep{Renzo:2019}.

\begin{table}[h]
    \caption{Space velocity estimates of the \vel system.}
    \renewcommand{\arraystretch}{1.1}
    \label{tab:space_velocities}
    \centering
    \begin{tabular}{lr}
    \hline\hline
    Reference & space velocity \\ 
    \hline
    \citet{Kaper:97} / \citet{Turon:92HIC}  & $89\pm 40$ km s$^{-1}$ \\
    \citet{Kaper:97} / \citet{Sahu:92PhD}   & $98\pm 26$ km s$^{-1}$\\
    \citet{Gvaramadze:2018}                 & $54.25\pm 0.55$ km s$^{-1}$ \\
    \hline
    \end{tabular}
    \renewcommand{\arraystretch}{1}
\end{table}

As found in the previous section, the distance to \vel based on \textit{Gaia} data is not fully consistent with the commonly assumed mean distance to the Vela~OB1 association. Therefore we also used \textit{Gaia} EDR3 astrometric data to revisit Vela~OB1 and obtain our own estimate of its distance.

Recently, \cite{Melnik+Dambis:2020a} cross-matched the catalog of \cite{Blaha:1989} with \textit{Gaia} DR2 using a matching radius of 3 arcsec and a magnitude tolerance of 3~mag. They identified 46 Vela~OB1 sources with \textit{Gaia} DR2 counterparts. We cross-matched these 46 sources with \textit{Gaia} EDR3 and obtained a counterpart for all of them. However, we rejected 7 sources with poor astrometric solution (renormalized unit weight error, RUWE,\footnote{https://www.cosmos.esa.int/web/gaia/dr2-known-issues}$>$1.4).  

We assumed that all Vela~OB1 members have similar distances (parallaxes) and kinematic (proper motions). We therefore applied an iterative 3$\sigma$ clipping in parallax, proper motion in RA, and proper motion in DEC. This reduced the sample to 30 \textit{Gaia} EDR3 sources with compatible parallaxes and proper motions that are considered to belong to Vela~OB1. They are shown in Fig.~\ref{fig:OB1} and listed in Table~\ref{tab:Vela-OB1}. This table contains the \textit{Gaia} EDR3 ID, the coordinates of the source, and the parallax and proper motion used in this work. 

Using this reduced sample, we estimated a mean parallax for Vela~OB1 of 0.53$\pm$0.07 mas. This corresponds to a mean distance of 1.88$\pm$0.24 kpc and is based on the inverse of the parallax and error propagation alone.
%

The distance to Vela~OB1 is therefore compatible with a possible origin of Vela~X-1 in this stellar association. However, the estimated distances to Vela~OB1 have large uncertainties that for the moment prevent us from reaching a firm conclusion about the origin of Vela~X-1. It is expected that future \textit{Gaia} data releases will allow us a better determination of the distance to Vela~OB1 and Vela~X-1.

\begin{table*}
\renewcommand{\arraystretch}{1.05}
\caption{Orbital parameters of the Vela X-1/HD~77581 system as derived by a selection of authors 
over the years. We also indicate where subsequent publications directly reused previous results.
Different definitions of the time of phase zero have been used over time: Mid-eclipse time ($T_\mathrm{ecl}$)
or time of mean longitude equal to $\pi/2$ ($T_{\pi/2} \equiv T_\mathrm{90}$) are the most
common, but some authors also use periastron time ($T_\mathrm{peri}$ or $\tau$) or
specific zero times \protect{\citep{Hutchings:74}}.}
\label{tab:orbit}
\centering
\begin{small}
\begin{tabular}{llr@{$\pm$}lr@{$\pm$}lr@{$\pm$}lr@{$\pm$}lr@{$\pm$}l}
\hline\hline
Reference & Time & 
\multicolumn{2}{c}{Phase zero} & 
\multicolumn{2}{c}{Orbital period} & 
\multicolumn{2}{c}{semimajor axis} & 
\multicolumn{2}{c}{eccentricity} & 
\multicolumn{2}{c}{longitude of} \\
& basis & \multicolumn{2}{c}{[MJD]} &
\multicolumn{2}{c}{$P_\mathrm{orb}$ [d]} & 
\multicolumn{2}{c}{$a_\mathrm{X}\sin i$ [lt-sec]} & 
\multicolumn{2}{c}{$e$} & 
\multicolumn{2}{c}{periastron $\omega$}\\
\hline
\citet{Forman:73}           & $T_\mathrm{ecl}$  &   41446.04 & 0.07   &     8.95 & 0.02     & \multicolumn{2}{c}{--} &  0.14 & 0.05            & \multicolumn{2}{c}{--}\\
\citet{Hutchings:74} (H74)  & $T_\mathrm{phot}$ &   41593.40 & 0.56   &    8.966 & 0.001    & \multicolumn{2}{c}{--} & \multicolumn{2}{c}{--} & \multicolumn{2}{c}{--}\\
\citet{Rappaport:76}        & $T_\mathrm{ecl}$  &  42611.501 & 0.12   & \multicolumn{2}{c}{(H74)} & 111.4 & 3.3    & 0.126 & 0.041 & 146 & 23 \\ 
\citet{Ogelman:77} (Ö77)    & $T_\mathrm{ecl}$  &   42446.03 & 0.06   &   8.9643 & 0.0005   & 113.0 & 4.0 & \multicolumn{2}{c}{--} & 162 & 25 \\
\citet{vanParadijs:77b}     &  --            & \multicolumn{2}{c}{--} &   8.9681 & 0.0016   & \multicolumn{2}{c}{--}  & 0.136 & 0.046 & \multicolumn{2}{c}{--}\\
\citet{WatsonGriffiths:77}  & $T_\mathrm{ecl}$  &   42620.30 & 0.05       &   8.9640 & 0.0005   & \multicolumn{2}{c}{--} & \multicolumn{2}{c}{--} & \multicolumn{2}{c}{--}\\
\citet{Rappaport:80}        & $T_\mathrm{peri}$ &   42822.90 & 0.13       &   8.9649 & 0.0002   & 113.0 & 0.8 & 0.092 & 0.005 & 154 & 5 \\
\citet{Nagase:83}           & $T_\mathrm{ecl}$  &   41446.02 & 0.05       &   8.9640 & 0.0006   & \multicolumn{2}{c}{--} & \multicolumn{2}{c}{--} & \multicolumn{2}{c}{--}\\
\citet{Nagase:84ApJ}        & $T_\mathrm{peri}$ &   45329.05 & 0.06   &   8.9642 & 0.0006   & 114.10 & 0.50 & 0.080 & 0.006 & 157.3 & 2.1 \\
\multicolumn{1}{p{0.18\textwidth}}{\citet{vanderKlisBonnet-Bidaud:84}} & $T_{\pi/2}$&  42727.750 & 0.024  & \multicolumn{2}{c}{(Ö77)} & \multicolumn{2}{c}{(Ö77)} & 0.116 & 0.022 & 166 & 7 \\
-- ditto --                  & $T_{\pi/2}$       &  42966.628 & 0.019  & \multicolumn{2}{c}{(Ö77)} & 113.4 & 1.7 & 0.085 & 0.022 & 121 & 16 \\
-- ditto --                  & $T_{\pi/2}$       &  44170.937 & 0.021  & \multicolumn{2}{c}{(Ö77)} & 113.20 & 1.4 & 0.102 & 0.024 & 177 & 15 \\
\citet{Boynton:86}          & $T_{\pi/2}$       & 43821.3604 & 0.0056 &  8.96443 & 0.00022  & 112.70 & 0.47 & 0.0881 & 0.0036 & 152.8 & 2.2 \\
\citet{Deeter:87a}          & $T_{\pi/2}$       & 43955.8261 & 0.0048 & 8.964353 & 0.000063      & 112.66 & 0.41 & 0.0896 & 0.0031 & 152.8 & 1.8 \\
\citet{Deeter:87b}          & $T_{\pi/2}$       & 44278.5466 & 0.0037 & 8.964416 & 0.000049 & 112.98 & 0.35 & 0.0885 & 0.0025 & 150.6 & 1.8 \\
\citet{Bildsten:97} (B97)   & $T_{\pi/2}$       & 48895.2186 & 0.0012 & 8.964368 & 0.00004  & 113.89 & 0.13 & 0.0898 & 0.0012 & 152.59 & 0.92 \\
\citet{Kreykenbohm:2008}    & $T_{\pi/2}$       &  52974.001 & 0.012  & 8.964357 & 0.000029 & \multicolumn{2}{c}{(B97)} & \multicolumn{2}{c}{(B97)} & \multicolumn{2}{c}{(B97)} \\
-- ditto --                  & $T_\mathrm{ecl}$  &  52974.227 & 0.007  & \multicolumn{2}{c}{\textit{same as above}} \\
\citet{Falanga:2015}        & $T_{\pi/2}$       &  42611.349 & 0.013  & 8.964427 & 0.000012 & \multicolumn{2}{c}{(B97)} & \multicolumn{2}{c}{(B97)} & \multicolumn{2}{c}{(B97)} \\
-- ditto --                  & $T_\mathrm{ecl}$  & 42611.1693 & 0.013  & 8.9644061 & 0.0000064 \\
\hline
\end{tabular}
\end{small}
\renewcommand{\arraystretch}{1.0}
\end{table*}%

\subsection{Orbital parameters and ephemerides}\label{sec:system:orbit}
Since the first clearly determined orbital period \citep{Forman:73}, the orbital parameters have been further refined in the following years and decades by various different groups based on eclipse timing, the analysis of Doppler shifts in the X-ray pulsations, or RVs from optical spectroscopy, for example. The values have essentially converged for the main parameters. Table~\ref{tab:orbit} presents an overview of this evolution and gives the references. In the more recent literature, the orbital parameters are frequently taken from \citet{Bildsten:97} or from the updated values from \citet{Kreykenbohm:2008}, who largely built on the previous work. \citet{Falanga:2015} have undertaken a systematic study of eclipsing HMXBs and derived new orbital parameters for Vela~X-1, which slightly contradict \citet{Bildsten:97} or \citet{Kreykenbohm:2008}; see Table~\ref{tab:orbit} and Figure~\ref{fig:periods}.

\begin{figure}[hbt]
\centerline{\includegraphics[width=1.0\linewidth]{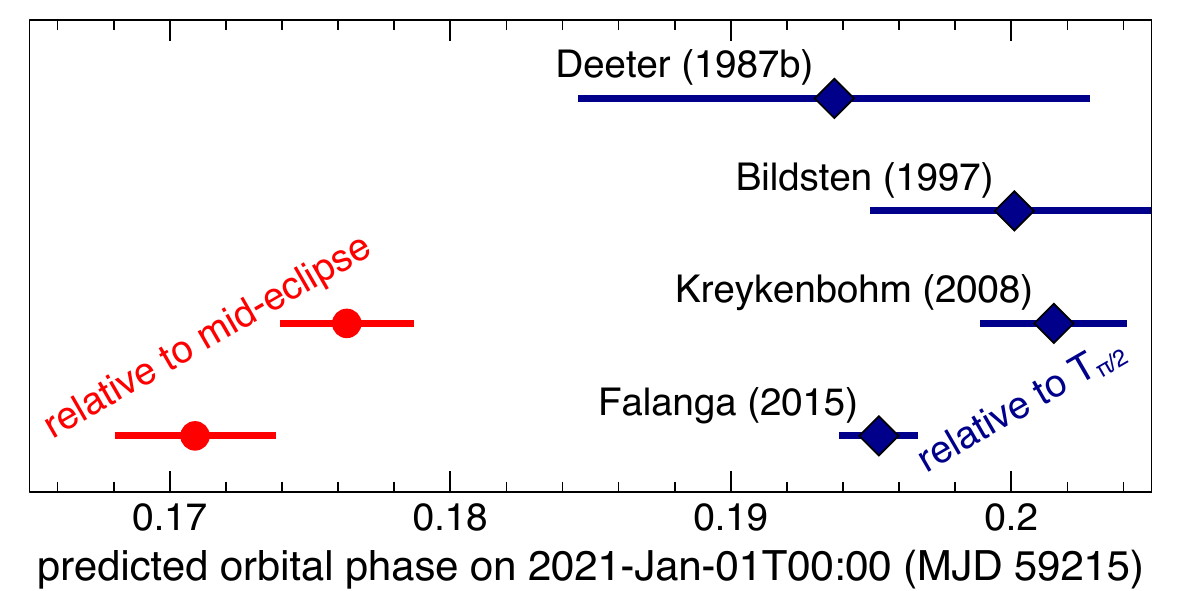}}
\caption{Predicted orbital phase for a given date, demonstrating the small differences between different ephemerides. In contrast, the difference in phase between mid-eclipse time and $T_{\pi/2}$ as zero-point, as shown for \protect{\citet{Kreykenbohm:2008}} and \protect{\citet{Falanga:2015},} is significant.}
\label{fig:eph}
\end{figure}

\begin{figure}[htbp]
\centerline{\includegraphics[width=1.0\linewidth]{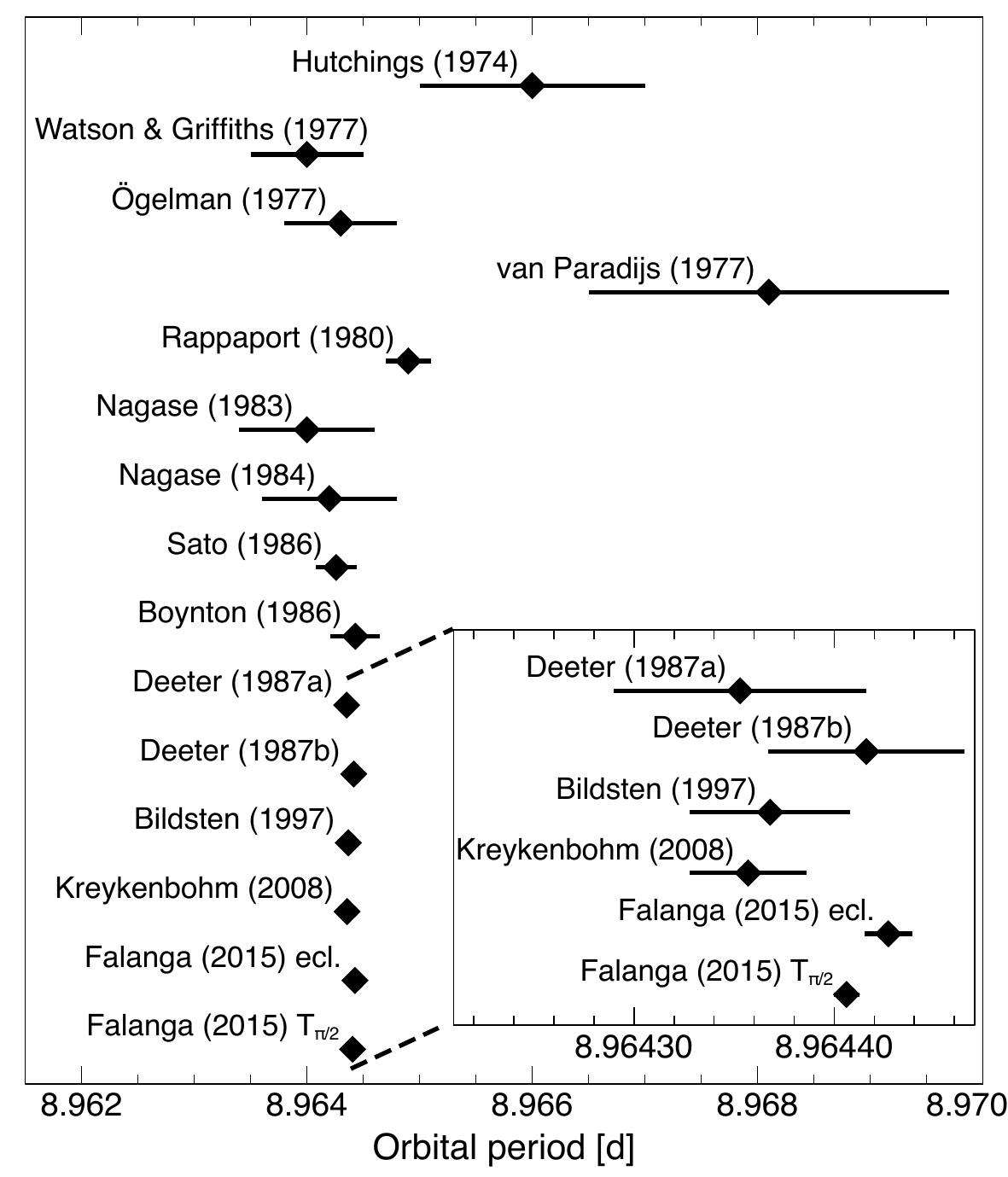}}
\caption{Evolution of orbital period determinations since \protect{\citet{Hutchings:74},} 
converging on very similar values by different groups.}
\label{fig:periods}
\end{figure}

Although the remaining differences are small, the orbital phases calculated using ephemerides published in the past three decades \citep{Deeter:87b,Bildsten:97,Kreykenbohm:2008,Falanga:2015} agree very well, at a level of a few times $10^{-3}$ in orbital phase all the way back to the detection of Vela X-1. The systematic phase shift induced by choosing either $T_\mathrm{ecl}$ or $T_{\pi/2}$ is significantly larger, about 0.025$\pm$0.01 \citep{Kreykenbohm:2008,Falanga:2015}. Figure~\ref{fig:eph} illustrates these facts for a recent date. Care should be taken in any discussion of orbital variations that it is clearly indicated which ephemeris and zero time is used.

The orbit of Vela X-1 is very stable and shows no indication of orbital decay. As the most recent study, \citet{Falanga:2015}, combining results from back to \citet{Forman:73} with data up to 2011 derived a period decay consistent with zero ($\dot{P}_\mathrm{orb}/P_\mathrm{orb}=0.1(3)\times10^{-6}\,\mathrm{yr}^{-1}$). They reported indications for a slight apsidal advance ($\dot{\omega} = 0.41(27) \mathrm{yr}^{-1}$), in contrast with previous studies \citep[][and references therein]{Deeter:87a}.

\subsection{Masses, radii, and inclination} \label{sec:system:mass}

\begin{table*}[ht]
\renewcommand{\arraystretch}{1.25}
\caption{Estimated masses of the two binary components of the Vela~X-1/HD~77581 system as given by various authors, together with the assumed inclination $i$. Where provided, the estimated radius of the optical star is also given. Errors or uncertainty ranges are given as in the original publications. See text for the multiple solutions provided by \citet{vanKerkwijk:95} or \citet{Stickland:97}. The latter underestimated $K_\textrm{opt}$ according to \protect{\citet{Barziv:2001}} due to a data reduction error. \citet{Quaintrell:2003} give a range of values, with the two extreme cases tabulated.}
\label{tab:masses}
\begin{center}
\begin{small}
\begin{tabular}{l@{}l@{}rrrrr} 
\hline\hline
\multicolumn{2}{l}{Reference} & 
\multicolumn{1}{r}{$i$} & 
\multicolumn{1}{r}{$M_\mathrm{NS}$ [$M_\sun$]} & 
\multicolumn{1}{r}{$M_\mathrm{opt}$ [$M_\sun$]} & 
\multicolumn{1}{r}{$R_\mathrm{opt}$ [$R_\sun$]} &
\multicolumn{1}{r}{$K_\mathrm{opt}$ [km s$^{-1}$]} \\
\hline
\cite{Hutchings:74} &            & 90       & 1.5--2.7       & --          & --         & $20.2\pm1.1$  \\
-- ditto --          &            & 70       & 2.1--2.9       & --          & --         & $20.2\pm1.1$         \\
\citet{MikkelsenWallerstein:74} &           & 90       & 1.0--3.0       & 10--30      & $\sim 30$  & --         \\
-- ditto -- (plausible $M_\mathrm{Opt}$) &   & 90       & 1.5--3.0       & 20--25      & $\sim 30$  & --         \\
\citet{Zuiderwijk:74}  &                    & 90       & $\ge2.5\pm0.3$ & $\ge30\pm5$ & --         & $26\pm0.7$ \\
\citet{vanParadijs:76} &                    & 74--90   & $1.61\pm0.27$  & $21\pm2.6$  & --         & $20\pm1$ \\
\citet{Lamers:1976}   &                       & --       & 2.0            & 22          & 27         & --         \\
\citet{Rappaport:76}  &                     & $\ge$70  & $1.45\pm0.16/(\sin i)^3$ & $21\pm0.9/(\sin i)^3$ & -- & -- \\
\citet{Vidal:76}      &          & 90       & 1.56--2.1      & 18.1--23.0  & -- & -- \\
-- ditto --            &          & 80       & 1.64--2.2      & 19.0--23.7  & -- & -- \\
-- ditto --            &          & 70       & 1.9--2.55      & 20.7--27.7  & -- & -- \\
\citet{vanParadijs:77b} &         & $>74$    & $1.67\pm0.12$  & $20.5\pm0.9/(\sin i)^3$  & -- & $21.75\pm1.15$ \\
\citet{Conti:78}        &       & 74       & 1.6            & 24          & 35 & -- \\
\citet{Vanbeveren:93ADS} &       & --       & --             & 21.6--26.5  & 31--35 & -- \\
\citet{JossRappaport-Review} & (JR84)  & $>73$    & $1.85^{+0.35}_{-0.30}$ & $23.0^{+3.5}_{-1.5}$ &   $31^{+4}_{-3}$ & --  \\
\citet{Bulik:95}    &            & --       & $1.34\pm0.16$  & --          & --  & -- \\
\citet{vanKerkwijk:95} & \#1      & $>74$    & $1.88^{+0.69}_{-0.47}$ & $23.5^{+2.2}_{-1.5}$ &   $30.0^{+1.9}_{-1.9}$ & $17.0-29.7$  \\
-- ditto --             & \#2      & $>73$    & $1.75^{+0.34}_{-0.33}$ & $23.3^{+2.5}_{-1.1}$ &   $30.2^{+1.7}_{-2.2}$ & $20.8\pm1.7$  \\
-- ditto --             & \#3      & $>75$    & $2.08^{+0.46}_{-0.43}$ & $23.6^{+1.9}_{-2.2}$ &   $29.9^{+1.4}_{-1.8}$ & $24.6\pm2.3$  \\
\citet{Stickland:97}   & (a)      & $>73$    & $1.39\pm0.14/(\sin i)^3$ & $21.41\pm0.22/(\sin i)^3$ & -- & $17.8\pm1.6$ \\
-- ditto --          & (b)         & $>73$    & $1.33\pm0.13/(\sin i)^3$ & $21.42\pm0.22/(\sin i)^3$ & -- & $17.8\pm1.6$  \\
-- ditto --          & (c)         & $>73$    & $1.45\pm0.14/(\sin i)^3$ & $21.42\pm0.22/(\sin i)^3$ & -- & $17.8\pm1.6$  \\
\citet{Barziv:2001}   &          & $>73$    & $1.86\pm0.32$            & $23.8^{+2.4}_{-1.0}$      & $30.4^{+1.6}_{-2.1}$ & $21.7\pm1.6$ \\
\citet{Quaintrell:2003} &        & $70.1\pm2.6$ & $2.26\pm0.32$        & $27.9\pm1.3$  & $32.1\pm0.6$ &  $21.2\pm0.7$  \\
-- ditto --          &            & 90       & $1.88\pm0.13$            & $23.1\pm0.2$  & $26.8\pm0.9$ & $22.6 \pm 1.5$ \\
\citet{Rawls:2011} (analytic) &   & $83.6\pm3.1$  & $1.788\pm0.157$     & --             & --          & --  \\
-- ditto --  (numerical)      &   & $78.8\pm1.2$  & $1.770\pm0.083$     & $24.00\pm0.37$ & $31.82\pm0.37$ & --  \\
\citet{KrtickaKubatKrtickova:2015}$^\dag$ & &   --   & --                 & 23.5        & 30     & -- \\ 
\citet{Falanga:2015}         &   & $72.8\pm0.4$  & $2.12\pm0.16$   & $26\pm1$  & $29\pm1$ & --  \\
\citet{Gimenez-Garcia:2016}  &   & --       & --             & $21.5\pm4.0$  & (JR84) & --  \\
\hline
\multicolumn{7}{l}{$^\dag$ Refer to \citet{vanKerkwijk:95}}
\end{tabular}
\end{small}
\end{center}
\renewcommand{\arraystretch}{1.0}
\end{table*}%

\begin{figure}[h]
\centerline{\includegraphics[width=1.0\linewidth]{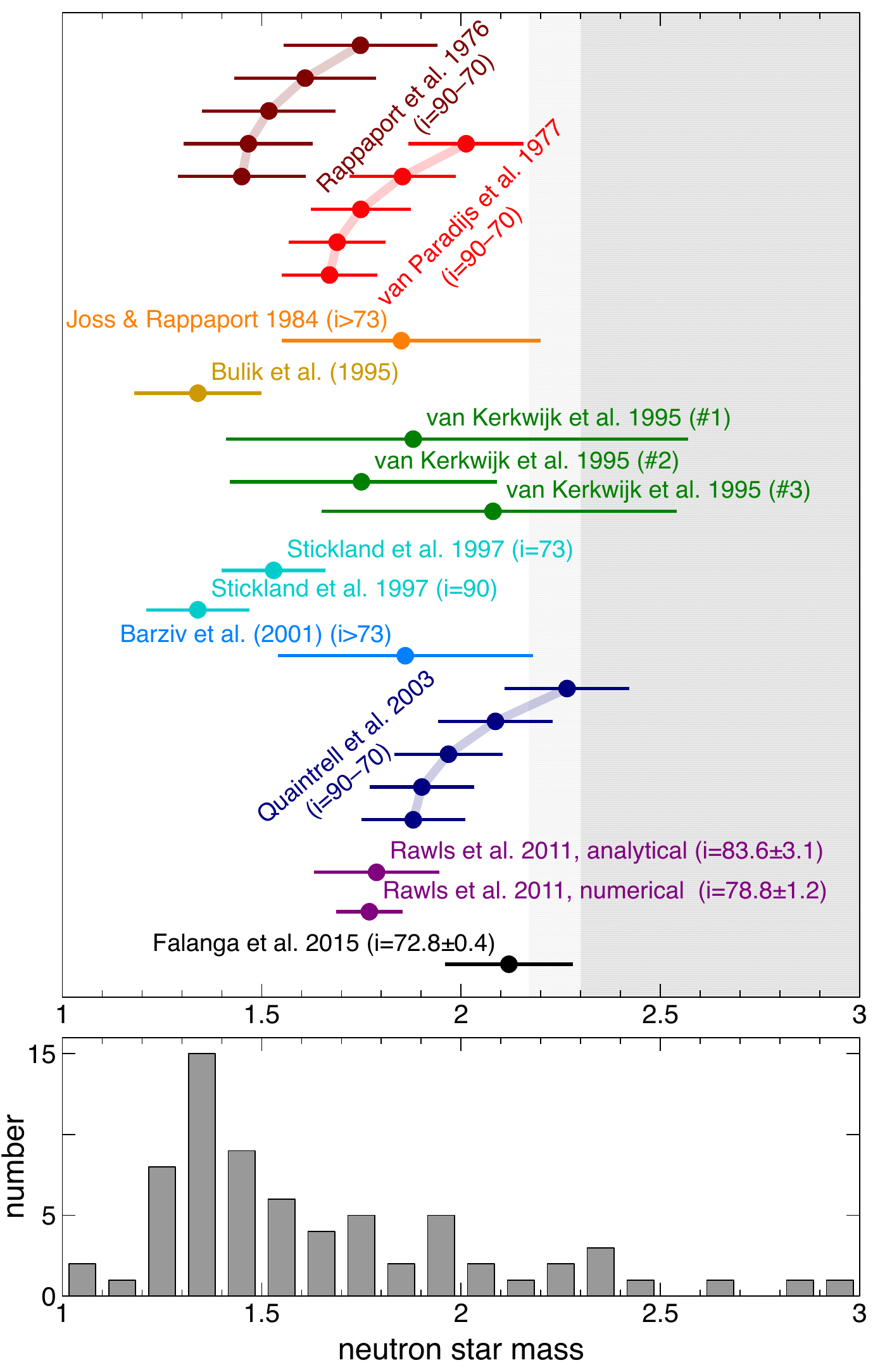}}
\caption{Upper panel: Selected estimates of the mass of the neutron star in Vela X-1 taken from the literature. 
The estimates are for specific assumed inclinations (see the original papers for how these where chosen) or plotted for a range of values with $i=90^\circ$ as the lowest mass point and then $5^\circ$ steps to $i=70^\circ$, to show the $\sin(i)$ dependence.
See text for the multiple solutions provided by \protect{\citet{vanKerkwijk:95}}.
The shaded areas on the right indicate possible maximum masses for neutron stars based on multi-messenger observations of GW170817 \protect{\citep[see][and references therein]{MargalitMetzger:2017}}.
To set them into perspective, the lower panel shows the distribution of derived masses of other neutron stars in the literature taken from \protect{\citet{Alsing:2018}} for comparison.
\label{fig:nsmass}}
\end{figure}

Most commonly, the masses of the two components of the binary have been determined from fits to RV curves (see Sect.~\ref{sec:obs:uvopt:radial}). In some cases, the mass and radius of HD~77581 have been estimated by other means. \citet{Lamers:1976}, for example, used half the orbital separation as the stellar radius estimate because this agreed with the mean value for these supergiants, and matched the temperature and evolutionary tracks to determine the mass of the donor star.

Early RV observations \citep{Hiltner:1972} were interpreted as indications of a possible  black hole in the binary system \citep{Wickramasinghe:74}, but with more measurements, this interpretation was soon changed to a high-mass neutron star system instead, with possible masses between 1.5 and 2.9~$M_\sun$ \citep{Hutchings:74}. In the following years, various other authors have undertaken to determine the masses of the two partners, sometimes using the same basic data sets, but arriving at different results.

As examples of how the results of the analysis may depend on assumptions, \citet{vanKerkwijk:95} derived three different sets of solutions for different ranges of the RV amplitude $K_\mathrm{opt}$ derived under different assumptions: The first set is for a conservative range, set 2 is for the best fit using all their data, and set 3 for $K_\mathrm{opt}$ derived excluding orbital phase 0.65--0.85, in which they found the strongest indications for systematic effects. As inclination constraints for these three cases, they used $i>74$, $i>73$ and $i>75$ degrees, respectively. Later, \citet{Stickland:97} used the same data set as \citet{vanKerkwijk:95}, but a different fitting procedure. They derived systematically lower estimates for the neutron star mass and reported three different possible solutions for (a) an unconstrained orbital fit, (b) using the orbital solution from \citet{Deeter:87a}, and (c) with the spectra of the comparison star taken with a small aperture.

An interesting complementary approach was used by \citet{Bulik:95}, who fit pulse-phase resolved X-ray spectra with a model of X-ray emission from polar caps in an inhomogeneous highly magnetized neutron star atmosphere. They included general relativistic effects to derive constraints on the neutron star mass and radius. Their results depend on assumptions about the emission region, however, and they reported systematic differences for fits made to different energy ranges. 

Table~\ref{tab:masses} compiles different mass estimates, and Figure~\ref{fig:nsmass} visualizes a selection of results for the neutron star mass in comparison with the mass distribution of other neutron stars with existing estimates. 

In summary, while there are clear indications for a relatively high-mass neutron star in Vela X-1, with a mass of about 1.8~$M_\sun$ or even higher, the systematic uncertainties in the derivation of the mass still allow a rather wide range of masses. The only moderately constrained inclination is a significant factor that by itself is responsible for a difference of $\sim$15\% between the lowest and highest mass of a given study, but there is larger scatter from the different approaches. The radius of the neutron star has not been derived from observations. It will depend on the actual mass and the equation of state of the neutron star. Discussing the many solutions for the equation of state and the corresponding mass-radius relations is beyond the scope of this article \citep[see, e.g.,][for a review]{Lattimer+Prakash:2007}, but based on the tracks shown in \citet{Maselli:2020} for  models compatible with current constraints, we arrive at possible radii between 11 and 12.5~km for a plausible range in neutron star masses in the Vela X-1 system.
Similarly, the mass of HD~77581 is effectively constrained to a range 20--30~$M_\sun$ , while its radius is mostly found to be $\sim 30 R_\sun$.

Finally, indirect theoretical arguments and numerical approaches can also be used in favor of certain values for the mass of the neutron star. For instance, \cite{ManousakisWalterBlondin:2012} proposed a mass estimate based on the geometry of the accretion flow as deduced from the absorption column density profile: All things being equal, the higher the mass of the neutron star, the more distorted the wind they obtained in their hydrodynamical simulations, leading to different column density profiles. Although they considered another HMXB, IGR J17252$-$3616, as an example case, their method would be applicable to Vela X-1 as well. We also note that for a mass ratio lower than $\sim$12, the absence of RLOF (see Section~\ref{sec:elements:mass_transfer}) sets a more stringent lower limit on the orbital inclination than the eclipse duration.


\begin{table*}[htb]
\renewcommand{\arraystretch}{1.1}
\caption{Parameters of HD~77581 derived by different authors. See Sect.~\ref{sec:obs:uvopt:spectra} for notes on the different approaches used. For \citet{Gimenez-Garcia:2016} and \citet{Sander:2018}, the temperature we report here as $T_\star$ corresponds to the parameter $T_{2/3}$ as defined in the PoWR code used by these publications.
For mass and radius estimates, see Table~\ref{tab:masses}.}
\label{tab:stellarpar}
\begin{small}
\centering
\begin{tabular}{@{}l@{}rrrrrl@{}} 
\hline\hline
Reference & 
\multicolumn{1}{c}{$\log L$} & 
\multicolumn{1}{c}{$E_{B-V}$} &
\multicolumn{1}{c}{$T_\star$} &
\multicolumn{1}{c}{$v_\mathrm{rot} \sin i$} &
\multicolumn{1}{c}{$\log g_\star$} \\ 
 & 
\multicolumn{1}{c}{[$L_\sun$]} & 
\multicolumn{1}{c}{[mag]} &
\multicolumn{1}{c}{[kK]} &
\multicolumn{1}{c}{[km s$^{-1}$]} &
\multicolumn{1}{c}{[cgs]} &
Comments\\
\hline
\citet{Wickramasinghe:74}       & --           & 0.75            & 20--28        & $\sim90$      & 2.5--3.0 & RV, spectrum and reddening \\
\citet{MikkelsenWallerstein:74} (MW74) &  5.2  & 0.75            & 22            & 130           & 2.64--2.80  & \\
\citet{Lamers:1976}             & --           & --              & 26.3          & --            & --       & \\
\citet{Ammann+Mauder:1978}      & --           & --              &  --           & $89\pm5$      & --       & HeI line fitting \\
\citet{Conti:78}  (C78)         & 5.7          & 0.73            & 26            & (MW74)        & --       & \\
\citet{Dupree:80}               & (C78)        & 0.7             & (C78)         & --            & --       & \\
\citet{Sadakane:85}             & 5.53         & --              & $25\pm 1$     & --            & --       & \\
\citet{Vanbeveren:93ADS}        & 5.5--5.7     & --              & 26            & 130           & 2.75--2.76 \\
\citet{Zuiderwijk:95}           & --           & --              & --            & $116\pm6$     & --       & HeI line profiles\\
\cite{Howarth:1997}             & --           & --              & --            & 114           & --       & cross-correlating IUE spectra \\
\citet{Fraser:2010} (F10)       & --           & --              & 26.5          & 56            & 2.90     & Arrive at mass of 40\msun \\
\citet{Falanga:2015}            & --           & --              & --            & $130.3\pm0.2$ & --       & From system geometry \\  
\citet{KrtickaKubatKrtickova:2015} & 5.63      & --              & 27            & --            & --       & \\
\citet{Gimenez-Garcia:2016}     & $5.5\pm 0.1$ & $0.77\pm 0.05$  & $24.4\pm 1.0$ & (F10)         & $2.86\pm 0.10$ & Uses $d=2.0\pm0.2$~kpc\\
\citet{Sander:2018}, no X-rays  & 5.485        & --              & 23.5          &  (F10)        & 2.84     & same distance as above \\ 
-- ditto --, moderate X-ray illum. & -- ditto -- & --            & 23.5          &  -- ditto --   & --       & \\
-- ditto --, strong X-ray illum.  & -- ditto --  & --            & 23.8          &  -- ditto --   & --       & \\
\textit{this work (Section~\ref{sec:system:chorizos})}    & 5.8--6.2     & $0.689\pm 0.018$ & $33.7\pm5.2$ & --            & $3.28\pm 0.19$ & Uses $d=1.99$~kpc \\
\hline
\end{tabular}
\end{small}
\renewcommand{\arraystretch}{1.0}
\end{table*}%

\subsection{Revising the spectral classification of HD~77581}
\label{sec:system:spectralclass}

 \href{http://simbad.u-strasbg.fr/simbad/}{SIMBAD} lists different spectral classifications of HD 77581. They range in spectral subtype from B0 to B0.5 and in luminosity class from Ib to Ia. This means that the different authors agree that it is an early-type B supergiant, but disagree on the precise spectral classification. This disagreement is quite common for OB stars because many objects were classified in the 1950s to 1970s using photographic plates, and they have not been revisited with modern digital data since then. For this reason, the Galactic O-Star Spectroscopic Survey (GOSSS, \citealt{Maizetal11}) was started a decade ago. It was originally restricted to the modern spectral classification of O stars and was later extended to B stars. A GOSSS spectrogram of HD~77581 was presented in \citet{MaizApellaniz:2018} and given the classification of B0.5~Ia. Since that paper, the GOSSS team (of which the PI is a coauthor here) has revisited the spectral classification criteria for B supergiants and is currently working on a new publication on this issue. The two main criteria for the horizontal classification of early-B stars are the ratio of Si\,{\sc iii} to Si\,{\sc iv} lines and the progressive disappearance of He\,{\sc ii} lines in the spectrum with evolution from B0 to B1. Originally, He\,{\sc ii} lines were only seen in O stars, but with improvements in S/N and the advent of digital data, the current criteria establish that for supergiants, He\,{\sc ii}~$\lambda$4542 disappears at B0.5 (and is just visible at B0.2) and that He\,{\sc ii}~$\lambda$4686 disappears at B1 (and is just visible at B0.7). HD~77581 shows a weak He\,{\sc ii}~$\lambda$4542 absorption, and the ratio of the Si\,{\sc iii} to Si\,{\sc iv} lines is intermediate between those at B0 and B0.5. On the other hand, the ratio of the Si\,{\sc iii} to He\,{\sc i} lines and the H$\beta$ to H$\delta$ profiles indicates a high luminosity class of Ia (but not Ia+ because H$\beta$ does not show a P-Cygni profile). Therefore, we revise the GOSSS spectral classification from B0.5~Ia to B0.2~Ia.

\subsection{New determination of stellar parameters} 
\label{sec:system:chorizos}

\begin{figure}
    \centering
    \includegraphics[width=1.0\linewidth]{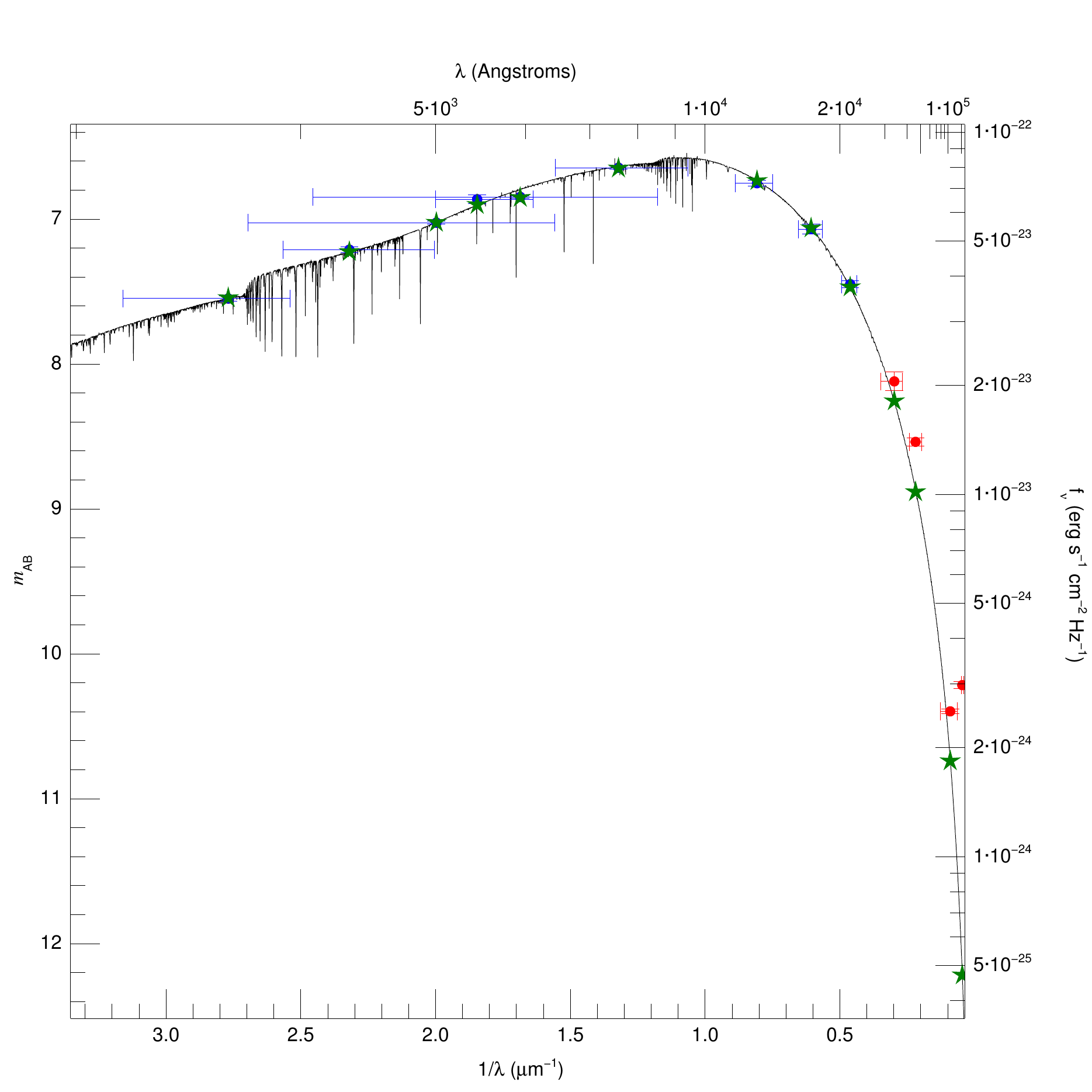}
    \caption{CHORIZOS mode SED. Blue error bars are used for the input photometry that was used for the fit (\textit{Gaia}~DR2 $G+G_{\rm BP}+G_{\rm RP}$, Johnson $U+B+V$, and 2MASS $J+H+K_{\rm s}$) and red error bars for the discarded input photometry (WISE $W1+W2+W3+W4$). In each case, the vertical error bar is the photometric uncertainty, and the horizontal error bar is a representation of the wavelength coverage of each filter. Green stars are used for the model photometry.}
    \label{fig:CHORIZOS}
\end{figure}

Based on newer data and our distance estimate (see Section~\ref{sec:system:distance}), we used the latest version of CHORIZOS \citep{Maiz04c} to again derive some of the parameters of HD~77581. CHORIZOS is a Bayesian photometric code that computes the likelihood that a set of photometric data is compatible with a family of SEDs.

\textit{Photometric data.} We collected the magnitudes from \textit{Gaia}~DR2 $G+G_{\rm BP}+G_{\rm RP}$ (\citealt{Evanetal18a} with the corrections and calibration from \citealt{MaizWeil18}), Johnson $U+B+V$ (\citealt{Mermetal97} with the calibration from \citealt{Maiz06a,Maiz07a}), 2MASS $J+H+K_{\rm s}$ (\citealt{Skruetal06} with the calibration from \citealt{MaizPant18}), and WISE $W1+W2+W3+W4$ \citep{Cutretal13}. As the object has a significant IR excess, we eliminated the four photometric points with the longest effective wavelengths from the runs (the four WISE bands), but included them in the analysis to evaluate the excesses.

\textit{Models.} We used the $T_{\rm eff}$-luminosity class grid with solar metallicity from \citet{Maiz13a}. The temperature of interest lies in the region for which the grid uses the TLUSTY models \citep{LanzHube07} in the optical and the Munari models \citep{Munaetal05} in the NIR. The reason for using these combined models is that the TLUSTY SEDs for B stars yield incorrect photometric values by up to several hundredths of a magnitude in the NIR. The Munari models do not have this problem.

\textit{Parameters.} For our run we fixed the value of the logarithmic distance $\log d$ to 3.299, and we left four free parameters: $T_{\rm eff}$, luminosity class, $E(4405-5495)$ (amount of extinction), and $R_{5495}$ (type of extinction). We used the family of extinction laws of \citet{Maizetal14a}, see \citet{Maiz04c,Maiz13b} and \citet{MaizBarb18} for an explanation of why monochromatic quantities need to be used instead of band-integrated quantities to characterize extinction. Because we fit nine photometric points, we had five degrees of freedom.

\textit{Results.} The results of the CHORIZOS run are given in Table~\ref{tab:CHORIZOS}, and the mode SED is shown in Fig.~\ref{fig:CHORIZOS}. 
The reduced $\chi^2$ of the mode SED is 0.75, indicating an excellent fit of the model to the data. This is shown in Fig.~\ref{fig:CHORIZOS}, where all green stars up to $K_{\rm s}$ are within the blue error bars or very close to them. The first five lines of Table~\ref{tab:CHORIZOS} give the results for the model parameters (mean, uncertainty, and mode), and the last nine lines list the results for the derived parameters (mean and uncertainty). We discuss the model parameters first.

\begin{table}[]
    \caption{Detailed results of the CHORIZOS analysis.}
    \label{tab:CHORIZOS}
    \renewcommand{\arraystretch}{1.05}
    \centering
    \begin{tabular}{lcr@{.}lr@{.}lr@{.}l}
    \hline\hline
     Quantity       & Units        & \mcii{Mean} & \mcii{Unc.}  & \mcii{Mode} \\
     \hline
     \multicolumn{8}{l}{Model parameters} \\
     \hline
     $T_{\rm eff}$  & kK           &   33&7      &   5&2        &    30&9     \\
     Lum. class     & ---          &    0&65     &   0&31       &     0&84    \\
     $R_{5495}$     & ---          &    3&505    &   0&071      &     3&519   \\
     $E(4405-5495)$ & mag          &    0&689    &   0&018      &     0&681   \\
     $\log d$       & log pc       &    3&299    & \mcii{fixed} &     3&299   \\
     \hline
     \multicolumn{8}{l}{Derived parameters from evolutionary tracks} \\
     \hline
     Luminosity     & $10^6$ solar &    1&05     &   0&46       &  \mcii{}    \\
     $\log g$       & log cgs      &    3&28     &   0&19       &  \mcii{}    \\
     ZAMS mass      & solar        &   71&2      &  22&6        &  \mcii{}    \\
     age            & Myr          &    2&75     &   0&78       &  \mcii{}    \\
     $A_{5495}$     & mag          &    2&367    &   0&064      &  \mcii{}    \\
     $A_{V_J}$      & mag          &    2&445    &   0&063      &  \mcii{}    \\
     $E(B-V)$       & mag          &    0&691    &   0&018      &  \mcii{}    \\
     $V_{J,0}$      & mag          &    4&478    &   0&057      &  \mcii{}    \\
     $M_{V_{J,0}}$  & mag          & $-$7&017    &   0&057      &  \mcii{}    \\
   \hline
    \end{tabular}
    \renewcommand{\arraystretch}{1.0}
\end{table}

$T_{\rm eff}$ is poorly constrained because the main discerning criterion is the Balmer jump, which is determined almost exclusively by the Johnson $U-B$ \citep{Maizetal14a}. The value is above those obtained from spectroscopy, but within 2$\sigma$. The luminosity class is the photometric equivalent to the spectral luminosity class; it varies from 0.0 (hypergiants) to 5.5 (ZAMS). The value found by CHORIZOS (which is dependent on the fixed distance) indicates it is a bright supergiant, which is consistent with the Ia spectral luminosity classification. The extinction parameters indicate that the value of $R_{5495}$ is above the canonical value of 3.1, which is relatively frequently the case for sightlines with $E(4405-5495)$ \citep{MaizBarb18}, and that $E(4405-5495)$ [or $E(B-V)$] is slightly lower than previous values. We note, however, that the higher value of $R_{5495}$ leads to only a small change in $A_{5495}$ (or in $A_{V_J}$) compared to a measurement of $E(4405-5495)$ of 0.75 assuming a value of $R_{5495}$ of 3.1, for example. 

\begin{figure*}
    \centering
    \includegraphics[width=1.0\linewidth]{{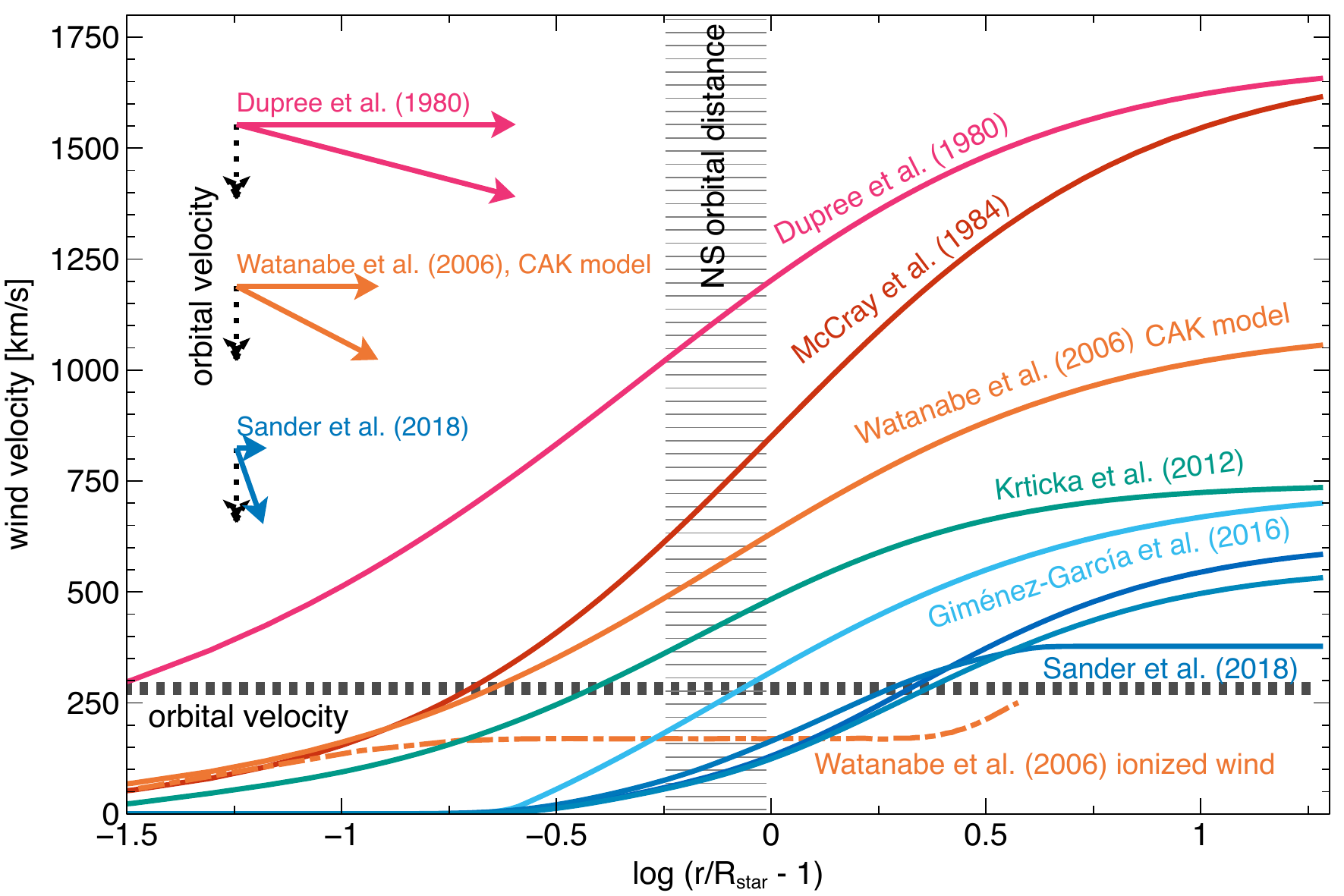}}
    \caption{Comparison of different derived wind velocity profiles from the literature (Tab.~\ref{tab:wind}). See text for more details of the models. These are mostly the solutions for the wind \emph{\textup{without}} accounting for the ionization of the neutron star, except in the case of two of the models by \citet{Sander:2018} and, at the bottom, the velocity profile calculated by \citet{Watanabe:2006} along the line connecting donor and neutron star. The range of possible distances covered by the neutron star on its eccentric orbit \citep{Bildsten:97} and the orbital velocity range, mainly driven by the uncertainty in inclination, are indicated as well. In the upper left corner we indicate approximately how wind (shown as horizontal lines) and orbital velocity (vertical) would combine at the mean neutron star distance from HD~77581 for three of the reported velocity profiles. These differences  have a strong effect on the expected accretion flow close to the neutron star (see Sect.~\ref{sec:models:wind}).}
    \label{fig:windspeeds}
\end{figure*}

The derived parameters in Table~\ref{tab:CHORIZOS} were obtained with the help of the Geneva \citep{LejeScha01} evolutionary tracks used by the \citet{Maiz13a} grid and assuming the value of $\log d$ given there. The luminosity we found is close to one million in solar units with a relatively large uncertainty caused by the relatively large uncertainty in $T_{\rm eff}$ (and, hence, on the bolometric correction). The values for $\log g$, ZAMS mass, and age depend in a complex way on distance, and if different distance were to be used, they would change in a nontrivial way. In any case, taken at face value and assuming the validity of the used evolutionary tracks, the CHORIZOS results indicate that HD~77581 is a very massive star that was born less than three million years ago and that initially would have been a very early type O star or even a WNh star. Based on the existence of the neutron star companion and on the mass transfer that must have taken place in the past, the previous initial state and evolution of the star are likely to have been different. Moreover, the uncertainty in the absolute magnitude of the star in the Johnson $V$ band is very low. This is so because (a) the extinction parameters are very well determined, (b) for this quantity we are not affected by the bolometric correction (as we are for the luminosity), and (c) we considered the distance as a fixed quantity. For a different distance, we would need to correct for the different value of $5\log d$.

Finally, we analyzed the behavior of the SED in the WISE bands in the MIR. There are clear excesses there that we quantify as 0.135$\pm$0.066~mag ($W1$), 0.346$\pm$0.035~mag ($W2$), 0.346$\pm$0.027~mag ($W3$), and 2.007$\pm$0.035~mag ($W4$). We note that the Munari models do not include stellar winds. 

\subsection{Stellar wind parameters} \label{sec:system:wind}

\begin{table*}[htb]
\renewcommand{\arraystretch}{1.2}
\caption{Parameters of the stellar wind in the Vela X-1 system as derived by different authors,  see Sects.~\ref{sec:obs:uvopt:spectra} and~\ref{sec:obs:uvopt:uv_resonance}.
When a value was taken from a previous publication, we mark this by a shorthand, e.g., D80 for \citet{Dupree:80}.
Numbers in \textit{\textup{italics}} in the table have been set as fixed in the respective publications. \citet{Krticka:2012} and \citet{Sander:2018} did not assume a basic $\beta$-law for wind acceleration, and the
value given in this table corresponds to an approximate $\beta$-law fit to their results for comparison. 
For other parameters, see Table~\ref{tab:masses} and Table~\ref{tab:stellarpar}.}
\label{tab:wind}
\begin{center}
\begin{small}
\begin{tabular}{lrrrrl}    
\hline\hline
Reference & 
\multicolumn{1}{r}{$v_\mathrm{esc}$ [km s$^{-1}$]} &
\multicolumn{1}{r}{$v_{\infty}$ [km s$^{-1}$]} &
\multicolumn{1}{r}{$\beta$} &
\multicolumn{1}{r}{$\dot{M}$ [$10^{-6} M_\sun \mathrm{yr}^{-1}$]} &
\multicolumn{1}{l}{Comments} \\
\hline
\citet{Lamers:1976}  (L76)      & --          & --                  & --             &   0.5 & X-ray fluxes \& wind accretion \\
\citet{Hutchings:1976} (H76)    & --          & --                  & --             &     7 & spectroscopy, mass-loss model \\      
\citet{Conti:78}                & 519         & 1600                & --             & (H76)  & Similar study to L76 \\
\citet{Dupree:80} (D80)         & 489         & $\sim$1700         & \textit{0.5}   & $\sim$1 & \\
\citet{McCray:1984} (M84)       & --          & (D80)               & \textit{1}     & 4  & \\
\citet{Sadakane:85}             & --          & $\sim$1500         & --             & (D80) & multiple components with different velocities \\
\citet{Sato:86PASJ}                 & --          & \textit{1600}$^{(a)}$ & (D80)        & $2.4\pm1.2$ & \\
\citet{Prinja:90}               & (590)       & 1105                & --             &  --   & $v_\mathrm{esc}$ is mean for spectral type \\
\citet{Sako:99}                 & --          & (D80)               & 0.79$\pm$0.23  & $0.265^{+0.065}_{-0.050}\,^{(b)}$  & X-ray DEM modeling \\
\citet{vanLoon:2001}            & --          & $600\pm100$         & --             & --      & Modeling of UV lines \\ 
\citet{Watanabe:2006}           & --          & \textit{1100}$^{(c)}$ & \textit{0.8} & 1.5\,--\,2.0 & X-ray line MC modeling \\
\citet{Krticka:2012}            &             & 750                 & ($\sim$0.8)    & 1.5 & see reference for details \\
\citet{Falanga:2015}            & --          & $640\pm190$         & --             & $<$(1\,--\,5.3) & see reference for details \\
\citet{ManousakisWalter:2015b}  & --           & (D80)              & 0.55   & (M84) & Simulating hard X-ray light curve \\
\citet{Gimenez-Garcia:2016}     & $436\pm 65$ & $700^{+200}_{-100}$ & \textit{1.0}    & 0.4\,--\,1 &  \\
\citet{Sander:2018}, no X-rays    & --        & 532                 & ($\sim$2.2)  &  0.65 & see reference for details \\
-- ditto --, moderate X-ray illum. & --        & 584                 & ($\sim$2.2)  &  0.66 \\
-- ditto --, strong X-ray illumination   & --  & 378                 & ($\sim$2.2)  &  0.85 \\
\hline
\multicolumn{6}{l}{$^{(c)}$ Referring to \citet{Dupree:80}.}\\
\multicolumn{6}{l}{$^{(b})$ Argued to be the mass fraction in hot ionized gas, while most mass would be contained in cool clumps.}\\
\multicolumn{6}{l}{$^{(c)}$ Referring to \citet{Prinja:90}.}\\
\end{tabular}
\end{small}
\end{center}
\renewcommand{\arraystretch}{1.0}
\end{table*}%

Over the years, many different approaches have been made to determine essential parameters of HD~77581 related to its strong stellar wind. They are collected in Tables~\ref{tab:masses}, \ref{tab:stellarpar}, and \ref{tab:wind}.
As summarized in section~\ref{sec:obs:uvopt:spectra}, one common approach based on the study of massive stars in general is through (quantitative) spectroscopy. This has ranged from the study of specific lines that  were sometimes compared with samples of example stars \citep[e.g.,][]{Wickramasinghe:74,Ammann+Mauder:1978,Sadakane:85,Prinja:90,Zuiderwijk:95,Howarth:1997}, to a detailed SED analysis against grids of stellar model spectra created with increasingly refined wind model codes \citep{Vanbeveren:93ADS,Fraser:2010,Gimenez-Garcia:2016,Sander:2018};  see also section~\ref{sec:models:wind}. 

Other authors explicitly included the X-ray source in the derivation, for example, by comparing the observed X-ray luminosity with that expected from a model of stellar wind accretion (see section~\ref{sec:elements:mass_transfer} \citep{Lamers:1976,Conti:78} ) or full hydrodynamic simulations \citep{ManousakisWalter:2015a,ManousakisWalter:2015b}. Others modeled the Hatchett-McCray effect on line features that vary with orbital phase \citep{Dupree:80,vanLoon:2001}.
Yet another approach is based on the modeling of high-resolution X-ray spectra (section~\ref{sec:obs:xray:lines}), assuming an ionized stellar wind of a certain structure, with some parameters taken as given from previous work and others then derived from the spectral modeling \citep{Sako:99,Watanabe:2006}.

We also note that most of the efforts start from a smooth, nonclumped wind, even though other physical effects that are taken into account then may break the original full symmetry. 

When we compare all these results, it is clear that the basic parameters of the mass donor and its stellar wind are still only known within a significant range of uncertainty. Care must be taken when astrophysical conclusions are derived from a comparison of model results with any specific value. 

For the rotational velocity, the derived values vary by a factor $>2$, although in all cases, HD~77581 rotates significantly more slowly than the orbital movement: by a factor of roughly one-third to two-thirds, based on the values in Table~\ref{tab:stellarpar}. \citet{Gimenez-Garcia:2016}, who adopted the low value of 56~km s$^{-1}$ from \citet{Fraser:2010}. We note that the high estimates of rotational velocity are not consistent with optical lines that are observed to be unblended. Additional broadening of lines by macroturbulence \citep[e.g.,][]{Ryans:2002} may play a role in these discrepancies.

The mass-loss rate is known to within a factor of a few when more recent estimates are compared. The spread in values is within an order of magnitude when a wider range of estimates and their uncertainties from the literature are considered. 
 
The terminal velocity $v_{\infty}$ shows an apparent trend with rather high values preferred in earlier studies and lower values in more current ones, even when we consider that \citet{Gimenez-Garcia:2016} and \citet{Sander:2018} were based on the same underlying model code. The $\beta$ parameter has usually not been derived, but was set to a specific value based on assumptions about the physics of line-driven winds. 

Figure~\ref{fig:windspeeds} compares different descriptions of the stellar wind velocity around HD~77581. 
The profiles based on \citet{Dupree:80}, \citet{McCray:1984}, and \citet{Watanabe:2006} are $\beta$-laws (Eq.~\ref{eq:beta-law}). \citet{Krticka:2012} used a more complex polynomial expansion in their study. In the case of \citet{Gimenez-Garcia:2016}, a $\beta$-law is connected to an atmosphere solution, while in the three solutions given by \citet{Sander:2018}, the velocity profile is determined as a solution of a hydrodynamically consistent atmosphere model describing the wind stratification.


\subsection{Wake structures} \label{sec:system:wake}
\citet{Charles:1976} reported observations of \vel with the MSSL X-ray instrument. They found indications in their data for increased absorption in a trailing accretion wake and possibly in a leading bow shock, noting that these structures varied from cycle to cycle and probably during each cycle. Referring to these results, \citet{Conti:78} noted that a wake like this would mean that the wind velocity at the distance of the source would be on the order of the orbital velocity, that is, at much lower values than assumed in some later publications (see Fig.~\ref{fig:windspeeds}). The existence of this wake is also the commonly accepted main contribution to the high column densities observed at late orbital phases (Sect.~\ref{sec:obs:xray:abs} and Fig~\ref{fig:nh_orbit}).

Indications for a photoionization wake (see Sect.~\ref{sec:ion_struct}) have been found by \citet{Kaper:94} from  high-resolution optical spectroscopy of the hydrogen Balmer lines and He lines and by \citet{vanLoon:2001} from orbital modulation through additional absorption of UV lines (see Sect.~\ref{sec:obs:uvopt:uv_resonance}). \citet{Barziv:2001} also explained part of their spectroscopic data with the presence of this wake (see Sect.~\ref{sec:obs:uvopt:radial}). 
From their simulations of stellar wind disruption, \citet{Blondin:90} noted that for a photoionization wake to contribute significantly to the integrated column density, the wind would have to be ionized most of the way to the companion star.

In summary, the trailing wakes caused by the movement of the neutron star through the dense stellar wind and the added effect of the ionizing X-ray radiation play an important role in the modulation of the observed emission at many wavelengths. When models are compared to data, it is important to keep in mind that these structures can be expected to significantly vary from one orbital cycle to the next, and that the photoionization wake is also strongly affected by the significant variations in the X-ray flux (Sect.~\ref{sec:obs:xray:flux}).

\subsection{Strength of the magnetic field of the neutron star} \label{sec:system:B-field}

When we identify the 25\,keV line that is visible in the X-ray spectrum of \vel (Section~\ref{sec:obs:xray:cyclo}) as the fundamental cyclotron line, we can use Eq.~\ref{eq:Ecyc} to infer a magnetic field strength in the line-forming region of $2\times(1+z) 10^{12}$\,G. As has been shown by \citet{Fuerst:2010} and \cite{LaParola:2016}, the first harmonic (at about 55\,keV) shows a correlation with luminosity, indicating that the line-forming region changes in altitude above the surface of the neutron star. Using this correlation and theoretical calculations by \citet{Becker:2012}, \citet{Fuerst:2010} inferred a surface magnetic field of $B_\star=2.6\times10^{12}$\,G. However, this value should be taken only as an estimate because it depends on the exact physical conditions within the accretion column, in particular, on the dominating process of decelerating the in-falling material, as well as on the mass and radius of the neutron star.


In principle, we could also infer constraints on the magnetic field strength from the observed torques, as has been done for other accreting X-ray sources. This type of estimate depends on many model assumptions (see Section~\ref{sec:models:acc-ind_torq}), however, especially for a source with a potentially variable accretion geometry, and thus does not add information for \vel.

\begin{table*}[hbt]
\renewcommand{\arraystretch}{1.15}
    \centering
    \begin{small}
    \caption{Selected estimates of the maximum spin period or frequency changes reported in the literature, as given in the reference. For comparison, they are converted into $\dot{P}$ in ms per day.}
    \label{tab:system:pdot}
\begin{tabular}{lrrlrr}  
\hline\hline
Reference & \multicolumn{2}{c}{Values given in reference} & Units used in & \multicolumn{2}{c}{Expressed as $\dot{P}$ in ms / day} \\ 
          & Spin-up & Spin-down                 & reference & Spin-up & Spin-down                  \\
\hline
\citet{Nagase:84ApJ} & $-4\times 10^{-3}$ & $+3\times 10^{-3}$ & $\dot{P}/P$ yr$^{-1}$ & $-3.1$ & $+2.3$  \\
\citet{Boynton:84} & \multicolumn{2}{c}{up to $\pm 5.8(1.4)\times 10^{-3}$} & $\dot{\Omega}/\Omega$ yr$^{-1}$ &  \multicolumn{2}{c}{up to  $\pm 4.5(1)$} \\
\citet{Deeter:89}    & +2.3(0.5) & $-$2.4(0.6) & $\dot{\Omega}$ in pico-rad s$^{-1}$ & $-$2.5(0.6) & +2.6(0.7) \\
\citet{Inam:00}      & $+7.1(2.2) \times 10^{-9}$ & $-6.7(2.2) \times 10^{-9}$ & $\dot{\nu}$ in Hz\,day$^{-1}$ & $-$0.6(0.2) & +0.5(0.2) \\
\citet{Liao:2020} & $+2.3(1.3)\times 10^{-13}$ & $-1.5(8)\times 10^{-13}$ & $\dot{\nu}$ in s$^{-2}$ & $-$1.6(0.9) & +1.0(0.6) \\
\hline
\end{tabular}
\end{small}
\renewcommand{\arraystretch}{1}
\end{table*}

\subsection{Observed torques on the neutron star} \label{sec:system:Pdot}

Changes in the observed spin period, or equivalently, the spin frequency (Section\,\ref{sec:obs:xray:pulse}), can in principle be used to derive information on the torques acting on the neutron star through interactions in the magnetosphere (Section\,\ref{sec:models:acc-ind_torq}). It is important to note, however, that the reported periods or frequencies are often average values over timescales of hours, days, and sometimes many weeks, which averages out short-term fluctuations. The reported changes are thus at least on the same timescales, or between data points even farther apart. In contrast, model predictions are instantaneous. A comparison between observed and predicted changes therefore implies that the characteristic timescale of a change in spin-period differences is on the same timescale as the observations. Otherwise, even for a correct model, the absolute values of the observed differential change will always be lower than the differential change inferred from the model.  
On the other hand, attempts to derive pulse periods on shorter time intervals within long observations will be affected by the pulse-to-pulse variations in the observed pulse profile (Section\,\ref{sec:obs:xray:pulseprofiles}). 


Table~\ref{tab:system:pdot} collects a number of reported spin-period or frequency changes in the literature, without aiming at completeness. It demonstrates rather consistent values with the largest observed changes in pulse period on the order of  a few milliseconds per day.

\section{Outlook to future observations or studies}
\label{sec:future}

In this section we consider possible future observations or studies that would help understand the \vel system better and thus allow deeper insights into the physics of this archetype, which may shed light on many other systems as well. The following ideas do not claim to be an exhaustive list, but reflect topics raised while compiling this overview.

\subsection{Evolutionary scenario and origin} \label{sec:future:evolution}

The precise evolutionary stage of the donor star is still unclear. Is the blue supergiant expanding toward becoming a red supergiant? Or is the donor already evolving blueward and might become a Wolf-Rayet star? The precise determination of the stellar properties will be key to answering these questions and constrain evolutionary modeling efforts. These will need to be provided by analyzing the donor with hydrodynamically consistent models that sufficiently reproduce the spectrum, but also take the X-ray irradiation of different parts of the wind into account. The numerical costs currently prevent full 3D scenarios from being developed, but sufficient approximations using multiple cones could become available in the near future and extend the first efforts by \cite{Krticka:2012} and \cite{Sander:2018}.

In order to test the association of \vel with the Vela~OB1 association (Sect.~\ref{sec:system:VelaOB1}), it might be attempted to retrace the system trajectory using a Galactic potential. The difficulty lies in an accurate determination of the center-of-mass velocity of the binary, which requires low uncertainties for both the tangential proper motion and the absolute RV. While the first can be obtained with high precision from \textit{Gaia} data, the wide absorption lines affected by line infilling from the stellar wind and pulsations make the second a challenge that requires multiepoch high-resolution spectroscopy.

\subsection{Multiwavelength observations} \label{sec:future:multiwavelength}

It is essential to coordinate systematic multiwavelength observations to follow variations on timescales of days or faster. Excellent data exist at all wavelengths, but for the most part, they were not coordinated in the past and are thus not able to connect the short-term variations across the spectral bands.

Another angle that could be further explored are possible correlations between the longer-term X-ray variations that have been captured for decades now by X-ray monitors, and the known significant variations in optical brightness around the mean curve determined by the orbit. In the area of robotic survey telescopes, obtaining regular short snapshots of a star as bright as HD~77581 should be rather easy. We also note that Vela X~1 has occasionally been in the field of view of the \textit{TESS}\footnote{\url{https://heasarc.gsfc.nasa.gov/docs/tess/}} mission with its excellent photometry.

The advances of optical and near-IR interferometry, especially recently with VLTI/Gravity \citep{GRAVITY:2017}, allow resolving HMXBs in the Galaxy at submilliarcsecond resolution and thus at the spatial scale where accretion takes place \citep{Waisberg:2017}. Spectral differential interferometry can provide spatial information on scales as small as $\sim$1 to 10~$\mu$as, but the only published interferometric results for \vel \citep[][see Sect~\ref{sec:obs:uvopt:ir}]{Choquet:2014} were unable to detect differential visibility signatures beyond the noise level. At the distance of \vel, a resolution of 10~$\mu$as would translate into approximately 10\% of the orbital separation, resolving the scale of the accretion radius (see Fig.~\ref{fig:radii}) and the level of the uncertainty of the radius of HD~77581 (see Tab.~\ref{tab:masses}). Reaching the scale of the magnetosphere would require further improvements by at least another order of magnitude, which appears beyond the scope of improvements in interferometry foreseen in the nearer future \citep{Eisenhauer:2019GRAVITYplus}.

In X-rays, no interferometry mission is in direct planning. However, mission concepts exist, and high-mass X-ray binaries feature prominently among the possible science cases \citep{Uttley:2019}. In particular, even a single spacecraft interferometer could reach a resolution of $\sim$100\,$\mu$as and thus resolve the components of a bright high-mass X-ray binary such as \vel in an observation of a few hours \citep{Uttley:2020}. Higher resolution would likely require formation flight of multiple spacecrafts, but would allow us to determine the location of different emission components.

In order to study the winds on timescales of seconds, data from the fast and sensitive optical spectrographs available today will be required from observations that ideally would cover different orbital phases and also longer timescales.

At the X-ray energy range, X-ray calorimeters such as that on board the X-ray Imaging and Spectroscopy Mission (\textit{XRISM}) \citep{XRISM:2020} or the X-ray Integral Field Unit \citep{2018SPIE_XIFU} of the \textit{Athena} observatory \citep{2017Athena} will definitely usher in a new era on the X-ray line diagnostics of the source in combination with more precise atomic data from laboratory studies. Compared to the current grating spectrometers, they will allow studying lines in detail on shorter timescales and thus will enable us to better follow the dynamics.

The advent of X-ray polarimetry with the upcoming Imaging X-ray Polarimetry Explorer (\textit{IXPE})\footnote{\url{https://ixpe.msfc.nasa.gov/}} mission \citep{Sgro:2019IXPE} and later in the decade, the planned enhanced X-ray Timing and Polarimetry (\textit{eXTP}) mission \citep{eXTP:2019}, will offer a complementary opportunity of obtaining information about the geometry of the intrinsic X-ray emission, as well as about the system geometry and structures in the stellar wind when compared with detailed models \citep[e.g.,][see also below]{Kallman:2015,Caiazzo+Heyl:2021a}.
Additional information can be obtained from hard X-ray polarimetry, such as has been proposed for the balloon experiment XL-Calibur \citep{Abarr:2021}, which will also cover the crucial CRSF energy range.

The unexpected detection of \vel as a radio source has opened a tantalizing new window to explore the system properties. At the time when this article was completed, further radio observations have taken place that were coordinated with X-ray observatories, but the results remain to be examined.

\subsection{Tracks for future models} \label{sec:future:models}

Future modeling efforts are still required to account for the features observed in Vela X-1 and distinguish those that are found in other wind-fed HMXBs from those that are specific to this system. This will be achieved both by neatly tailoring the spatial scales together, from the stellar photosphere all the way down the accretion columns at the neutron magnetic poles, and by including additional physical ingredients at each level.

In Vela~X-1, the orbital eccentricity has long been considered as negligible by modelers. However, the increasingly improving performances reached by time-resolved spectroscopy invite us to investigate the effect it could have on the accretion flow (e.g., presence of a wind-captured disk, shape of the accretion wake, and mass transfer rate). Eccentricity induces a periodic modulation at the orbital period that might produce systematic differences as a function of the orbital phase. This is worth looking for in observations and simulations.

The gravitational pull of the neutron star is an essential factor in shaping the wind flow. Thus, improved simulations of the flow compared to observational diagnostics,  expanding on the approach of \citet{ManousakisWalterBlondin:2012} (see Sect.~\ref{sec:system:mass}), may be able to provide alternative, albeit model-dependent constraints on the masses of the system components.

Owing to the Roche potential and to its own spin, the donor star is not spherical, which means that the temperature of its photosphere is not uniform. For stars with a radiative envelope such as HD~77581, gravity darkening is well described by the von Zeipel law, which connects local effective gravity and temperature \citep{VonZeipel:24}. Given the high stellar filling factor, if the star were in synchronous rotation, the local effective temperature of the photosphere near the inner Lagrangian point could be up to $\sim$50\% lower than at the poles \citep{EspinosaLara:12}. Nonsynchronous rotation mitigates this effect, but given how dependent line-driven winds are on the radiative field, the local mass-loss rate probably varies throughout the stellar photosphere \citep{Friend:82,Hadrava:12,Cechura:15}. An accurate description of the 3D wind-launching mechanism requires taking the nonuniform conditions in the photosphere into account, along with clump formation. This research track echoes the questions brought up by the multiple evidence in favor of tidally induced nonradial oscillations of the stellar photosphere.

Another aspects this relates to is the pressing need for multidimensional models of radiative transfer for UV (e.g., for resonant line-absorption of stellar photons by the outflowing wind) and for X-rays (e.g., for photoionization and for the altered conditions in the X-ray irradiated stellar photosphere). The computational cost of radiative transfer in 3D is very high, however, therefore it does not seem feasible, even in the longer term, to fully couple radiation and matter in time-dependent radiative-hydrodynamics simulations without resorting to simplifying assumptions. Laboratory X-ray spectroscopic data will be an invaluable compass to determine which hypotheses are the least harmful \citep{Hell:16}. Notwithstanding the radiative feedback on the flow dynamics, radiative computations could be performed on simulation snapshots obtained by using an elementary representation of radiation-matter coupling (e.g., for wind launching and for the inhibited line-acceleration beyond a certain critical ionization parameter). It will be of tremendous importance to interpret the absorbing column density observed by the upcoming generation of X-ray instruments. The process could even be iterated to obtain reasonably self-consistent radiative-hydrodynamics simulations.

In the light of the upcoming X-ray polarimetry instruments, another promising perspective is the possibility of gaining insights into the geometry and properties of the system based on detailed modeling of polarized radiation \citep[see, e.g.,][further described in Sect.~\ref{sec:obs:xray:polarization}]{Kallman:2015, Caiazzo+Heyl:2021a}.   Future studies may build on these approaches, considering also the effect of the emitted X-rays on the ambient medium in multiscale X-ray radiative transfer computations in order to provide an accurate view of the polarization of the emitted radiation and the system parameters driving its observed variations. Polarization signatures like this can be probed with upcoming missions such as \textit{IXPE} and \textit{eXTP} (see Sect.~\ref{sec:future:multiwavelength}).

In order to obtain an accurate description of the accretion process itself, we eventually need better magnetohydrodynamics models, guided by the observed pulse profiles and the changing neutron star spin period. The latter encapsulates the information on the coupling between the accreted plasma and the magnetosphere, while the former tells us about the X-ray emission and scattering processes in the immediate vicinity of the neutron star surface. Regarding the accretion-induced torques, significant progress has been made for disk-magnetosphere coupling around millisecond pulsars \citep{Parfrey:17} and around T~Tauri stars and magnetized white dwarfs in cataclysmic variables \citep{Romanova:03a,Zanni:09}. They pave the way for a future model adapted to a system in which the accretor is a fast-spinning highly magnetized neutron star and in which wind-captured disks are transient at best. Alternatively, the quasi-spherical subsonic settling models introduced by \cite{Shakura:2012} might help to connect the evolution of the X-ray luminosity and of the spin period of the neutron star. 

In order to connect X-ray luminosity and mass accretion rate, the accretion column model needs to be compared to the pulse profiles and their variations in Vela X-1, however. The complex pulse profile structure and the wealth of observational data at various luminosity levels promise significant diagnostic power, first to distinguish plausible emission geometries, and in a further step, to connect observed short-term variations to changes in the emission regions and/or in the intervening material.

\subsection{Key parameters for future studies}
\label{sec:future:keypar}

\begin{table*}
\renewcommand{\arraystretch}{1.1}
    \centering
    \begin{small}
    \caption{Key parameters of the Vela~X-1 system for future studies. Because the scatter in the values found in the literature is often significant, we refrain from giving specific values here, but refer to the discussion in the corresponding section(s) and provide a rough assessment of the robustness of the knowledge of the parameters. We also note sections describing system elements, observational diagnostics or modeling, where these parameters are relevant.}
    \label{tab:keypar}
\begin{tabular}{llll}  
\hline\hline
Parameter & Current knowledge   & See also sections  & Robustness, uncertainty \\
\hline
\multicolumn{4}{c}{\textit{System}} \\
\hline
Distance & Sect.~\ref{sec:system:distance} & \ref{sec:obs:xray:spectrum}, \ref{sec:obs:uvopt:spectra} & Quite robust, known to $\sim$5\% \\
Orbital phase parameters  ($T_\textrm{ecl}$ or $T_{\pi/2}$, $P_\textrm{orb}$) & Sect.~\ref{sec:system:orbit}, Tab.~\ref{tab:orbit} & \ref{sec:obs:xray:abs}, \ref{sec:obs:xray:flux}, \ref{sec:obs:uvopt:photometry}, \ref{sec:obs:uvopt:radial}, \ref{sec:obs:uvopt:uv_resonance} & Very robust, uncertainty $<$1\% \\
Eccentricity $e$, longitude of periastron $\omega$ &  Sect.~\ref{sec:system:orbit}, Tab.~\ref{tab:orbit} & \ref{sec:obs:uvopt:photometry}, \ref{sec:models:mass_transf}, \ref{sec:abs_var}, \ref{sec:int_var} & Robust, known to $\sim$1\%,  \\
Inclination $i$ & Sect.~\ref{sec:system:mass}, Tab.~\ref{tab:masses} & \ref{sec:elements:mass_transfer}, \ref{sec:models:mass_transf}, Fig.~\ref{fig:nsmass} & Within $\sim$20$^\circ$ wide interval\\
\hline
\multicolumn{4}{c}{\textit{HD~77581}} \\
\hline
Stellar mass  $M_\mathrm{opt}$ & Sect.~\ref{sec:system:mass}, Tab.~\ref{tab:masses} & \ref{sec:obs:uvopt:radial}, \ref{sec:models:evolution} & Systematics often 10–-20\%  \\
Stellar radius $R_\mathrm{opt}$ & Sect.~\ref{sec:system:mass}, Tab.~\ref{tab:masses} & \ref{sec:elements:starwind}, \ref{sec:models:mass_transf} & Known to $<$10\% \\
Eff. temperature $T_\star$ and surface gravity $g_\star$ & Sect.~\ref{sec:system:chorizos}, Tab.~\ref{tab:stellarpar} & \ref{sec:elements:starwind}, \ref{sec:obs:uvopt:spectra} & Dep.\ on spectral model codes \\
Mass-loss rate $\dot{M}$ & Sect.\,\ref{sec:system:wind}, Tab.~\ref{tab:wind} & \ref{sec:elements:starwind}, \ref{sec:obs:radio}, \ref{sec:abs_var}, \ref{sec:models:evolution}  & Dep.\ on distance \& spectral models \\
Terminal wind velocity $\varv_\infty$ & Sect.\,\ref{sec:system:wind}, Tab.~\ref{tab:wind} & \ref{sec:elements:starwind}, \ref{sec:obs:uvopt:spectra}, \ref{sec:obs:radio}  & Known to $\sim$5\%, phase-dependent \\
\hline
\multicolumn{4}{c}{\textit{Neutron star}} \\
\hline
Mass $M_\mathrm{NS}$ & Sect.~\ref{sec:system:mass}, Tab.~\ref{tab:masses}, Fig.~\ref{fig:nsmass} & \ref{sec:elements:mass_transfer}, \ref{sec:elements:column}, \ref{sec:obs:uvopt:radial}, \ref{sec:models:mass_transf} & Only known to a few 10\% \\
Magnetic field $B$ & Sect.~\ref{sec:system:B-field} & \ref{sec:elements:magnetosphere}, \ref{sec:elements:column}, \ref{sec:obs:xray:cyclo}, \ref{sec:models:acc-ind_torq}, \ref{sec:models:CRSF} & $\sim$10\% in CRSF formation zone\\
\hline
\end{tabular}
\end{small}
\renewcommand{\arraystretch}{1}
\end{table*}

Finally, we give an overview of key parameters of the \mbox{Vela~X-1} system that it would be useful to study and constrain further, in our opinion. It is important to note that these parameters are typically not independent of each other. Refining existing determinations or having independent constraints for any of these parameters will affect the ranges of several other parameters and thus lead to a better-defined baseline for the physics of the system. The parameters are summarized in Tab.~\ref{tab:keypar}.

The distance to the Vela~X-1 system is now better determined based on \textit{Gaia}~EDR3 data and consistent with other methods (Sect.~\ref{sec:system:distance}). Still, the range determined in this work leads to a $\sim$30\% uncertainty in absolute luminosities just from the distance factor.

More recent orbital ephemerides agree very well in general, but there is a slight difference between the last detailed values in the literature \citep{Kreykenbohm:2008,Falanga:2015} that is significant at the quoted uncertainties (Sect.~\ref{sec:system:orbit}, Tab.~\ref{tab:orbit}). An updated ephemeris with a recent zero-point might resolve this difference and would avoid the systematic uncertainty from extrapolating over many years.
The orbital eccentricity, which is frequently ignored in modeling efforts, plays an important role for the varying shape of the mass donor (Sect.~\ref{sec:obs:uvopt:photometry}) and the mass transfer in this system, where the donor star is close to filling its Roche lobe (Sect.~\ref{sec:models:mass_transf}). For an exact description of these effects, the longitude of periastron $\omega$ also needs to be known well.

While it is well-constrained for an X-ray binary, the inclination $i$ of the system remains a strong source of uncertainty for the determination of absolute values for the masses of the two partners and the radius of HD~77581 (Sect.~\ref{sec:system:mass}, Tab.~\ref{tab:masses}). Conversely, if  the supergiant size could be tightly constrained, the inclination and thus other system parameters would be better constrained.

The stellar parameters may also be improved in two ways. First, through detailed SED modeling of high-resolution spectroscopic data, which would yield more accurate and precise values of the effective temperature and gravity and would investigate possible composition anomalies. Second, using well-calibrated photometry and/or spectrophotometry to independently estimate the effective temperature from the Balmer jump.

With its multiple constraints from the orbit on the one hand and the spectra and photometric data on the other hand, HD~77581 acts as a testbed for our understanding of the winds and mass loss of hot, massive stars. To launch a stellar wind, the ratio of luminosity to mass as well as the radius or the effective temperature of HD~77581 are vital ingredients. With the considerable change in the presumed luminosity of the donor star due to the revised spectral classification, the absolute mass-loss rate needs to be redetermined, while the terminal wind velocity must be known much better due to its direct spectral imprint. New quantitative studies with hydrodynamically consistent atmosphere models building on these updated values are required to determine whether a coherent solution for the observed appearance and the current concept of radiation-driven winds for OB stars can be found. The solution for HD~77581 will also mark an important constraint for the still highly uncertain winds of binary evolution products.

Because Vela~X-1 is frequently quoted as an example of a heavy neutron star, having a firmer constraint of $M_\mathrm{NS}$ would be of high interest beyond the studies of accreting X-ray pulsars per se. As Tab.~\ref{tab:masses} and Fig.~\ref{fig:nsmass} in Sect.~\ref{sec:system:mass} demonstrate, the effective uncertainty of comparing different determinations remains high, however, and the values depend on other assumptions about the system, especially the inclination $i$.

The magnetic field of the neutron star is another key parameter that drives the accretion physics (Sect.~\ref{sec:elements:magnetosphere}) and the physics of the emission region (Sect.~\ref{sec:elements:column}), among other elements. While a good estimate of the magnetic field strength in the main X-ray emission region exists from cyclotron resonance scattering features (Sect.~\ref{sec:obs:xray:cyclo}) constrained to about 10\%, it is not clear if this value can be trivially extrapolated to apply everywhere based on a simple dipole geometry, however.

\section{Summary}\label{sec:summary}
In this extensive review of a well-known X-ray binary system that shares many properties with other often less well studied systems, we have first described the essential elements and basic physics of this archetypical system. Second, we summarized the observational diagnostics and results from the radio to the hard X-rays, also including caveats that need to be considered. Third, we discussed the options and challenges of modeling this X-ray binary in detail and reviewed the model efforts untertaken so far. Fourth, we summarized the known properties of the system, which often have a larger spread in their reported values than is commonly assumed, including essential parameters such as the neutron star mass or the velocity of the stellar wind. Fifth, we demonstrated that the system may be at least close to RLOF at periastron and that the eccentricity is high enough to imply a significant variation of mass transfer with orbital phase. Sixth, we derived an updated distance estimate to the \vel system based on the latest \textit{Gaia} Data Release\footnote{\textit{Gaia} EDR3, \url{https://www.cosmos.esa.int/web/gaia/release}}. It has tighter margins than before. This distance estimate is closer to previous estimates and corrects for the systematic shift that affects the distance given in \citet{Bailer-Jones:2018}. Seventh, we revisited the proposed connection to the Vela~OB1 association and derived an updated \textit{Gaia} distance estimate to this association. Eight, we newly determined the stellar parameters of HD~77581, finding B0.2~Ia as the most probable spectral type and luminosity class. Ninth, we provided an overview of opportunities for future observations and modeling efforts to improve our knowledge of the physics of this specific system and by extrapolation of related, less well studied systems, together with an overview of especially relevant key parameters. We encourage similar efforts to group and discuss the actual knowledge about system parameters also for other well-known systems that are used as prototypes in the literature in order to create a reliable framework for future studies.

\begin{acknowledgements} 
We thank the referee for positive and succinct, but very pertinent comments that identified some remaining gaps in the presented material and led us to several additions to the text of this publication.

IEM has received funding from the Research Foundation Flanders (FWO), from the European Union's Horizon 2020 research and innovation program under the Marie Sk\l odowska-Curie grant agreement No 665501 and from the European Research Council (ERC) under the European Union’s Horizon 2020 research and innovation programme (grant agreement No 863412). SMN acknowledges funding by the Spanish Ministry MCIU under project RTI2018-096686-B-C21 (MCIU/AEI/FEDER, UE), co-funded by FEDER funds and by the Unidad de Excelencia María de Maeztu, ref. MDM-2017-0765. VG is supported through the Margarete von Wrangell fellowship by the ESF and the Ministry of Science, Research and the Arts Baden-W\"urttemberg. IEM and PK acknowledge support from the ESA/ESAC Faculty Visiting Scientist Programme. JMA acknowledges support from the Spanish Government Ministerio de Ciencia through grant PGC2018-\num{095049}-B-C22. FJE acknowledges support from ESCAPE - The European Science Cluster of Astronomy \& Particle Physics ESFRI Research Infrastructures, which received funding from the European Union's Horizon 2020 research and innovation programme under Grant Agreement no. 824064. JvdE is supported by a Lee Hysan Junior Research Fellowship awarded by St. Hilda's College. 
\\

The simulations were conducted on the Tier-1 VSC (Flemish Supercomputer Center funded by Hercules foundation and Flemish government).
\href{http://www.arizona-software.ch/graphclick/}{GraphClick} (\textcopyright\ Arizona Software) and \href{https://automeris.io/WebPlotDigitizer/}{WebPlotDigitizer} (\textcopyright\ Ankit Rohatgi)
have been used to digitise data from figures in older publications.\\

We thank Eva Laplace for very helpful comments on stellar evolution in binary systems. PK thanks Mercedes Hainzl for help in digitizing older data.
\end{acknowledgements}

\bibliographystyle{aa} 
\bibliography{VelaX1} 

\onecolumn
\begin{appendix}
\section{Appendix}

\begin{figure}[b!]
    \centering
    \includegraphics[width=0.9\linewidth]{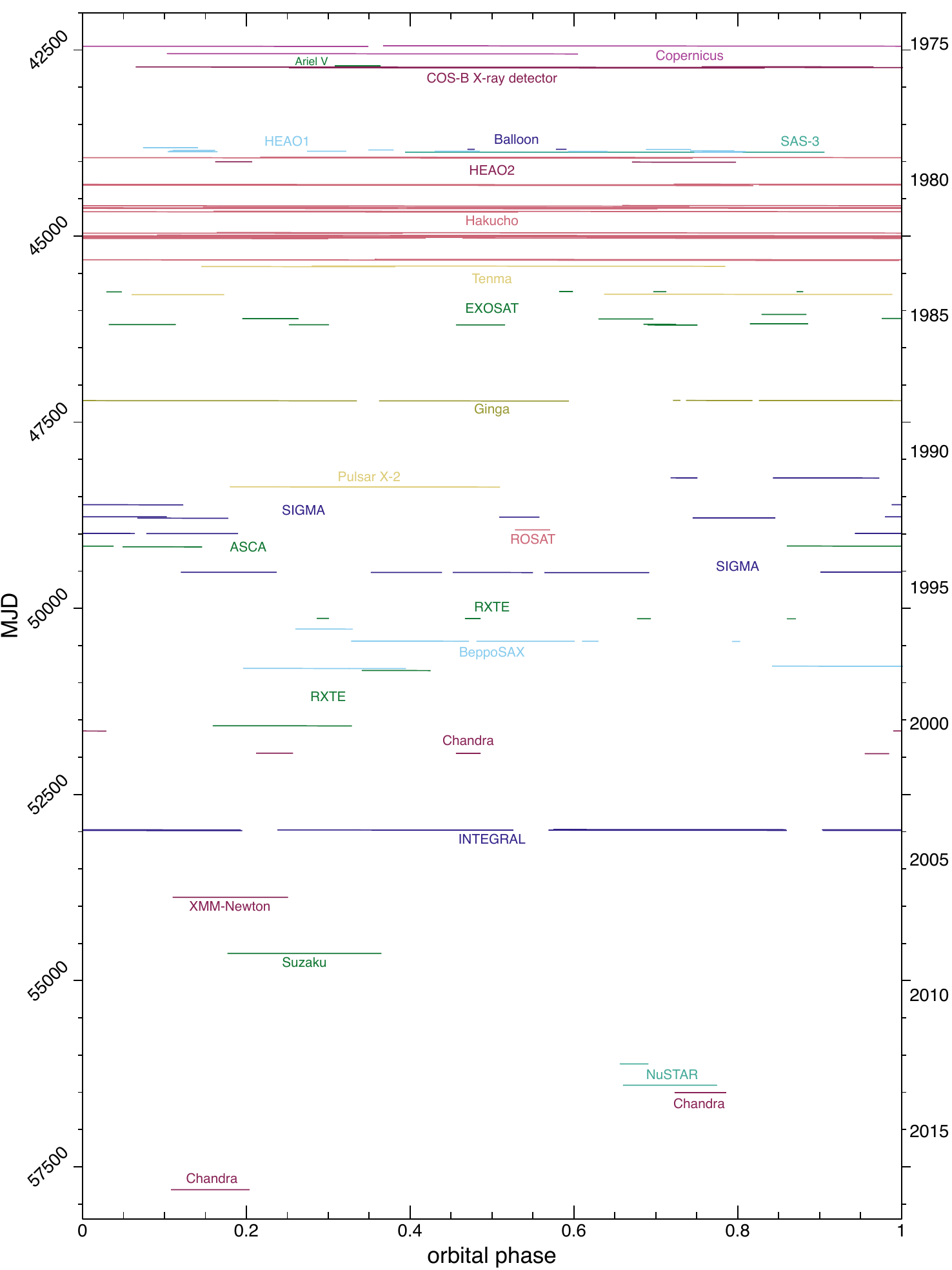}
    \caption{X-ray observations of \vel as reported in the literature and listed in Table~\ref{tab:observations_xray}.}
    \label{fig:obsphase}
\end{figure}
\clearpage
\setlength{\LTcapwidth}{0.85\textwidth}
\renewcommand{\arraystretch}{1.2}

\begin{longtable}{llrrrrr}
\caption{\label{tab:observations_xray}Selected X-ray observations of Vela~X-1 by different space missions as discussed in the literature, excluding X-ray monitors. The information is approximately ordered by time, and similar cases are grouped together. For observations covering multiple binary orbits, the information is split to show each consecutive data set within one orbital period. The orbital phase range here is calculated based on the mid-eclipse time in \citet{Kreykenbohm:2008}, see Table~\ref{tab:orbit}, and may accordingly differ from the values given in the original publications. The uncertainty in this conversion is typically 2--3\,$\times 10^{-3}$. The same observations have sometimes also been used in further publications that are not explicitly listed in this table. As examples, \textit{Tenma} observations have also been shown in \citet{Nagase:86}, while the \textit{Chandra} observations used by \citet{Goldstein:2004} have been also used in \citet{Watanabe:2006}, \citet{Grinberg:2017}, and \citet{Rahin+Behar:2020}. }\\
\hline\hline
Reference                   & Mission          & \multicolumn{1}{c}{Energy range}       & \multicolumn{1}{l}{MJD}      & \multicolumn{1}{l}{MJD}      & \multicolumn{2}{c}{orbital phase} \\
 &  & \multicolumn{1}{c}{[keV]} & \multicolumn{1}{l}{start} & \multicolumn{1}{l}{end} & \multicolumn{1}{c}{start} & \multicolumn{1}{c}{end} \\
\hline
\endfirsthead
\caption{Selected X-ray observations of Vela~X-1 -- continued.}\\
\hline\hline
Reference                   & Mission          & \multicolumn{1}{c}{Energy range}       & \multicolumn{1}{l}{MJD}      & \multicolumn{1}{l}{MJD}      & \multicolumn{2}{c}{orbital phase} \\
 &  & \multicolumn{1}{c}{[keV]}  & \multicolumn{1}{l}{start} & \multicolumn{1}{l}{end} & \multicolumn{1}{c}{start} & \multicolumn{1}{c}{end} \\
\hline\endhead
\hline
\endfoot
\small
\citet{Charles:1978}   & \textit{Copernicus}   & 2.5–-7.5           & 42444.40 & 42450.07 & 0.367 & 1.000  \\
                       &                       &                    & 42450.07 & 42453.20 & 0.000 & 0.349  \\
                       &                       &                    & 42549.60 & 42554.10 & 0.103 & 0.605  \\
                       \cline{2-7}
                       & \textit{Ariel V}      & 0.3--40            & 42712.80 & 42713.30 & 0.308 & 0.364  \\
\hline
\citet{Ogelman:77}     & \textit{COS-B} X-ray detector  & 2--12     & 42725.78 & 42727.66 & 0.756 & 0.966  \\
                            &                  &                    & 42728.55 & 42736.95 & 0.065 & 1.002  \\
                            &                  &                    & 42739.19 & 42744.40 & 0.252 & 0.833  \\
\hline
\citet{Staubert:80}    & Balloon X-ray detector & 15--150           & 43834.80 & 43834.88 & 0.470 & 0.479  \\
                            &                  &                    & 43835.76 & 43835.88 & 0.578 & 0.591  \\
\hline
\citet{Boynton:86}     & \textit{HEAO-1}/A-1 module 3 & 0.7--43.3       & 43813.32 & 43813.92 & 0.074 & 0.141  \\
                       & \textit{HEAO-1}/A-1 module 3 & 0.7--47.5       & 43836.75 & 43837.24 & 0.688 & 0.743  \\
                       & \textit{HEAO-1}/A-2 MED, HED & 2--20, 2--60    & 43842.68 & 43842.95 & 0.349 & 0.380 \\
                       & \textit{HEAO-1}/A-1 module 7 & 0.1--18.9       & 43849.50 & 43849.96 & 0.110 & 0.162 \\
                       & \textit{HEAO-1}/A-2 MED, HED & 2--20, 2--60    & \multicolumn{4}{c}{-- ditto --}      \\          
                       & \textit{HEAO-1}/A-2 MED, HED & 2--20, 2--60    & 43855.16 & 43855.65 & 0.742 & 0.796 \\
                       & \textit{HEAO-1}/A-2 MED, HED & 2--20, 2--60    & 43858.43 & 43858.90 & 0.106 & 0.159 \\
                       & \textit{HEAO-1}/A-1 module 7 & 0.1--18.9       & \multicolumn{4}{c}{-- ditto --}       \\ 
                       & \textit{HEAO-1}/A-2 MED, HED & 2--20, 2--60    & 43859.93 & 43860.36 & 0.274 & 0.322 \\
                       & \textit{HEAO-1}/A-2 MED, HED & 2--20, 2--60    & 43861.33 & 43861.82 & 0.430 & 0.485 \\
                       & \textit{HEAO-1}/A-1 module 7 & 0.1--18.9       & 43861.71 & 43861.76 & 0.472 & 0.478 \\
                       & \textit{HEAO-1}/A-2 MED, HED & 2--20, 2--60    & 43862.79 & 43863.22 & 0.593 & 0.641 \\
                       & \textit{HEAO-1}/A-1 module 7 & 0.1--18.9       & \multicolumn{4}{c}{-- ditto --}       \\ 
                       & \textit{HEAO-1}/A-2 MED, HED & 2--20, 2--60    & 43864.26 & 43864.74 & 0.757 & 0.810 \\
                       & \textit{HEAO-1}/A-1 module 7 & 0.1--18.9       & \multicolumn{4}{c}{-- ditto --}       \\ 
                       & \textit{HEAO-1}/A-2 MED, HED & 2--20, 2--60    & 43867.37 & 43867.92 & 0.104 & 0.165 \\
                       & \textit{HEAO-1}/A-1 module 7 & 0.1--18.9       & \multicolumn{4}{c}{-- ditto --}       \\ 
                       & \textit{HEAO-1}/A-2 MED, HED & 2--20, 2--60    & 43873.14 & 43873.67 & 0.747 & 0.806 \\
                       \cline{2-7}
                       & \textit{SAS-3} HTC, XTC &  1--8, 8--35  & 43887.90 & 43892.49 & 0.394 & 0.906 \\
\hline
\citet{Nagase:84ApJ}        & \textit{Hakucho} &        1--22    & 43940.10 & 43947.12 & 0.217 & 1.000  \\
                            &                  &                    & 43947.12 & 43953.80 & 0.000 & 0.745  \\
                            &                  &                    & 44303.20 & 44305.69 & 0.722 & 1.000  \\
                            &                  &                    & 44305.69 & 44313.00 & 0.000 & 0.815  \\
                            &                  &                    & 44313.10 & 44314.66 & 0.826 & 1.000  \\
                            &                  &                    & 44314.66 & 44322.00 & 0.000 & 0.819  \\
                            &                  &                    & 44589.50 & 44592.55 & 0.659 & 1.000  \\
                            &                  &                    & 44592.55 & 44599.20 & 0.000 & 0.741  \\
                            &                  &                    & 44611.80 & 44619.45 & 0.147 & 1.000  \\
                            &                  &                    & 44619.45 & 44622.90 & 0.000 & 0.385  \\
                            &                  &                    & 44623.00 & 44628.41 & 0.396 & 1.000  \\
                            &                  &                    & 44628.41 & 44634.70 & 0.000 & 0.702  \\
                            &                  &                    & 44665.70 & 44673.23 & 0.160 & 1.000  \\
                            &                  &                    & 44673.23 & 44678.00 & 0.000 & 0.532  \\
                            &                  &                    & 44952.60 & 44960.09 & 0.164 & 1.000  \\
                            &                  &                    & 44960.09 & 44963.60 & 0.000 & 0.391  \\
                            &                  &                    & 44987.80 & 44995.95 & 0.091 & 1.000  \\
                            &                  &                    & 44995.95 & 44998.80 & 0.000 & 0.318  \\
                            &                  &                    & 44999.60 & 45004.91 & 0.407 & 1.000  \\
                            &                  &                    & 45004.91 & 45008.60 & 0.000 & 0.411  \\
                            &                  &                    & 45008.70 & 45013.88 & 0.422 & 1.000  \\
                            &                  &                    & 45013.88 & 45018.40 & 0.000 & 0.504  \\
                            &                  &                    & 45019.40 & 45022.84 & 0.616 & 1.000  \\
                            &                  &                    & 45022.84 & 45026.60 & 0.000 & 0.419  \\
                            &                  &                    & 45027.00 & 45031.81 & 0.464 & 1.000  \\
                            &                  &                    & 45031.81 & 45034.50 & 0.000 & 0.300  \\
                            &                  &                    & 45312.90 & 45318.67 & 0.357 & 1.000  \\
                            &                  &                    & 45318.67 & 45327.60 & 0.000 & 0.997  \\        
\hline
\citet{KallmanWhite:82}     & \textit{HEAO-2} / &         0.5--4 /  & 44002.38 & 44001.99 & 0.162 & 0.207 \\
                            & SSS+MPC           &         2--10  & 44006.92 & 44008.06 & 0.671 & 0.798 \\
\hline
\citet{Ohashi:84}           & \textit{Tenma} GSPC  &     1--35   & 45401.85 & 45406.39 & 0.280 & 0.785 \\
                            &                  &                    & 45409.61 & 45411.73 & 0.145 & 0.382 \\
\hline
\citet{Sato:86PASJ}             & \textit{Tenma} GSPC  & 1--35   & 45781.56 & 45784.72 & 0.637 & 0.989 \\
                            &                  &                    & 45785.35 & 45786.37 & 0.060 & 0.173 \\
\hline
\citet{HaberlWhite:90}      & \textit{EXOSAT} ME & 1--50        & 45745.21 & 45745.36 & 0.582 & 0.599 \\
                            &                  &                   & 45746.24 & 45746.38 & 0.697 & 0.713 \\
                            &                  &                   & 45747.81 & 45747.88 & 0.872 & 0.880 \\
                            &                  &                   & 45749.22 & 45749.38 & 0.029 & 0.048 \\
                            &                  &                   & 46052.21 & 46052.70 & 0.829 & 0.884 \\
                            &                  &                   & 46107.31 & 46107.72 & 0.976 & 1.021 \\
                            &                  &                   & 46109.28 & 46109.90 & 0.195 & 0.264 \\
                            &                  &                   & 46113.17 & 46113.78 & 0.630 & 0.697 \\
                            &                  &                   & 46177.58 & 46178.23 & 0.815 & 0.886 \\
                            &                  &                   & 46185.38 & 46185.75 & 0.685 & 0.725 \\
                            &                  &                   & 46188.50 & 46189.23 & 0.032 & 0.114 \\
                            &                  &                   & 46190.47 & 46190.91 & 0.252 & 0.301 \\
                            &                  &                   & 46192.30 & 46192.83 & 0.456 & 0.516 \\
                            &                  &                   & 46194.39 & 46194.94 & 0.690 & 0.751 \\
\hline
\citet{Lewis:92}            & \textit{Ginga}   & 1.5--37        & 47207.65 & 47207.73 & 0.721 & 0.730  \\
                            &                  &                   & 47207.79 & 47208.51 & 0.737 & 0.818  \\
                            &                  &                   & 47208.58 & 47210.15 & 0.826 & 1.000  \\
                            &                  &                   & 47209.90 & 47213.15 & 0.000 & 0.335  \\
                            &                  &                   & 47213.39 & 47215.47 & 0.362 & 0.594  \\
\hline
\citet{Loznikov:92a}        & \textit{Pulsar X-2} & 2--25       & 48368.16 & 48371.11 & 0.180 & 0.510  \\
\hline
\citet{Haberl:94a}          & \textit{ROSAT}   & 0.1--2.4       & 48945.00 & 48945.39 & 0.528 & 0.571  \\  
\hline
\citet{Nagase:94}           & \textit{ASCA}    & 1--10          & 49163.12 & 49164.38 & 0.860 & 1.000  \\
                            &                  &                & 49164.38 & 49164.72 & 0.000 & 0.038  \\
                            &                  &                & 49173.78 & 49174.65 & 0.049 & 0.146  \\
\hline
\citet{Laurent:95}          & \textit{SIGMA}   & 30--1300       & 48247.48 & 48247.78 & 0.718 & 0.751 \\
                            &                  &                & 48248.60 & 48249.77 & 0.843 & 0.973 \\
                            &                  &                & 48608.48 & 48608.59 & 0.988 & 1.000 \\
                            &                  &                & 48608.59 & 48609.69 & 0.000 & 0.123 \\
                            &                  &                & 48769.76 & 48769.94 & 0.980 & 1.000 \\
                            &                  &                & 48769.94 & 48770.87 & 0.000 & 0.103 \\
                            &                  &                & 48774.51 & 48774.95 & 0.509 & 0.558 \\
                            &                  &                & 48785.59 & 48786.49 & 0.745 & 0.846 \\
                            &                  &                & 48788.47 & 48789.47 & 0.067 & 0.178 \\
                            &                  &                & 48993.54 & 48994.05 & 0.943 & 1.000 \\
                            &                  &                & 48994.05 & 48994.63 & 0.000 & 0.064 \\
                            &                  &                & 48994.75 & 48995.76 & 0.078 & 0.190 \\
                            &                  &                & 49513.10 & 49513.99 & 0.901 & 1.001 \\
                            &                  &                & 49515.06 & 49516.11 & 0.120 & 0.237 \\
                            &                  &                & 49517.14 & 49517.92 & 0.352 & 0.439 \\
                            &                  &                & 49518.04 & 49518.92 & 0.452 & 0.550 \\
                            &                  &                & 49519.04 & 49520.19 & 0.564 & 0.692 \\
\hline
\citet{Kreykenbohm:99}      & \textit{RXTE}    & 2--30 /           & 50135.09 & 50135.23 & 0.286 & 0.301  \\
                            &                  & 20--200           & 50136.71 & 50136.88 & 0.467 & 0.486  \\
                            &                  &                   & 50138.60 & 50138.74 & 0.677 & 0.694  \\
                            &                  &                   & 50140.24 & 50140.34 & 0.860 & 0.871  \\
\hline
\citet{Kretschmar:CGRO99}   & \textit{RXTE}    &                   & 50834.81 & 50835.56 & 0.341 & 0.425  \\
\hline
\citet{Kreykenbohm:2002}    & \textit{RXTE}    &                   & 51577.21 & 51578.74 & 0.159 & 0.329  \\
\hline
\citet{LaBarbera:2003}      & \textit{BeppoSAX} & 0.1--200      & 50278.29 & 50278.91 & 0.260 & 0.330  \\
                            &                  &                   & 50440.25 & 50440.52 & 0.328 & 0.358  \\
                            &                  &                   & 50440.47 & 50441.54 & 0.352 & 0.472  \\
                            &                  &                   & 50441.63 & 50442.70 & 0.481 & 0.601  \\
                            &                  &                   & 50442.78 & 50442.96 & 0.610 & 0.630  \\
                            &                  &                   & 50444.43 & 50444.52 & 0.793 & 0.803  \\
                            &                  &                   & 50776.55 & 50777.98 & 0.842 & 1.002  \\
                            &                  &                   & 50806.61 & 50807.77 & 0.196 & 0.326  \\
                            &                  &                   & 50807.76 & 50808.39 & 0.325 & 0.395  \\ 
\hline
\citet{Goldstein:2004}      & \textit{Chandra} & 0.2--10           & 51945.23 & 51945.63 & 0.212 & 0.257  \\
                            &                  &                   & 51947.41 & 51947.68 & 0.456 & 0.486  \\
                            &                  &                   & 51951.89 & 51952.16 & 0.955 & 0.985  \\
\hline
\citet{Kreykenbohm:2008}    & \textit{INTEGRAL}/ISGRI & 20--100    & 52970.41 & 52972.97 & 0.575 & 0.859  \\
                            &                  &                   & 52973.36 & 52974.23 & 0.903 & 1.000  \\
                            &                  &                   & 52974.23 & 52975.96 & 0.000 & 0.193  \\
                            &                  &                   & 52976.36 & 52978.95 & 0.238 & 0.526  \\
                            &                  &                   & 52979.33 & 52981.94 & 0.569 & 0.860  \\
                            &                  &                   & 52982.33 & 52983.19 & 0.904 & 1.000  \\
                            &                  &                   & 52983.19 & 52984.94 & 0.000 & 0.195  \\ 
\hline
\citet{Martinez-Nunez:2014} & \textit{XMM-Newton}  & 0.15--12      & 53880.61 & 53881.88 & 0.110 & 0.251 \\
\hline
\citet{Fuerst:2014}         & \textit{NuSTAR}  & 3--79             & 56117.63 & 56117.94 & 0.656 & 0.691 \\
                            &                  &                   & 56404.53 & 56405.56 & 0.660 & 0.775 \\
\hline
\citet{Doroshenko:2011}     & \textit{Suzaku}  & 0.2--12, 10--600  & 54634.22 & 54635.90 & 0.177 & 0.365 \\
\hline
\citet{Liao:2020}           & \textit{Chandra} & 0.2--10           & 57806.98 & 57807.85 & 0.108 & 0.204 \\
\hline
\citet{Rahin+Behar:2020}    & \textit{Chandra} & 0.2--10           & 51647.41 & 51647.50 & 0.990 & 1.000 \\
\textit{also used previously} &                &                   & 51647.50 & 51647.76 & 0.000 & 0.029 \\
\textit{listed Chandra data}         &                  &                   & 56503.70 & 56504.27 & 0.723 & 0.786 \\
\end{longtable}

\newpage

\setlength{\LTcapwidth}{0.68\textwidth}

\begin{longtable}{lrrl}
\caption{\label{tab:pulse_periods}Pulse-period estimates taken from the literature, especially \protect{\citet{Nagase:89}}. The date and duration of the observation is indicated together with the derived pulse period.}\\ 
\hline\hline
MJD       & Duration [d] & Pulse Period [s] & References \\
\hline
\endfirsthead
\caption{continued.}\\
\hline\hline
MJD       & Duration [d] & Pulse Period [s] & References \\
\hline
\endhead
\hline
\endfoot
42448.8   &  8.8         & 282.9083(12)     & \citet{Charles:1978}\\ 
42551.8   &  4.5         & 282.9010(37)     & \citet{Charles:1978} \\ 
42600.1   & 36.4         & 282.8916(2)      & \citet{Rappaport:76}\\ 
42713.0   & 0.5          & 282.937(22)      & \citet{Charles:1978}\\ 
42727.9   & 21.0         & 282.9108(12)     & \citet{Ogelman:77}\\ 
42749.8   & 10.0         & 282.919(3)       & \citet{Becker:78} \\
42899.5   & 3.0          & 282.870(4)       & \citet{Rappaport+Joss:77Accretion}\\ 
42919.5   & 14.0         & 282.869(3)       & \citet{Becker:78}\\ 
42996.6   & 28.0         & 282.8183(3)      & \citet{vanderKlisBonnet-Bidaud:84}\\ 
43111.5   & 4.0          & 282.838(12)      & \citet{Becker:78}\\ 
43639.51  & 3.18         & 282.80452(169)   & \citet{Deeter:89}\\ 
43645.12  & 3.65         & 282.80016(127)   & \citet{Deeter:89}\\ 
43650.54  & 5.15         & 282.79188(88)    & \citet{Deeter:89}\\ 
43653.5   & 3.0          & 282.787(4)       & \citet{Bautz:83}\\ 
43651.52  & 3.79         & 282.78806(127)   & \citet{Deeter:89} \\ 
43660.25  & 4.66         & 282.78884(99)    & \citet{Deeter:89} \\ 
43664.97  & 3.70         & 282.78899(127)   & \citet{Deeter:89} \\ 
43669.49  & 4.30         & 282.79104(110)   & \citet{Deeter:89} \\ 
43824.0   & 9.0          & 282.7484(4)      & \citet{Rappaport:80} \\ 
43825.28  & 23.5         & 282.74884(5)     & \citet{Deeter:89}\\ 
43836.3   & 1.0          & 282.80(4)        & \citet{Staubert:80}\\ 
43839.93  & 5.66         & 282.74606(27)    & \citet{Deeter:89} \\ 
43846.27  & 6.88         & 282.75233(2)     & \citet{Deeter:89} \\ 
43849.5   & 26.0         & 282.7513(5)      & \citet{Bautz:83} \\ 
43852.60  & 5.66         & 282.75604(19)    & \citet{Deeter:89} \\ 
43857.09  & 3.18         & 282.75640(37)    & \citet{Deeter:89} \\ 
43859.44  & 1.40         & 282.75881(87)    & \citet{Deeter:89} \\ 
43860.88  & 1.34         & 282.75240(108)   & \citet{Deeter:89} \\ 
43862.25  & 1.28         & 282.74948(94)    & \citet{Deeter:89} \\ 
43863.71  & 1.52         & 282.75004(68)    & \citet{Deeter:89} \\ 
43866.09  & 3.12         & 282.74611(29)    & \citet{Deeter:89} \\ 
43871.00  & 5.60         & 282.75346(21)    & \citet{Deeter:89} \\ 
43881.14  & 15.56        & 282.74632(14)    & \citet{Deeter:89} \\ 
43890.80  & 2.50         & 282.75020(88)    & \citet{Deeter:89} \\ 
43947.6   & 13.7         & 282.7337(9)      & \citet{Nagase:84ApJ} \\ 
44160.9   & 35.0         & 282.7809(3)      & \citet{vanderKlisBonnet-Bidaud:84} \\ 
44308.2   & 9.8          & 282.7977(6)      & \citet{Nagase:84ApJ} \\ 
44317.6   & 8.9          & 282.8174(7)      & \citet{Nagase:84ApJ} \\ 
44594.4   & 9.7          & 282.8693(6)      & \citet{Nagase:84ApJ} \\ 
44617.4   & 11.1         & 282.8608(5)      & \citet{Nagase:84ApJ} \\ 
44628.8   & 11.7         & 282.8872(6)      & \citet{Nagase:84ApJ} \\ 
44671.9   & 12.3         & 282.9085(4)      & \citet{Nagase:84ApJ} \\ 
44958.1   & 11.0         & 282.9545(1)      & \citet{Nagase:84ApJ} \\ 
44993.3   & 11.0         & 282.9451(7)      & \citet{Nagase:84ApJ} \\ 
45004.1   & 9.0          & 282.9315(1)      & \citet{Nagase:84ApJ} \\ 
45013.5   & 9.7          & 282.9350(13)     & \citet{Nagase:84ApJ} \\ 
45023.0   & 7.2          & 282.9287(22)     & \citet{Nagase:84ApJ} \\ 
45031.3   & 8.5          & 282.9337(12)     & \citet{Nagase:84ApJ} \\ 
45320.3   & 14.7         & 282.9293(5)      & \citet{Nagase:84ApJ} \\ 
45403.5   & 14           & 282.9306(3)      & \citet{Nagase:84PASJ} \\ 
45745.79  & 0.205        & 282.9450(9)      & \citet{Raubenheimer:90} \\ 
45784.5   & 5            & 282.912(5)       & \citet{Sato:86PASJ}\\ 
46109.56  & 0.562        & 282.9494(2)      & \citet{Raubenheimer:90} \\ 
46113.46  & 0.575        & 282.9549(1)      & \citet{Raubenheimer:90} \\ 
46185.57  & 0.339        & 282.9433(5)      & \citet{Raubenheimer:90} \\ 
46190.64  & 0.330        & 282.9441(7)      & \citet{Raubenheimer:90} \\ 
46192.56  & 0.512        & 282.9383(3)      & \citet{Raubenheimer:90} \\ 
46194.47  & 0.155        & 282.9331(8)      & \citet{Raubenheimer:90} \\ 
46954.5   & 3            & 283.09(1)        & \citet{Tsunemi:89} \\ 
47011.5   & 1            & 283.10(1)        & \citet{Tsunemi:89} \\ 
47035.5   & 5            & 283.07(1)        & \citet{Tsunemi:89} \\ 
47211.0   & 0            & 283.134(2)       & \citet{Nagase:89} \\ 
47485.82  & 0.20         & 283.28(5)        & \citet{Kretschmar:Dr} \\ 
47490.79  & 0.14         & 283.23(7)        & \citet{Kretschmar:Dr} \\ 
47491.84  & 0.08         & 283.12(12)       & \citet{Kretschmar:Dr} \\ 
47560.80  & 0.14         & 283.0845(7)      & \citet{Kretschmar:Dr} \\ 
47906.0   & 15           & 283.230(22)      & \citet{Lapshov:92} \\ 
48149.0   & 15           & 283.244(22)      & \citet{Lapshov:92} \\ 
48269.74  & 0.08         & 283.33(12)       & \citet{Kretschmar:Dr} \\ 
48301.0   & 14           & 283.26(13)       & \citet{Lapshov:92} \\ 
48368.1575& --           & 283.2487(31)     & \citet{Loznikov:92a}     \\ 
50278.0   & 0.556        & 283.201(1)       & \citet{Doroshenko:2017PhD} \\ 
50440.0   & 1.111        & 283.201(1)       & \citet{Doroshenko:2017PhD} \\ 
50441.0   & 1.146        & 283.271(1)       & \citet{Doroshenko:2017PhD} \\ 
50776.0   & 1.157        & 283.5549(1)      & \citet{Doroshenko:2017PhD} \\ 
50806.0   & 0.88         & 283.452(1)       & \citet{Doroshenko:2017PhD} \\ 
50807.0   & 0.88         & 283.452(1)       & \citet{Doroshenko:2017PhD} \\ 
50835.2   & 0.75         & 283.5(1)         & \citet{Kreykenbohm:2002}\\ 
51547.0   & 1.5          & 283.2(1)         & \citet{Kreykenbohm:2002}\\ 
52816.5   & 19.6         & 283.535(5)       & \citet{Staubert:2004IWS5}\\ 
52975.5   & 11.2         & 283.510(5)       & \citet{Staubert:2004IWS5}\\ 
54635.0   & 1.157        & 283.473(4)       & \citet{Doroshenko:2011} \\
\hline

\end{longtable}
\twocolumn

\begin{table*}
\caption[]{Vela-OB1 candidate members identified using \textit{Gaia}-EDR3 parallax and proper motion (see Sect.~\ref{sec:system:VelaOB1}).} 
\label{tab:Vela-OB1}
\begin{center}
\begin{tabular}[t]{cccccc}
\hline\hline\noalign{\smallskip}
\textit{Gaia}   & \multicolumn{2}{c}{Coordinates (J2000)} & Parallax & \multicolumn{2}{c}{Proper motion (mas·yr$^{-1}$)}\\
Source ID   & RA & Dec & (mas) & pmRA & pmDE \\
\hline\noalign{\smallskip}\noalign{\smallskip}
  5327601751793718784 & 137.10034591579 & -46.25365769648 & 0.633$\pm$0.014 & -5.496$\pm$0.015 & 4.400$\pm$0.015\\
  5328797401980675968 & 133.56361754098 & -47.49191953770 & 0.531$\pm$0.012 & -4.821$\pm$0.013 & 4.707$\pm$0.015\\
  5328798673290708480 & 133.75184216422 & -47.41593934303 & 0.455$\pm$0.017 & -5.27$\pm$0.02   & 4.72$\pm$0.02  \\
  5328819529652913024 & 132.88932643565 & -47.57162297804 & 0.608$\pm$0.015 & -5.432$\pm$0.016 & 4.995$\pm$0.016\\
  5329458895666400896 & 132.18660123905 & -47.76335069579 & 0.524$\pm$0.015 & -5.365$\pm$0.017 & 4.216$\pm$0.017\\
  5329509159652062976 & 132.20963371443 & -47.29378761077 & 0.493$\pm$0.016 & -4.201$\pm$0.017 & 5.107$\pm$0.019\\
  5329827708782254208 & 130.82630499257 & -46.44818006673 & 0.524$\pm$0.014 & -5.054$\pm$0.016 & 5.310$\pm$0.016\\
  5329859598936534528 & 130.90267763785 & -46.10146886011 & 0.480$\pm$0.015 & -4.532$\pm$0.018 & 4.003$\pm$0.015\\
  5329870972004887040 & 131.86648644183 & -46.45107834877 & 0.54$\pm$0.04   & -5.14$\pm$0.05   & 4.79$\pm$0.04  \\
  5329890866290156288 & 131.82848720377 & -46.15539593457 & 0.70$\pm$0.04   & -4.85$\pm$0.04   & 4.58$\pm$0.04  \\
  5329944020807311616 & 131.62724792307 & -45.91248510017 & 0.69$\pm$0.05   & -4.83$\pm$0.06   & 4.55$\pm$0.06  \\
  5330299918969106560 & 134.23263882998 & -47.04595937430 & 0.454$\pm$0.012 & -5.142$\pm$0.016 & 4.541$\pm$0.013\\
  5331167124402148736 & 133.34164824874 & -46.03576523452 & 0.523$\pm$0.014 & -5.185$\pm$0.015 & 4.817$\pm$0.015\\
  5331419737203575040 & 132.81874875513 & -45.67629125935 & 0.437$\pm$0.014 & -5.020$\pm$0.015 & 5.050$\pm$0.016\\
  5331435370884547328 & 132.56209322263 & -45.52269184007 & 0.453$\pm$0.012 & -5.089$\pm$0.012 & 5.024$\pm$0.014\\
  5331660839484919936 & 133.01719704082 & -44.00969251854 & 0.55$\pm$0.03   & -6.14$\pm$0.03   & 4.03$\pm$0.03  \\
  5521670158315787392 & 129.77377322364 & -46.24593591797 & 0.590$\pm$0.013 & -5.813$\pm$0.013 & 5.157$\pm$0.015\\
  5521670227035251328 & 129.76737341168 & -46.22678981187 & 0.596$\pm$0.015 & -5.449$\pm$0.016 & 4.948$\pm$0.016\\
  5521670227035254400 & 129.76910829443 & -46.22945375949 & 0.575$\pm$0.013 & -5.911$\pm$0.014 & 5.292$\pm$0.015\\
  5521672799702366080 & 129.74528665690 & -46.22707533220 & 0.542$\pm$0.014 & -5.747$\pm$0.015 & 5.036$\pm$0.017\\
  5521677923616934784 & 129.41644439982 & -46.28268509352 & 0.56$\pm$0.02   & -5.53$\pm$0.03   & 4.94$\pm$0.02  \\
  5521903220412978816 & 129.23431322745 & -45.52168566852 & 0.564$\pm$0.017 & -5.891$\pm$0.017 & 5.329$\pm$0.017\\
  5522000733351648768 & 130.21214055581 & -46.09591100078 & 0.536$\pm$0.011 & -5.089$\pm$0.012 & 4.802$\pm$0.013\\
  5522096008597568000 & 130.48708723543 & -45.41069183378 & 0.60$\pm$0.07   & -5.79$\pm$0.07   & 4.49$\pm$0.07  \\
  5522106767507224064 & 130.30656847735 & -45.27720464294 & 0.461$\pm$0.016 & -4.882$\pm$0.015 & 4.505$\pm$0.017\\
  5522333026384059904 & 129.33115011865 & -45.20719231581 & 0.48$\pm$0.03   & -4.00$\pm$0.03   & 4.34$\pm$0.04  \\
  5522363091152539776 & 129.42223778955 & -44.60312825733 & 0.441$\pm$0.016 & -4.559$\pm$0.016 & 5.042$\pm$0.019\\
  5522723696605828864 & 128.60966713499 & -45.02101522891 & 0.464$\pm$0.015 & -4.564$\pm$0.016 & 4.489$\pm$0.017\\
  5523208576927237120 & 129.15461570352 & -44.08002274422 & 0.541$\pm$0.015 & -5.270$\pm$0.017 & 5.087$\pm$0.018\\
  5523825888285399296 & 132.71682631647 & -43.83967512344 & 0.456$\pm$0.011 & -5.289$\pm$0.011 & 4.362$\pm$0.011\\
\noalign{\smallskip}\hline
\end{tabular} 
\end{center}
\end{table*}

\end{appendix}

\end{document}